\documentclass[11pt,a4paper]{article} 
\pdfoutput=1
\usepackage{jcappub} 
\usepackage{subfig}
\usepackage{graphicx}% Include figurefiles
\newcommand{\eq}{\begin{eqnarray}}
\newcommand{\en}{\end{eqnarray}}

\newcommand{\ba}[1]{\begin{eqnarray} \label{(#1)}}
\newcommand{\ea}{\end{eqnarray}}

\newcommand{\newc}{\newcommand}
\newc{\lra}{\leftrightarrow}
\newc{\beq}{\begin{equation}}
\newc{\eeq}{\end{equation}}
\newc{\barr}{\begin{eqnarray}}
\newc{\earr}{\end{eqnarray}}
  
\title {Direct dark matter searches - Test of the Big Bounce Cosmology}
 
\author[1]{Yeuk-Kwan~E.~Cheung}

\author[1,2,3]{J.D. Vergados}

\affiliation[1]{Department of Physics, Nanjing University,
22 Hankou Road, Nanjing, China 210093} 

\affiliation[2]{TEI of Western Macedonia, Kozani, Gr 501 00,  Greece and KAIST, 291 Daehak-r0, Yuesong-gu, Daejeon 305-701, Republic of Korea}

\affiliation[3]{University of Ioannina, Ioannina, Gr 451 10, Greece}

\emailAdd{cheung@nju.edu.cn} 
\emailAdd{vergados@uoi.gr} 

\abstract{
We  consider  the possibility of using dark matter particle's mass and its interaction cross section as a smoking gun signal of the existence of a Big Bounce at the early stage in the evolution of our currently observed universe.   
A study of dark matter production in the pre-bounce contraction and the post bounce expansion epochs of this universe reveals a new venue
for achieving the observed relic abundance of our present universe.
Specifically, it predicts a characteristic relation governing a dark matter mass and interaction cross section and  a factor of $1/2$ in thermally averaged cross section, as compared to the non-thermal production in standard cosmology, is needed for creating enough dark matter particle to satisfy the currently observed relic abundance because  dark matter is being created during the pre-bounce contraction, in addition to the post-bounce expansion.
As the production rate is lower than the Hubble expansion rate  information of the bounce universe evolution is preserved. Therefore once  the value of dark matter mass and interaction cross section are obtained by direct detection in laboratories, this alternative route becomes a signature prediction of the bounce universe scenario. 
This leads us to consider  a scalar dark matter candidate, which if it is light, has important implications on dark matter searches.}

\dedicated{To the memory of Tan Lu (1932.2.23 -- 2014.12.3).}
\keywords{Dark matter, WIMP,  direct  detection,  Bounce Universe, WIMP-nucleus scattering, event rates, modulation, Debris Flows}
\arxivnumber{1410.5710} 
\begin{document} 
\maketitle 
%%%%%%%%%%%%%%%%%%%%%%%%%%%%%%%%%%%%%%%%%%%%%%%%%%%%%%%%%%%%%%%%%%%%%
%\include{matrix}
\section{Introduction}

The combined MAXIMA-1 \cite{MAXIMA1},\cite{MAXIMA2},\cite{MAXIMA3}, BOOMERANG \cite{BOOMERANG1},\cite{BOOMERANG2}
DASI \cite{DASI02} and COBE/DMR Cosmic Microwave Background (CMB)
observations \cite{COBE} imply that the Universe is flat
\cite{flat01}
%, $\Omega=1.11\pm0.07$
and that most of the matter in
the Universe is Dark \cite{SPERGEL},  i.e. exotic. These results have been confirmed and improved
by the recent WMAP  \cite{WMAP06} and Planck \cite{PlanckCP13} data. Combining 
the data of these quite precise measurements one finds:
$$\Omega_b=0.0456 \pm 0.0015, \quad \Omega _{\mbox{{\tiny CDM}}}=0.228 \pm 0.013 , \quad \Omega_{\Lambda}= 0.726 \pm 0.015~$$
%$$\Omega_b=0.05, \Omega _{CDM}= 0.25, \Omega_{\Lambda}= 0.70$$ 
(the more  recent Planck data yield a slightly different combination $ \Omega _{\mbox{{\tiny CDM}}}=0.274 \pm 0.020 , \quad \Omega_{\Lambda}= 0.686 \pm 0.020)$. It is worth mentioning that both the WMAP and the Plank observations yield essentially the same value of $\Omega_m h^2$,
% namely  $\Omega_m h^2=0.1423\pm 0.0029$,
  but they differ in the value of $h$, namely $h=0.704\pm0.013$ (WMAP) and $h=0.673\pm0.012$ (Planck).
Since any ``invisible" non exotic component cannot possibly exceed $40\%$ of the above $ \Omega _{\mbox{{\tiny CDM}}}$
~\cite {Benne}, exotic (non baryonic) matter is required and there is room for cold dark matter candidates or WIMPs (Weakly Interacting Massive Particles).\\
Even though there exists firm indirect evidence for a halo of dark matter
in galaxies from the
observed rotational curves, see e.g. the review \cite{UK01}, it is essential to directly
detect such matter in order to 
unravel the nature of the constituents of dark matter. 

The possibility of such detection, however, depends on the nature of the dark matter constituents and their interactions.
%can be simply described as a Majorana fermion, a linear
%combination of the neutral components of the gauginos and
%higgsinos
% \cite{ref2a},\cite{ref2b},\cite{ref2c},\cite{ref2},\cite{ELLROSZ},\cite{Gomez},\cite{ELLFOR}
%  \cite{GOODWIT}-\cite{ref2}.
%In most calculations the
%neutralino is assumed to be primarily a gaugino, usually a bino.
% \section{The Essential Theoretical Ingredients  of Direct Detection.}

Since the WIMP's are  expected to be
% very massive, $m_{\mbox{{\tiny WIMP}}} > 10$ GeV,  and
extremely non-relativistic, with average kinetic energy $\langle T\rangle  \approx
50 \ {\rm keV} (m_{\mbox{{\tiny WIMP}}}/ 100 \ {\rm GeV} )$, they are not likely to excite the nucleus.
So they can be directly detected mainly via the recoiling of a nucleus
(A,Z) in elastic scattering. The event rate for such a process can
be computed from the following ingredients~\cite{LS96}: 
i) The elementary nucleon cross section.
% This most important parameter will not, however, be the subject of the present work. We will adopt the view that it   can be extracted from the data of event rates, if and when such data become available. From limits on the event rates, one can obtain exclusion plots on the nucleon cross sections as  functions of the WIMP mass. 
ii) knowledge of the relevant nuclear matrix elements
 obtained with as reliable as possible many
body nuclear wave functions, 
iii) knowledge of the WIMP density in our vicinity and its velocity distribution.
%In the present work we will considera Maxwellian distribution.\\

 % We do not know for sure what this distribution is, but  it is not expected to depend on the
%nature of the WIMP.   In other words all WIMPs are expected to have the same velocity
%distribution and matter density. The particle density, however, which enters the event rate, is expected
%to be inversely proportional to the WIMP mass. In the present work we will consider
%a Maxwellian distribution.
% differing in their characteristic velocities.
%\end{enumerate}
%%%%%%%%%%
In the standard nuclear recoil experiments, first proposed more than 30 years ago \cite{GOODWIT}, one has to face the problem that the reaction of interest does not have a characteristic feature to distinguish it
from the background. So for the expected low counting rates the background is
a formidable problem. Some special features of the WIMP-nuclear interaction can be exploited to reduce the background problems. Such are:

i) the modulation effect: this yields a periodic signal due to the motion of the earth around the sun. Unfortunately this effect, also proposed a long time ago \cite{Druck} and subsequently studied by many authors \cite{PSS88,GS93,RBERNABEI95,LS96,ABRIOLA98,HASENBALG98,JDV03,GREEN04,SFG06,FKLW11}, is small and becomes even smaller than  $2\%$ due to cancelations arising from nuclear physics effects,
%Furthermore for heavy targets the sign of the modulation amplitude is not known, since it depends on the magnitude of the unknown WIMP mass. In %addition
%one cannot exclude  the possibility of backgrounds with a seasonal variation.

ii) backward-forward asymmetry expected in directional experiments, i.e. experiments in which the direction of the recoiling nucleus is also observed. Such an asymmetry has also been predicted a long time ago \cite{SPERGEL88}, but it has not been exploited, since such experiments have been considered  very difficult to perform.
% but but it has not been exploited, since such experiments have been considered  very difficult to perform, but
Some progress has, however, has recently been made in this direction and   they now appear  feasible \cite{SPERGEL88,DRIFT,SHIMIZU03,KUDRY04,DRIFT2,GREEN05,Green06,KRAUSS,KRAUSS01,Alenazi08,Creswick010,Lisanti09,Giometal11}. In such experiments the event rate depends on the direction of observation. In the most favorable direction, opposite to the sun's direction of motion, is comparable to the standard event rate. The sensitivity of these experiments for various halo models has also been discussed \cite{GREEN05,Green06}. Furthermore we should mention that in such experiments \cite{DRIFT,DRIFT2,Giometal11} all events are counted. If some interesting events can be found, they can be established by further analyzing them by the direction of the observed recoils.  %\cite{SPERGEL88,DRIFT,SHIMIZU03,KUDRY04,DRIFT2,GREEN05,Green06,KRAUSS,KRAUSS01,Alenazi08,Creswick010,Lisanti09,Giometal11}.
%iii) transitions to excited states: in this case one need not measure nuclear recoils, but the de-excitation $\gamma$ rays. This can happen only in very special cases since the average WIMP energy is too low to excite the nucleus. It has, however, been found that in the special case of the target $^{127}$I such a process is feasible \cite{VQS04} with branching ratios around $5\%$,
%(iv) detection of electrons produced during the WIMP-nucleus collision \cite{VE05,MVE05} and
%v) detection of hard X-rays produced when the inner shell holes are filled\cite{MouVerE}.

  %For a target like Xe these X-rays are in the $30$ keV region with the rate of about 0.1 per recoil for a WIMP mass of $100$ GeV.
%\end{itemize}

% In connection with nuclear structure aspects, in a series of calculations, e.g. in \cite{JDV03,JDVSPIN04,VF07} and references there in, it has been shown that for the coherent contribution, due to the scalar interaction, the inclusion of the nuclear form factor is important, especially in the case of relatively heavy targets. They also showed that the nuclear spin cross sections  are characterized by a single, i.e. essentially isospin independent, structure function and two static spin values, one for the  proton and one for the neutron, which depend on the target.

 An essential ingredient in direct WIMP detection is the WIMP density in our vicinity and, especially, the WIMP velocity distribution. The dark matter in the solar neighborhood is commonly assumed to be smoothly distributed in space and to have a Maxwellian velocity distribution. Some of the calculations have considered various forms of phenomenological non symmetric velocity distributions~\cite{VEROW06,JDV09,TETRVER06,VSH08}~\cite{DRIFT2,GREEN04,GREEN05} and some of them even more exotic dark matter flows like
% Among those one should mention
the late infall of dark matter into  the galaxy, i.e caustic rings~\cite{SIKIVI1,SIKIVI2,Verg01,Green,Gelmini} and Sagittarius dark matter~\cite{GREEN02}.
 
In addition to the above models very recently it was found that the velocity distributions measured in high resolution numerical simulations exhibit deviations from the standard Maxwell-Boltzmann assumption, especially at large velocities~\cite{KUHLEN10,LAWW11}. Furthermore a distinction was  between a velocity structure that is spatially localized, such as 
streams~\cite{SBWMZ08, PKB09}, and that which is spatially homogenized, which was  designated as ``debris flow''~\cite{LisSper11}. 
Both streams  and debris flows arise from the disruption of satellites that fall into the Milky Way, but differ in the relative amount of phase-mixing that they have undergone. Implications of streams~\cite{streams11} and  the debris flows in direct dark matter searches have also been considered~\cite{spergel12}, ~\cite{VergF12}.
 
In the present paper  we will address the following points:
\begin{itemize}
\item The implications scalar WIMPs on the expected event rates. 
The interest in such a WIMP has recently been revived due to a new scenario of dark matter production in bounce cosmology~\cite{Li:2014era, Cheung:2014nxi} in which the authors point out the possibility of using dark matter as a probe of a  big bounce at the early stage of cosmic evolution. 
A model independent study of dark matter production in the
contraction and expansion phases of the Big Bounce reveals a new venue for achieving the observed relic abundance in which dark matter was produced completely out of chemical equilibrium~\footnote{
Note that in Standard Cosmology,  non-thermal production of  dark matter could also happen, which has been utilized in~\cite{Chung:1998ua} to test non-standard cosmologies proposed in that era. 
The relation among dark matter mass and cross section predicted in the standard cosmology is, however, generically different from the predictions from the bounce universe scenario.  
One of the major differences is that--for non-thermal production--in standard cosmology the relic abundance of dark matter depends substantially--at the leading order--on reheating temperature, $T_{RH}$, as was first pointed in~\cite{Chung:1998ua}.  Whereas  in the bounce universe scenario, the relic abundance does not depend on the  bounce temperature, $T_b$, at the leading order;  instead it appears at  the sub-leading orders of the relic abundance~\cite{Li:2014era}.
}. 
A characteristic relation, Fig.~\ref{fig:RelicEvolution}, 
\begin{figure}[!ht]
 \begin{center}
\rotatebox{90}{\hspace{-0.0cm} $\prec\sigma \upsilon\succ\rightarrow$cm$^3$s$^{-1}$}
\includegraphics[width=0.7\textwidth]{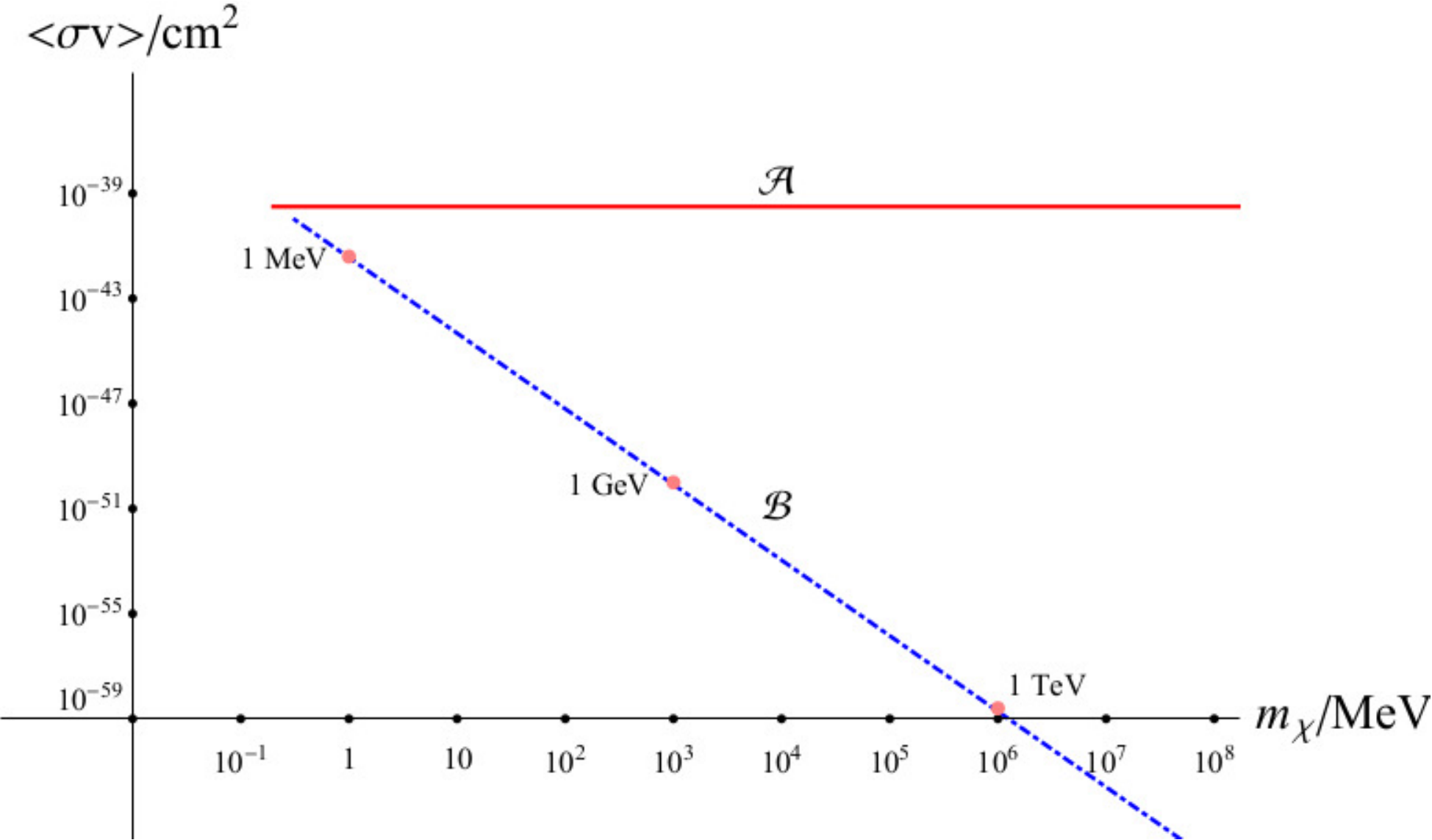}\\
\hspace{-3.0cm}$\rightarrow \log(m_{\chi}),\,(m_{\chi}$in GeV) \\
 \caption{The cross section $\prec\sigma \upsilon\succ$ as a function of the WIMP mass. In the standard cosmology it is a constant (solid line), but it varies considerably in the bounce universe scenario (dotted line)}
 \label{fig:cbplog}
 \end{center}
  \end{figure}
\begin{figure}[!ht]
\begin{center}
\includegraphics[width=0.8\textwidth]{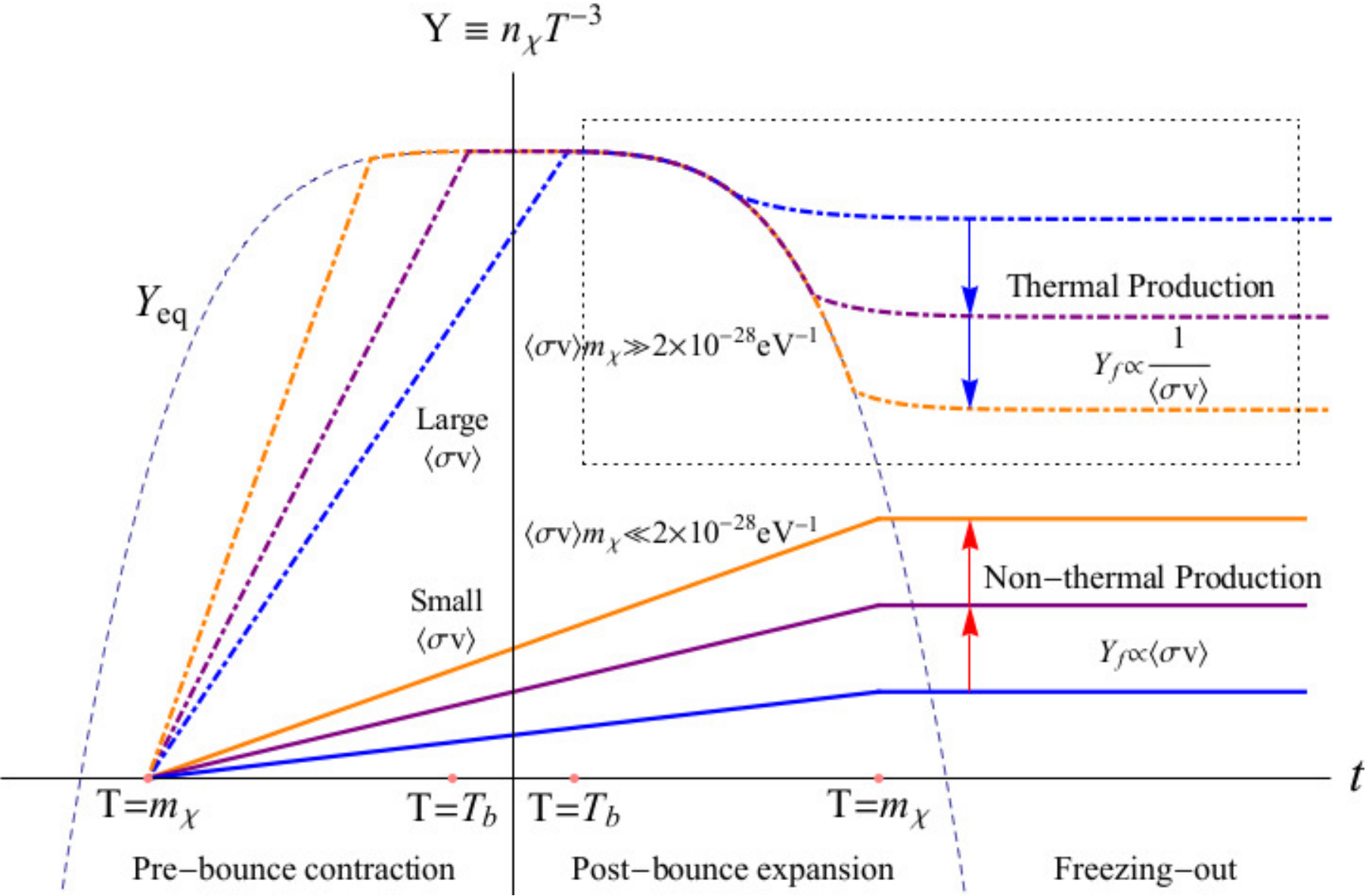}
%\hspace*{0.0cm}\tiny{$(T_A)_{th} \rightarrow$MeV }
 \caption{A schematic plot of the time evolution of dark matter in a generic bounce universe scenario. Two pathways of producing dark matter yet satisfying current observations
 thermal production (which is indistinguishable from standard cosmology) and non-thermal production (characteristic to bounce universe)  are illustrated. The horizontal axis indicates both the time, t, as well as the temperature, T, of the cosmological background.} 
\label{fig:RelicEvolution}
 \end{center}
  \end{figure}
comes out of the model independent analysis. 
This is to be contrasted with the straight line (cross-section being independent of dark matter mass) of the standard cosmology.\\
Once DM mass and its coupling constant with ordinary matter are extracted from   experimental data  we can check if they obey the predicted relation. In this way, this alternative route of dark matter production in bounce cosmology can be used to test the bounce cosmos hypothesis.  
\item In order to settle the issues raised above we will compute the differential and total event rates in a variety of targets such as those employed in XENON~\cite{XENON10, XENON100.11}, CoGENT~\cite{CoGeNT11}, DAMA~\cite{DAMA1,DAMA11}, LUX ~\cite{LUX11},  CDMSII ~\cite{CDMSII04}, CRESST ~\cite{CRESST} and PICASSO~\cite{PICASSO09,PICASSO11}. For this study we will consider not only the standard Maxwell Boltzmann distribution but also  the effects of  debris flows ~\cite{spergel12} on these rates including the modulation due to annual motion of the Earth~\cite{JDV12n}.

\end{itemize} 

In any case, regardless of the validity of the big bounce universe scenario, the scalar WIMPs have the characteristic feature that the elementary cross section in their scattering off ordinary quarks is increasing as the WIMPs get lighter, which leads to an interesting experimental feature, namely  it is expected to enhance the event rates at low WIMP mass. In the present calculation we will adopt  this view and study its implications in direct direct dark matter searches compared to other types of WIMPs, such as the neutralinos, which we will call standard.

Scalar WIMP's can occur in particle models. Examples are i) In Kaluza-Klein theories for models involving    universal extra dimensions (for applications to direct dark matter detection  see, e.g.,~\cite{OikVerMou}). In such models  the scalar WIMPs are characterized by ordinary couplings, but they are expected to be quite massive. ii) very light  particles~\cite{Fayet03} not relevant to the ongoing WIMP searches ii) Scalar WIMPS such  as those  considered previously in various extensions of the standard  model~\cite{Ma06}, which  can be quite  light and long lived protected by a discrete symmetry. Thus they are viable cold dark matter candidates.

%%%%%%%%%%%%%%%%%%%%%%%%%%%%%%%%%%%%%%%r%%%%%%
\section{The big bounce universe scenario}
Recently a stable as well as scale-invariant power spectrum of primordial density perturbations is finally obtained~\cite{Li:2011nj, Li:2013bha} in the bounce universe scenario. The
``Bounce Cosmology'' postulates that there exists  a phase
of matter-dominated contraction before the Big Bang~\cite{Wands:1998yp}
during which  the matter content of the universe comes into
thermal contact--resulting in a scale invariant spectrum--before a subsequent expansion after the big bounce.  
In view of this development we are motivated to work out
further experimental or observational predictions the Bounce Universe model~\cite{Li:2014era, Cheung:2014nxi} (See also~\cite{Cai:2011ci}.)~\footnote{Use of observational data from WMAP, Planck and BISEP2 has been made to test bounce models~\cite{Liu:2010fm, Cai:2011zx, 
Li:2014msi,%
Quintin:2014oea,%1406.6049
Wan:2014fra,%1405.2784
Cai:2014bea,%1405.1369
Liu:2014tda,%1405.1188
Li:2014qwa,%1405.0211
Cai:2014hja,%1404.6672
Cai:2014xxa,%1404.4364, 
Hu:2014aua, 
Li:2014cka,
Xia:2014tda, %PhysRevLett.112.251301, %1403.7623 
Cai:2014zga%1402.3009
}. }.
Our study is model independent of a particular bounce model and our predictions are of particle physics nature and can be tested independently at LHC or dark matter direct detections, outside of the cosmological context.

A signature prediction from the bounce universe: By
investigating the production process of dark matter in the
pre-bounce contraction and the post-bounce expansion
epochs of a generic bounce universe, we find that, in the
big bounce scenario, dark matter production can be extended beyond the Big Bang, as shown in Fig.~\ref{fig:cbplog} (compare the dotted  and solid lines). Furthermore an out-of-thermal-equilibrium production of dark
matter is allowed, which encodes information of early universe evolution, marked the non-thermal production'' in
Fig.~\ref{fig:RelicEvolution}. Specifically it predicts a relation governing a dark
matter mass and interaction cross section, depicted by the
solid line in Fig.~\ref{fig:cbplog} .
 This behavior reflects a mass dependence of the cross-section, characteristic of a scalar type WIMP.
% A model independent approach to dark matter study in bounce universe: 
As shown in Fig.~\ref{fig:RelicEvolution}, we divide the
bounce (See~\cite{NoBer08, Branden12} for recent reviews.) schematically
into three stages to facilitate a model independent analysis~\cite{Li:2014era, Cheung:2014nxi}.

\section{The particle model.}
If the WIMP is a scalar \cite{ZeeScal85,ZeeScal01,BentoRos01,BentoBero00}  particle $\chi$  interacting with another scalar $\phi$ via a quartic coupling  the cross section $\prec\sigma \upsilon\succ$ for the process:
 \beq
 \phi+\phi \rightarrow \chi+\chi
 \eeq 
 in the center of mass system is given by:
 \beq
\prec\sigma \upsilon\succ=\frac{1}{16 \pi}\frac{\lambda^2}{m_{\chi}^2}\frac{\sqrt{s-4 m_{\chi}^2}\sqrt{s}}{ 4 m_{\phi}^2},\quad \sqrt{s}\geq 2 m_{\chi}
 \eeq
 In the limit in which $m_{\phi}>>m_{\chi}$ and $\sqrt{s}\approx 2 m_{\phi}$ we find:
 \beq
 \prec\sigma \upsilon\succ\approx\frac{1}{16 \pi}\frac{\lambda^2}{m_{\chi}^2}
 \eeq
% which is in essential agreement with the expression after Eq. (4) given previously \cite{Edna14}, but with different assumptions.
We will assume in this work that $\phi$ is the Higgs scalar discovered at LHC.

\begin{figure}[!ht]
\begin{center}
\subfloat[]
{
\includegraphics[width=0.45\textwidth]{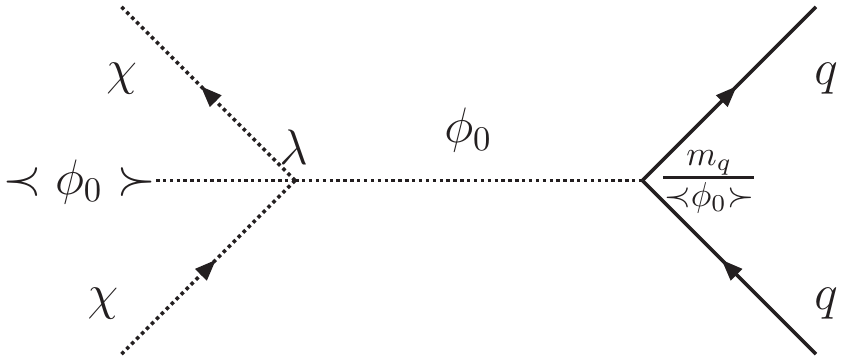}
%\hspace*{0.0cm}\tiny{$(T_A)_{th} \rightarrow$MeV }
}
\subfloat[]
{
\includegraphics[width=0.45\textwidth]{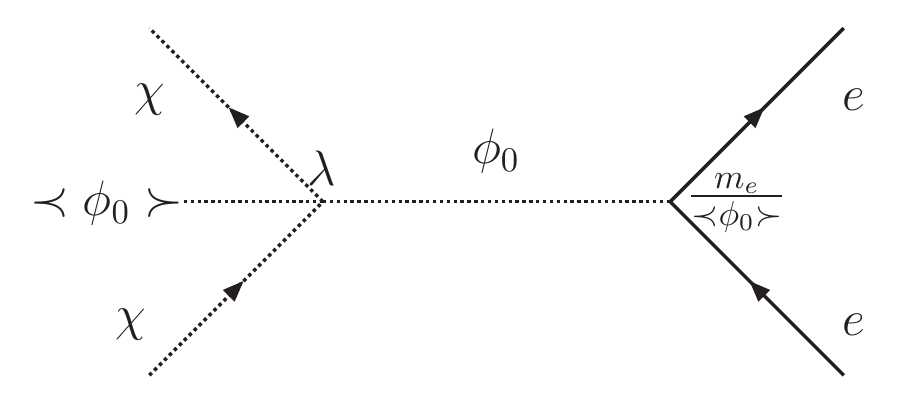}
%\hspace*{0.0cm}\tiny{$(T_A)_{th} \rightarrow$MeV }
}
\caption{The quark - scalar WIMP  scattering mediated by a scalar particle. Note that the amplitude is independent of the vacuum expectation value $\prec\phi_0\succ$ of the scalar (a). The corresponding diagram for electron scalar- WIMP
scattering (b)}
\label{fig:xxphiphiqe}
 \end{center}
  \end{figure}

If the WIMP is a scalar particle $\chi$  interacting with another scalar $\phi$ via a quartic coupling  the cross section $\prec\sigma \upsilon\succ$ for the process:
 \beq
 \phi+\phi \rightarrow \chi+\chi
 \eeq 
 in the center of mass system is given by:
 \beq
\prec\sigma \upsilon\succ=\frac{1}{16 \pi}\frac{\lambda^2}{m_{\chi}^2}\frac{\sqrt{s-4 m_{\chi}^2}\sqrt{s}}{ 4 m_{\phi}^2},\quad \sqrt{s}\geq 2 m_{\chi}
 \eeq
 In the limit in which $m_{\phi}>>m_{\chi}$ and $\sqrt{s}\approx 2 m_{\phi}$ we find:
 \beq
 \prec\sigma \upsilon\succ\approx\frac{1}{16 \pi}\frac{\lambda^2}{m_{\chi}^2}
 \eeq
 which is in essential agreement with the expression after Eq. (4) given previously \cite{Li:2014era, Cheung:2014nxi}, but with different assumptions.
 For the scalar WIMP- quark scattering  the relevant Feynman diagram is shown in Fig.~\ref{fig:xxphiphiqe}.
  %%%%%%%%%%%%%%%
\begin{figure}[!ht]
\begin{center}
\rotatebox{90}{\hspace{0.0cm} $\sigma_p\rightarrow$pb}
\includegraphics[width=0.8\textwidth]{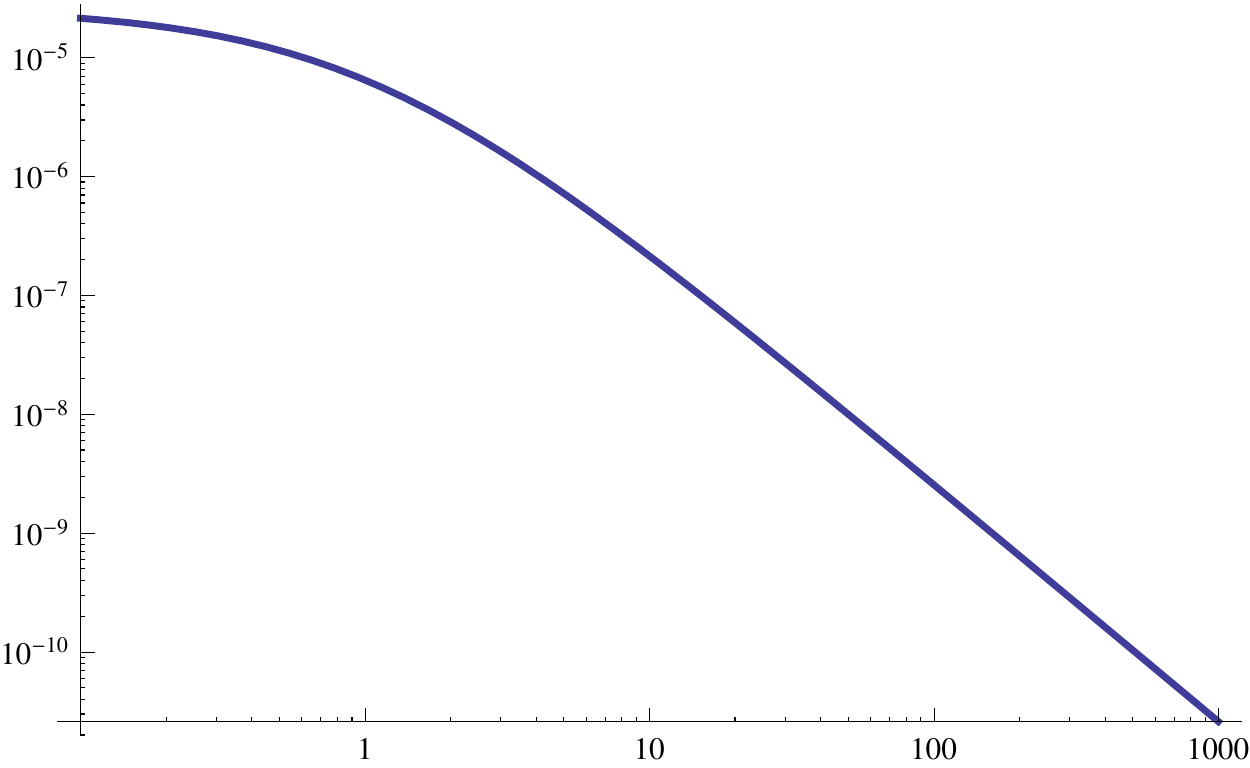}
{\hspace{0.0cm}$m_{\chi} \rightarrow$GeV }
\caption{The nucleon cross section as a function of the WIMP mass in the case the WIMP is a scalar particle. The overall scale was adjusted to fit the cross section of $10^{-8}$pb obtained from the exclusion plots of XENON100 at 50 GeV.}
 \label{fig:sigmap}
 \end{center}
  \end{figure}
	
  The resulting nucleon cross section is given by:
  \beq
  \sigma_p=\frac{1}{4 \pi}\frac{ \lambda^2 m_p^2 (\mu_r^2)}{m_{\phi}^4}\frac{1}{m_{\chi}^2}(\sum_q f_q)^2=\frac{1}{4 \pi}\frac{ \lambda^2 m_p^2}{m_{\phi}^4}\frac{1}{\left (1+m_{\chi}/m_p\right )^2}(\sum_q f_q)^2
  \eeq
  Note that the vacuum expectation value  $\prec\phi_0\succ$ in the quartic coupling is canceled by  the Yukawa coupling of the Higgs with the quarks. The parameter $f_q$ is related to the probability of finding the quark $q$ in the nucleon:
  \beq
  f_q=\frac{\prec m_q q \bar{q}\succ}{m_N}
  \eeq
  i.e. the heavy quarks become important, even though the probability of finding them in the nucleon is small.
  If the scalar is the Higgs particle discovered at LHC, $\lambda=1/2$, $m_{\phi}=126$ GeV, one finds:
  \beq
  \sigma_p= \sigma_0\left(1+\frac{m_{\chi}}{m_p }\right )^{-2}, \, \sigma_0=6\times 10^{-11}m_p^{-2}\left (\sum_q f_q \right )^2
	\label{Eq:sigma0}
	\eeq
	 The value of $\sum_q f_q$ , of course, can vary, but a reasonable, albeit rather optimistic,  value of 0.5 is acceptable \cite{Chen,Dree00},\cite{JDV06}. Thus
	\beq
	\sigma_0\approx 0.009 \mbox{pb}\rightarrow\sigma_p \approx 3 \times 10^{-6}\mbox{pb} \left(\frac{50}{m_{\chi}\mbox{(GeV)}}\right )^2, \mbox{ for } m_{\chi}>>m_p
  \eeq
  This for $m_{\chi}=50$GeV this value is quite a bit bigger than  the  limit extracted from the current experimental searches. 
	So in our treatment we have  fixed  the parameter $\sigma_0$  in the nucleon cross section so that for a WIMP mass of 50 GeV we get the limit extracted from experiments, e.g. $10^{-8}$ pb from  XENON100 ~\cite{XENON10012,XENON100.11}. The thus obtained cross section is exhibited in Fig. \ref{fig:sigmap}. It is interesting to compare the behavior of this cross section with that of the relic abundance of the BUS shown in  Fig.~\ref{fig:cbplog}. We note that this mass dependence of the cross section of scalar WIMPs, i.e. exhibiting an enhancement  in  the low WIMP mass regime, may favor the searches at low energy transfers.

	In the case of light WIMPs, another interesting domain of the BUS (Fig.~\ref{fig:cbplog}), one finds that WIMPs with energy less than 100 MeV  cannot produce  a detectable recoiling nucleus, but they could produce electrons \cite{MVE05}  with energies in the tens of eV, which  could be detected with current mixed phase detectors ~\cite{XENON14}. We are not, however, going to discuss further this possibility in this work. If the WIMP is a scalar particle, however, it can interact in a similar pattern with other fermions, e.g. electrons. The    relevant Feynman diagram is shown in Fig. \ref{fig:xxphiphiqe}.

	For WIMPs with mass in the range of  the electron mass, both the WIMP and the electron are not relativistic. So the expression for  elementary electron cross section is similar to that of hadrons , i.e. it is now given by:
	  \beq
  \sigma_e=\frac{1}{4 \pi}\frac{ \lambda^2 m_e^2}{m_{\phi}^4} \left (\frac{m_e m_{\chi}}{m_e+m_{\chi}}\right )^2\frac{1}{m_{\chi}^2}\approx 8.0 \times 10^{-7}\mbox{pb}\left (\frac{1}{1+m_{\chi}/m_e}\right )^2,
	%(\sum_q f_q)^2\approx\frac{1}{4 \pi}\frac{ \lambda^2 m_N^4}{m_{\phi}^4}\frac{1}{m_{\chi}^2}(\sum_q f_q)^2
  \eeq
	obtained using  the same values of $\lambda$ and $\phi$ as above. This is a respectable size cross section dependent on the ratio $m_{\chi}/m_e$. 
	%It is therefore smaller by a factor of
	%$$2 \left( \frac{m_e}{m_p}\right)^4\approx 10^{-6} \mbox{ down compared to the nucleon cross section.} $$
	%So it may not be practical.
	In this case one must consider electron recoils, but the highest possible electron energy is about 1.5 eV and the WIMP mass must greater than 0.3 electron masses. So the detection of  WIMPs with mass around the electron mass  requires another type of detector and it will not be discussed further in the present work.

 \section{The formalism for the WIMP-nucleus differential event rate}
\begin{figure}
\begin{center}
\rotatebox{90}{\hspace{0.0cm} $g(\upsilon_{min},\upsilon_E(\alpha))\times10^{3}\rightarrow$(km/s)$^{-1}$}
\includegraphics[height=0.4\textwidth]{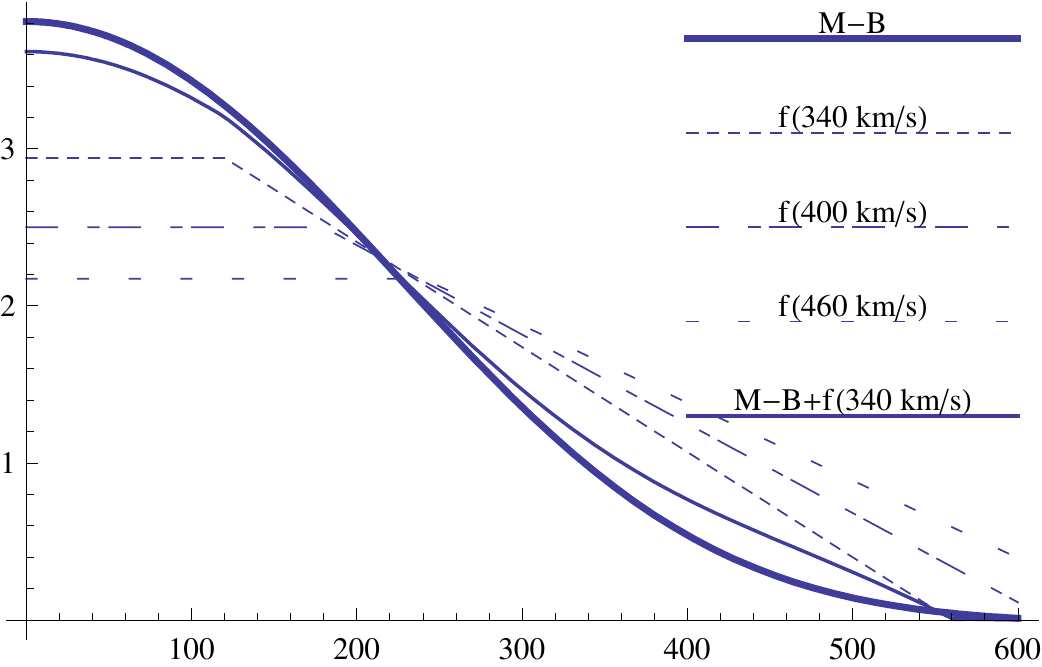}
\\
{\hspace{-2.0cm} $\upsilon_{min}\rightarrow$km/s}
\caption{ The  function $g(\upsilon_{min})$ as a function of $\upsilon_{min}$ in the local frame considered in this work in the case of  the traditional M-B distribution as well as the indicated velocity flows\cite{spergel12}.}
 \label{fig:flowv}
\end{center}
\end{figure}

The  most interesting quantity which depends on the velocity distribution is the quantity $g(\upsilon_{min})$. For the M-B distribution in the local frame it is defined as follows:
\beq
g(\upsilon_{min},\upsilon_E(\alpha))=\frac{1}{\left (\sqrt{\pi}\upsilon_0 \right )^3}\int_{\upsilon_{min}}^{\upsilon_{max}}e^{-(\upsilon^2+2 \upsilon . \upsilon_E(\alpha)+\upsilon_E^2(\alpha))/\upsilon^2_0}\upsilon d\upsilon d \Omega
\eeq 
For  isotropic debris flows \cite{spergel12} it is given by:
\beq
g(\upsilon_{min},\upsilon_E(\alpha))=\int_{\upsilon_{min}}\frac{f(\upsilon)}{\upsilon}d \upsilon, \, f(\upsilon)=\{\begin{array}{ll}\frac{\upsilon}{2 \upsilon_{flow}\upsilon_E(\alpha)},&\upsilon_{flow}-\upsilon_E(\alpha)<\upsilon<\upsilon_{flow}+\upsilon_E(\alpha)\\
0,&\mbox{otherwise}
\end{array}
\eeq
%where $f(\upsilon)$ is non zero only in the range  %$\upsilon_{flow}-\upsilon_E(\alpha)<\upsilon<\upsilon_{flow}+\upsilon_E(\alpha)$
These functions are  shown in Fig.~\ref{fig:flowv}.

In what follows we will find it useful to  expand $g(\upsilon_{min},\upsilon_E(\alpha))$ in  powers of $\delta$, the ratio of the Earth's velocity around the sun divided by the velocity $\upsilon_0$ of the sun around the galaxy (220km/s). Keeping terms up to linear in $\delta \approx 0.135$ and expressing everything in dimensionless variables we obtain:
\beq
\upsilon_0 g(\upsilon_{min},\upsilon_E(\alpha))=\Psi_0(x)+\Psi_1(x)\cos{\alpha}, \quad x=\frac{\upsilon_{min}}{\upsilon_{0}}
\eeq
where $\Psi_0(x$ represents the quantity relevant for the average rate , $\Psi_1(x$, which is proportional to $\delta$, represents the modulation and $\alpha$ is the phase of the Earth ($\alpha=0$ around June 3nd). In the case of the flows they were derived from the semi-analytic approximations of simulations as discussed by Spergel and co-workers \cite{spergel12}.\\
In the case of a M-B distribution these functions have been given  previously~\cite{JDVDF12}.
%take the following form:
%\beq
%\Psi_0(x)=\frac{1}{2}
 %  (\mbox{erf}(1-x)+\mbox{erf}(x+1)+\mbox{erfc}(1-y_{\mbox{\tiny{yesc}}})+\mbox{erfc}(y_{\mbox{\tiny{yesc}}}+1)-2)
%\eeq
%$$
%\frac{1}{4} \delta 
%   \left(-\text{erf}(1-x)-\text{erf}(x+1)-\text{erfc}(1-
%   \text{yesc})-\text{erfc}(\text{yesc}+1)+\frac{2
%   e^{-(x-1)^2}}{\sqrt{\pi }}+\frac{2
%   e^{-(x+1)^2}}{\sqrt{\pi }}-\frac{2
%   e^{-(\text{yesc}-1)^2}}{\sqrt{\pi }}-\frac{2
%   e^{-(\text{yesc}+1)^2}}{\sqrt{\pi }}+2\right)
%$$
%\barr
%\Psi_1(x)&=&\frac{1}{2} \delta 
 %  \left(\frac{ -\mbox{erf}(1-x)-\mbox{erf}(x+1)-\mbox{erfc}(1-y_{\mbox{\tiny{esc}}})-
  % \mbox{erfc}(y_{\mbox{\tiny{esc}}}+1)}{2} \right . \nonumber\\
 % && \left . +\frac{ e^{-(x-1)^2}}{\sqrt{\pi }}
  % +\frac{
   %e^{-(x+1)^2}}{\sqrt{\pi }}-\frac{ e^{-(y_{\mbox{\tiny{esc}}}-1)^2}}{\sqrt{\pi
  % }}-\frac{ e^{-(y_{\mbox{\tiny{esc}}}+1)^2}}{\sqrt{\pi }}+1\right)
%\earr
%where erf$(x)$ and erfc$(x)$ are the error function and its complement respectively $y_{esc}\approx2.8 \upsilon_0$ is the escape velocity. 
For  isotropic debris flows one finds:
\beq
\Psi_0(x)=\left \{\begin{array}{ll}\frac{1}{y_f}&0<x<y_f-1\\ \frac{1+y_f-x}{2 y_f}&y_f-1<x<1+y_f\\0&x>1+y_f\end{array}\right . ,\quad y_f=\frac{\upsilon_{flow}}{\upsilon_0}
\eeq
\beq
\Psi_1(x)=\delta \left \{\begin{array}{ll}0&0<x<y_f-1\\  \frac{x-y_f}{4 y_f} &y_f-1<x<1+y_f\\0&x>1+y_f\end{array}\right . ,\quad y_f=\frac{\upsilon_{flow}}{\upsilon_0}
\eeq
\\
We note that the variable $x$ depends on the nuclear recoil energy $E_R$ as well as the WIMP-nucleus reduced mass.
 As we shall see below there is an additional dependence of the rates on $E_R$ coming from the nuclear form factor.

\begin{figure}
\begin{center}
\rotatebox{90}{\hspace{0.0cm} $\epsilon(x)\rightarrow$}
\includegraphics[height=0.4\textwidth]{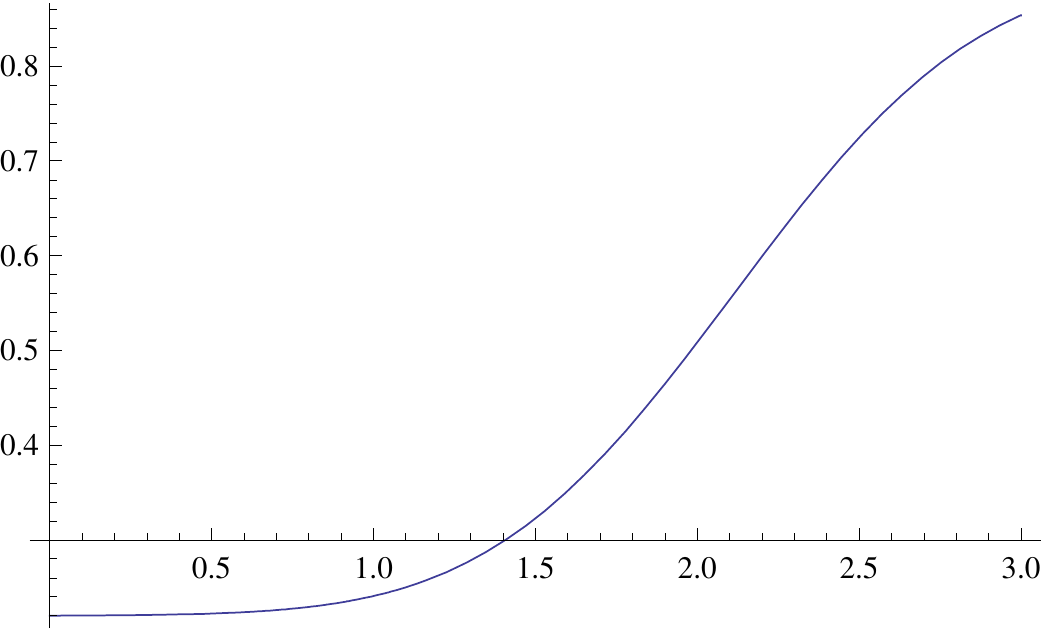}
\\
{\hspace{-2.0cm} $x=\frac{\upsilon_{min}}{\upsilon_{0}}\rightarrow$}
\caption{ The  function $\epsilon(x),\,x=\upsilon_{min}/\upsilon_0$ as a function of $x$, which gives a possible combination of a M-B distribution and  debris flows\cite{spergel12}.}
 \label{fig:epsilon}
\end{center}
\end{figure}

\begin{figure}
\begin{center}
\subfloat[]
{
\rotatebox{90}{\hspace{0.0cm} $\Psi_0(x)\rightarrow$}
\includegraphics[height=0.27\textwidth]{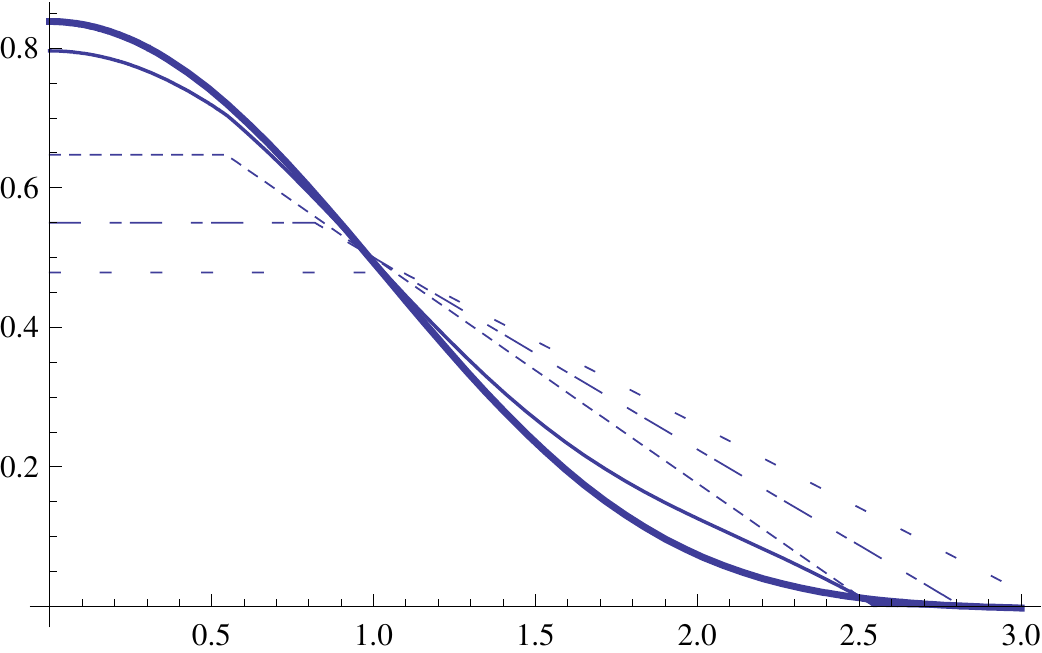}
}
\subfloat[]
{
\rotatebox{90}{\hspace{0.0cm} $\Psi_1(x)\rightarrow$}
\includegraphics[height=0.27\textwidth]{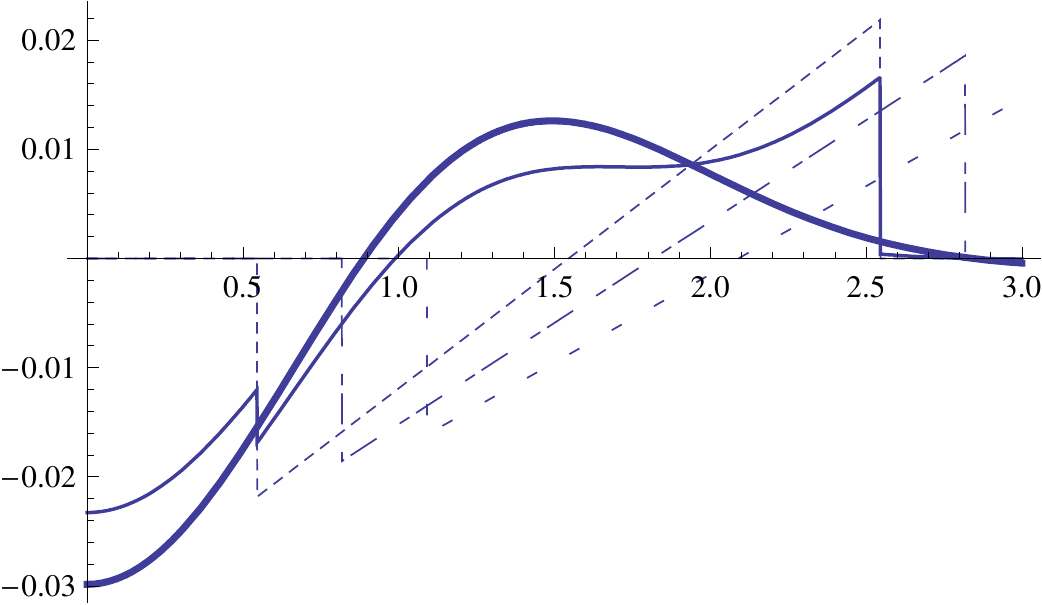}
}
\\
{\hspace{-2.0cm} $x=\frac{\upsilon_{min}}{\upsilon_{0}}\rightarrow$}
\caption{ The  functions $\Psi_0(x)$ and $\Psi_1(x)$ as a function of $x={\upsilon_{min}}/{\upsilon_{0}}$. Note that these functions have been computed at $\alpha=0$, i.e. the average local velocity. Note also that the variable $x$ depends on the nuclear recoil energy $E_R$ as well as the WIMP-nucleus reduced mass. Otherwise the labeling of the curves is the same as that of Fig. \ref{fig:flowv}.}
 \label{fig:Psi01}
\end{center}
\end{figure}

At Earth-frame velocities greater than 450 km/s,
debris flow comprises more than half of the dark matter at the Sun's location, and up to
$80\%$ at even higher velocities\cite{spergel12}. In the VL2 simulation, the combination of debris flows and standard M-B 
 is very well fit by the function
\beq
\epsilon(x)=0.22 + 0.34 \left (\mbox{erf}\left (x \frac{220}{185}-\frac{465}{185}\right ) + 1\right )
\eeq
This function is exhibited in Fig. \ref{fig:epsilon}.
In this case we find:
\beq
\Psi_i(x)\rightarrow\left (1-\epsilon(x)\right )\Psi^{MB}_i(x)+\epsilon(x)\Psi^{f}_i(x),\quad i=0,1
\eeq
The functions $\Psi_0(x)$ and $\Psi_1(x)$ are exhibited in Fig. \ref{fig:Psi01}. As expected in the case of the flows $\Psi_0(x)$ falls off linearly for large values of $x$. Note that in all cases  $\Psi_1(x)$  takes both positive and negative values, which affects the location of the maximum of the modulated rate as a function of $\alpha$, depending on the target and the WIMP mass. We will explore this effect of the different distributions in direct experiments searching any time dependence of the rates.

\begin{figure}
\begin{center}
\subfloat[]
{
\rotatebox{90}{\hspace{0.0cm} $F^2\rightarrow$}
\includegraphics[height=.17\textheight]{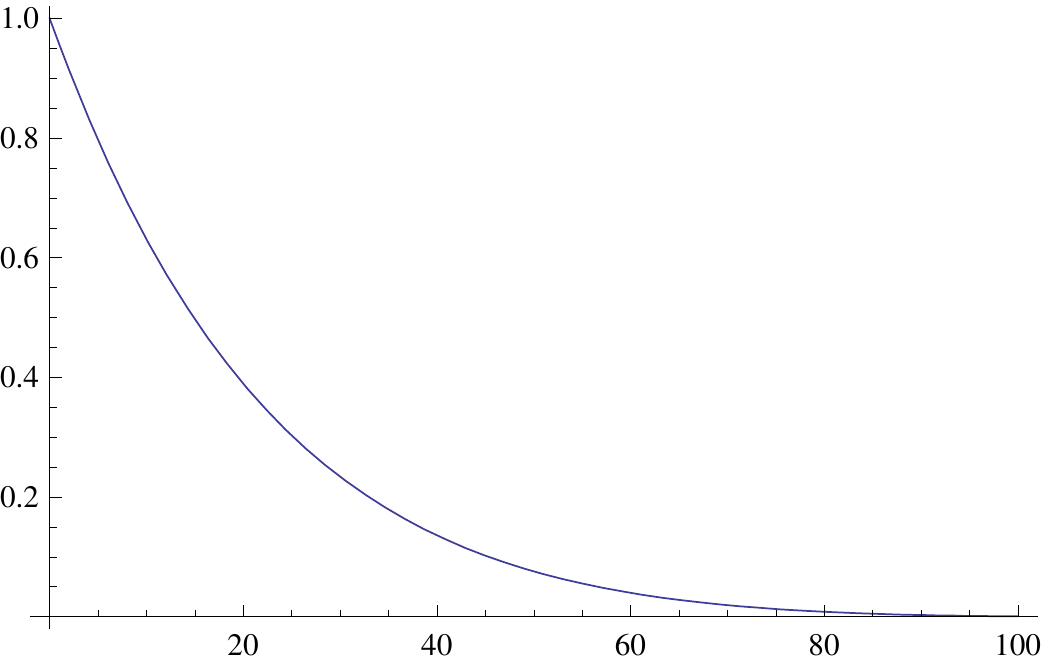}
}
\subfloat[]
{
\rotatebox{90}{\hspace{0.0cm} $F^2\rightarrow$}
\includegraphics[height=.17\textheight]{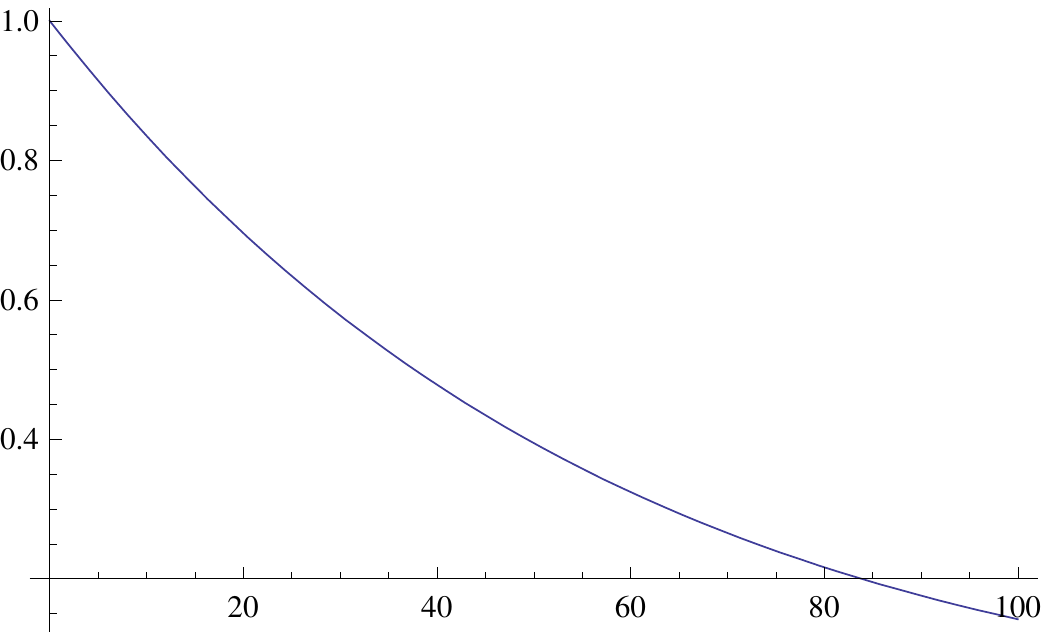}
}
\\
{\hspace{-2.0cm} $E_R\rightarrow$keV}
\caption{ The square of the nuclear form factor for a heavy target, e.g. $^{127}$I (a) and an intermediate target , e.g. $^{73}$Ge (b). For light targets the effect of the form factor is small.}
 \label{fig:FFsq}
\end{center}
\end{figure}

Once these functions are known the formalism to obtain the direct detection rates is fairly well known (see e.g. the recent reviews \cite{JDV06a,VerMou11}). So we will briefly discuss its essential elements here.
The differential event rate can be cast in the form:
\beq
\frac{d R}{ d E_R}|_A=\frac{dR_0}{dE_R}|_A+\frac{d{\tilde H}}{dE_R}|_A \cos{\alpha}
\eeq
where the first term represents the time averaged (non modulated) differential event rate, while the second  gives the time dependent (modulated) one due to the motion of the Earth (see below). Furthermore
\barr
\frac{d R_0}{ d E_R}|_A&=&\frac{\rho_{\chi}}{m_{\chi}}\frac{m_t}{A m_p} \sigma_n\left (\frac{\mu_r}{\mu_p} \right )^2 \sqrt{<\upsilon^2>} A^2\frac{1}{Q_0(A)}\frac{d t}{du},\nonumber\\
\frac{d {\tilde H}}{ d E_R}|_A&=&\frac{\rho_{\chi}}{m_{\chi}}\frac{m_t}{A m_p} \sigma_n\left (\frac{\mu_r}{\mu_p} \right )^2 \sqrt{<\upsilon^2>} A^2 \frac{1}{Q_0(A)} \frac{d h}{du}
\label{drdu}
\earr
with $\mu_r$ ($\mu_p$) the WIMP-nucleus (nucleon) reduced mass, $A$ is the nuclear mass number and $\sigma_n$ is the elementary WIMP-nucleon cross section. $ m_{\chi}$ is the WIMP mass and $m_t$ the mass of the target. 
%The first term gives the time averaged rate, while the second gives the modulated amplitude. 
Furthermore one can show that
\beq
\frac{d t}{d u}=\sqrt{\frac{2}{3}} a^2 F^2(u)   \Psi_0(a \sqrt{u}),\quad \frac{d h}{d u}=\sqrt{\frac{2}{3}} a^2 F^2(u) \Psi_1(a \sqrt{u})
\eeq
with $a=(\sqrt{2} \mu_r b \upsilon_0)^{-1}$, $\upsilon_0$ the velocity of the sun around the center of the galaxy and $b$ the nuclear harmonic oscillator size parameter characterizing the nuclear wave function.  $ u$ is the energy transfer $Q$ in dimensionless units given by
\begin{equation}
 u=\frac{E_R}{Q_0(A)}~~,~~Q_{0}(A)=[m_pAb^2]^{-1}=40A^{-4/3}\mbox{ MeV}
\label{defineu}
\end{equation}
and $F(u)$ is the nuclear form factor. In the present calculation they were obtained in context of the nuclear shell model in the spirit of\cite{DIVA00} (for the spin induced  process see,e.g.  \cite{Ress,DIVA00}). The form factor is important in the case of a heavy target and large WIMP mass, i.e. for large recoil energies (see Fig. \ref{fig:FFsq}). 

Note that the parameter $a$ depends both on the WIMP , the target and the velocity distribution. Note also that for a given energy transfer $E_R$ the quantity $u$ depends on $A$.\\
Sometimes one  writes the differential rate as:
\beq
\frac{d R}{ d E_R}|_A=\frac{\rho_{\chi}}{m_{\chi}}\frac{m_t}{A m_p} \sigma_n \left ( \frac{\mu_r}{\mu_p} \right )^2 \sqrt{<\upsilon^2>} A^2 \frac{1}{Q_0(A)}\left(\frac{d t}{du}(1+ H(a \sqrt{E_R/Q_0(A)}) \cos{\alpha}\right )
\label{dhduH}
\eeq
%\beq
%\frac{d t}{d u}=\sqrt{\frac{2}{3}} a^2 F^2(u) \Psi_0(a \sqrt{u})\left (1+H(a \sqrt{u}) \cos{\alpha}\right )
%\label{dhduH}
%\eeq
In this formulation $H(a \sqrt{E_R/Q_0(A)}) $, the ratio of the modulated to the non modulated differential rate, gives the relative differential modulation amplitude. It coincides with the ratio $\Psi_1(a \sqrt{E_R/Q_0(A)})/\Psi_0(a \sqrt{E_R/Q_0(A)})$, i.e. it is independent of the nuclear form factor and  depends only on the reduced mass and the velocity distribution. It is thus the same for both the coherent and the spin mode. Note that it can take both positive and negative values, which affects the location of the maximum of the modulated rate as a function of $\alpha$. For the convenience of the analysis of experiments, however, we will present our results in the form of Eq. \ref{drdu}.
%\begin{figure}
%\begin{center}
%\rotatebox{90}{\hspace{0.0cm} $H(a \sqrt{u})\rightarrow$}
%\includegraphics[height=.30\textheight]{apsiH.pdf}
%\\
%{\hspace{-2.0cm} $Q\rightarrow$keV}
%\caption{ The  function $H(a \sqrt{u})$ entering   the modulated differential rate as a function of the recoil energy for a heavy target, e.g. $^{127}$I. Note that this is independent of  the form factor. The solid, dotted, dot-dashed, dashed, long dashed and thick solid lines correspond to 5, 7, 10, 20, 50 and 100 GeV WIMP masses.
% \label{fig:apsiH}}
%\end{center}
%\end{figure}
\section{Some results on differential rates}
\begin{figure}
\begin{center}
\subfloat[]
{
\rotatebox{90}{\hspace{0.0cm} $dR/dQ\rightarrow$kg/(y keV)}
\includegraphics[height=.17\textheight]{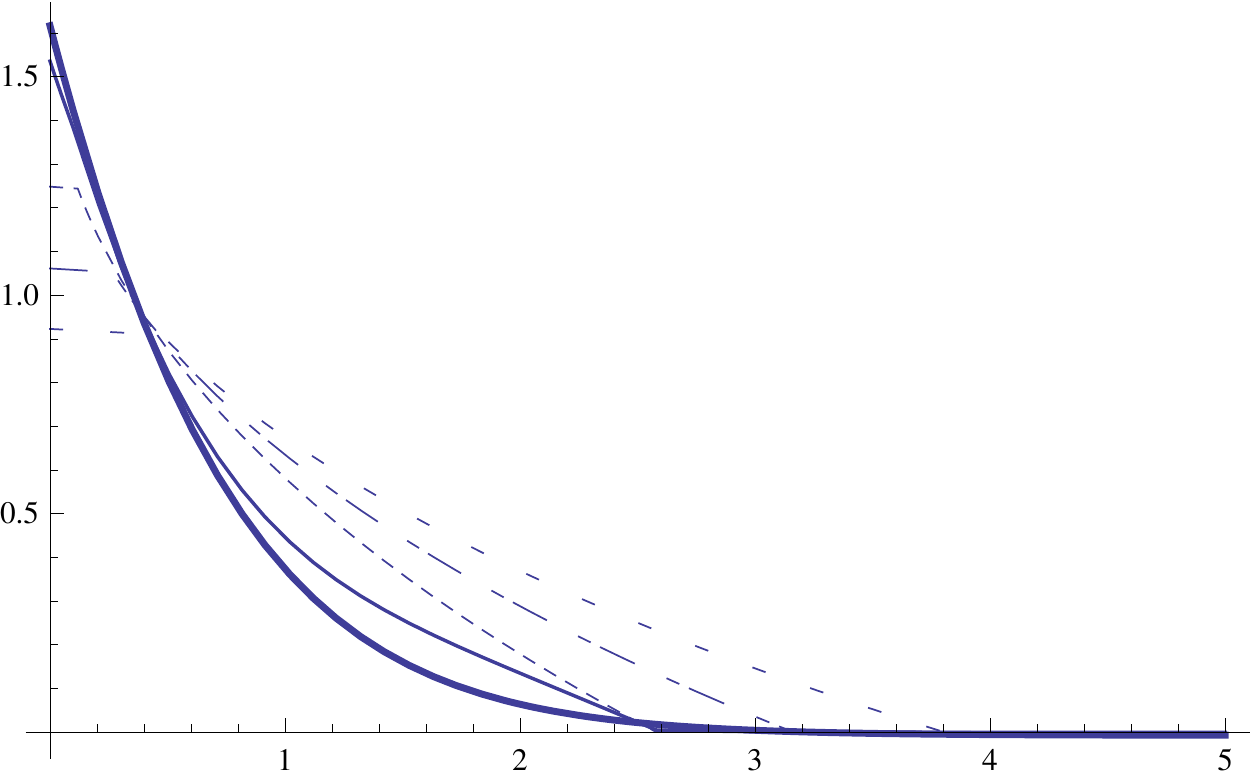}
}
\subfloat[]
{
\rotatebox{90}{\hspace{0.0cm} $dR/dQ\rightarrow$kg/(y keV)}
\includegraphics[height=.17\textheight]{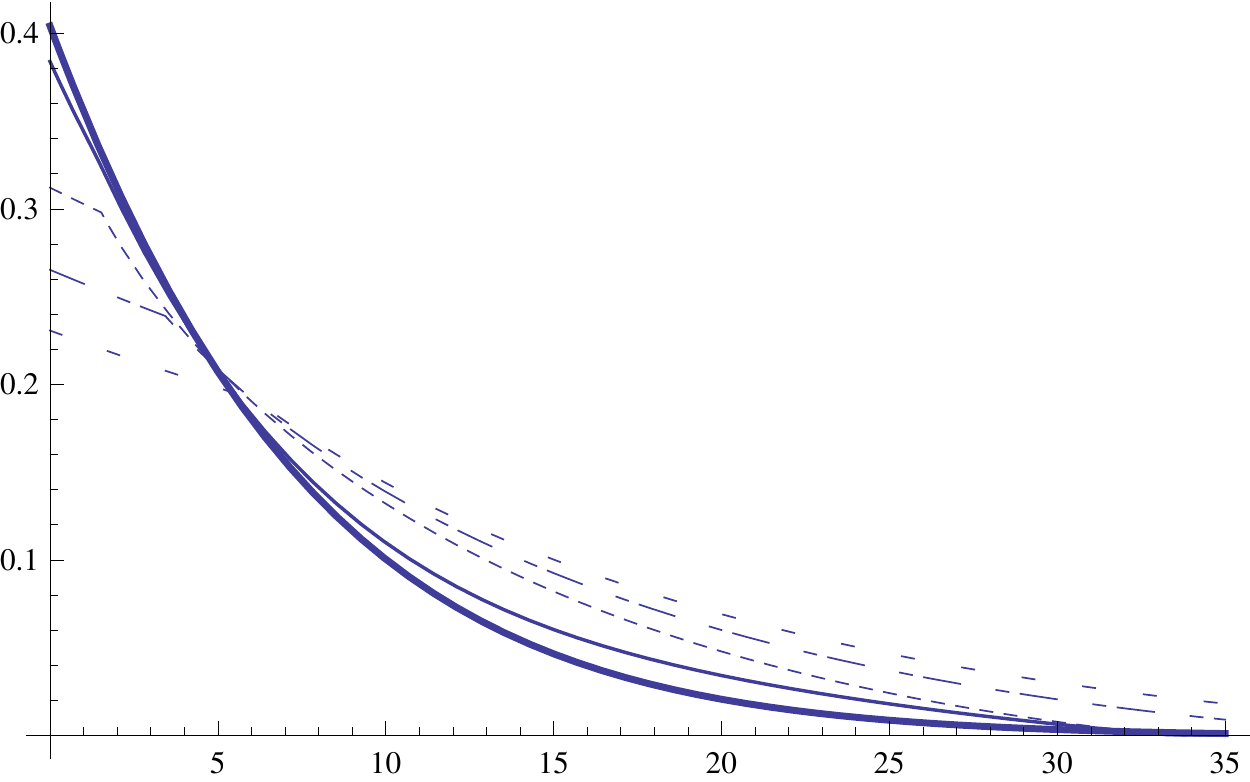}
}
\\
%{\hspace{-2.0cm} $Q\rightarrow$keV}
\subfloat[]
{
\rotatebox{90}{\hspace{0.0cm} $dR/dQ\rightarrow$kg/(y keV)}
\includegraphics[height=.17\textheight]{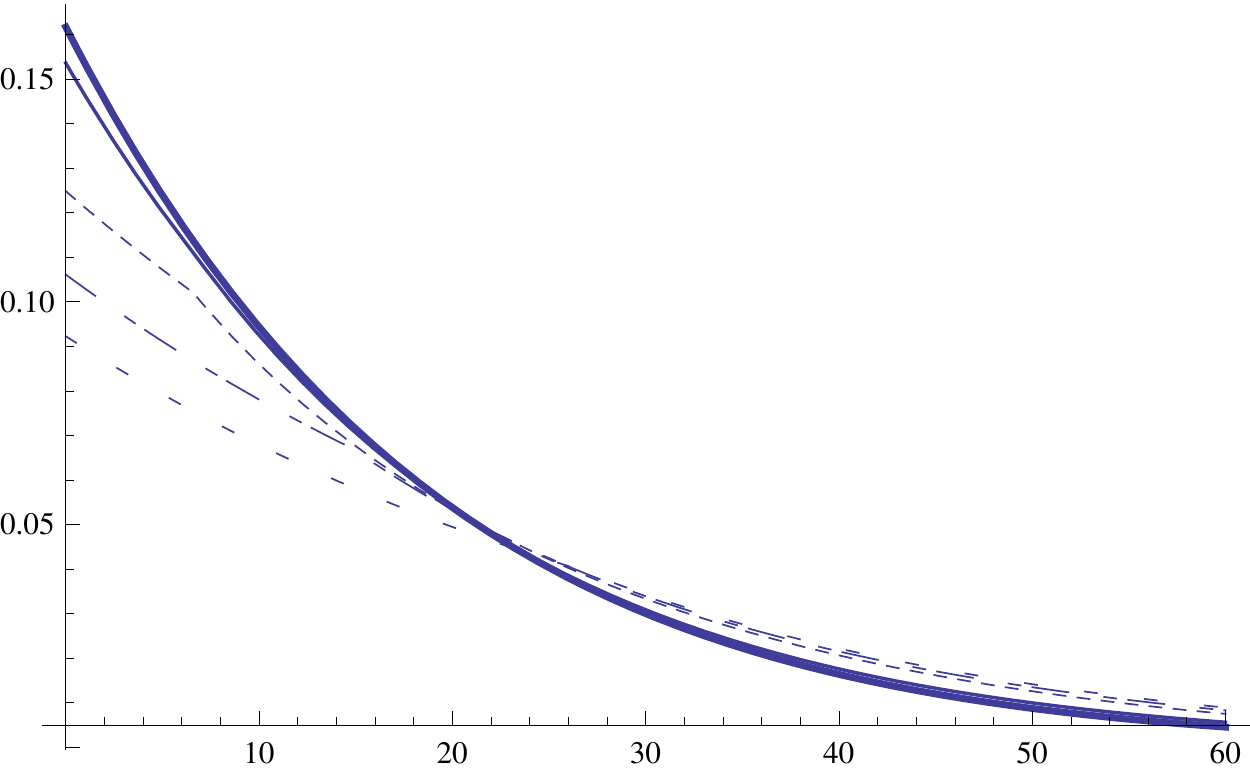}
}
\subfloat[]
{
\rotatebox{90}{\hspace{0.0cm} $dR/dQ\rightarrow$kg/(y keV)}
\includegraphics[height=.17\textheight]{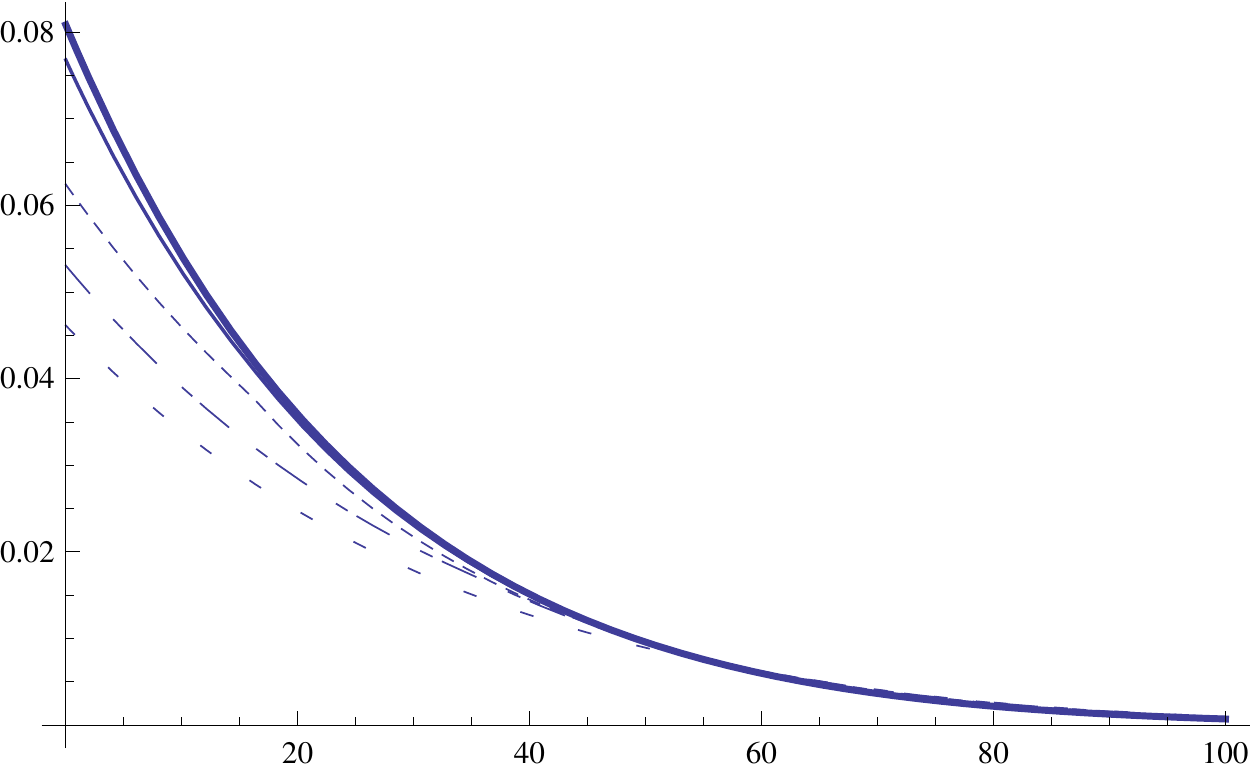}
}
\\
{\hspace{-2.0cm} $Q\rightarrow$keV}
\caption{ The differential rate $\frac{dR}{dQ}$,   as a function of the recoil energy for a heavy target, e.g. $^{127}$I assuming a nucleon cross section of $10^{-8}$pb. Panels (a) (b), (c) and (d) correspond to to 5, 20, 50 and 100 GeV WIMP masses. Otherwise the notation is the same as that of Fig. \ref{fig:flowv}.}
 \label{fig:dRdQ127}
\end{center}
\end{figure}

\begin{figure}
\begin{center}
\subfloat[]
{
\rotatebox{90}{\hspace{0.0cm} $dR/dQ\rightarrow$kg/(y keV)}
\includegraphics[height=.17\textheight]{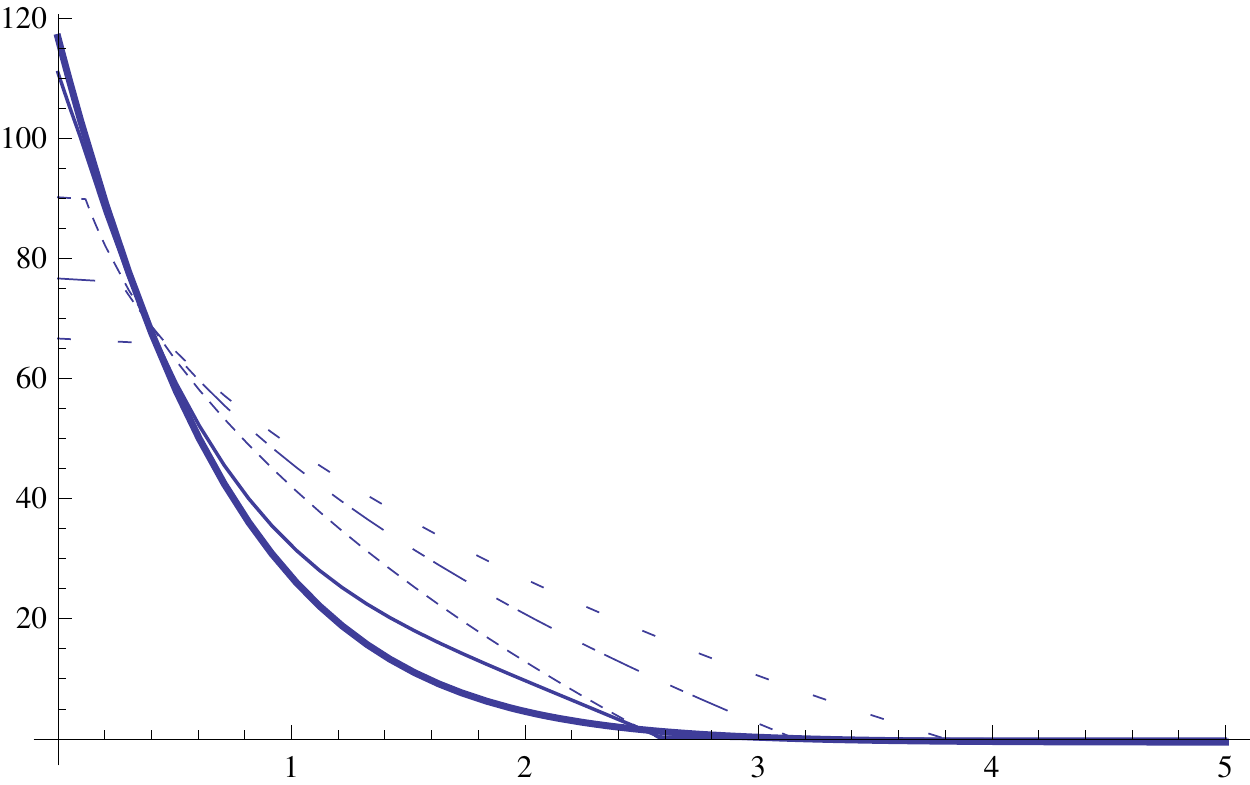}
}
\subfloat[]
{
\rotatebox{90}{\hspace{0.0cm} $dR/dQ\rightarrow$kg/(y keV)}
\includegraphics[height=.17\textheight]{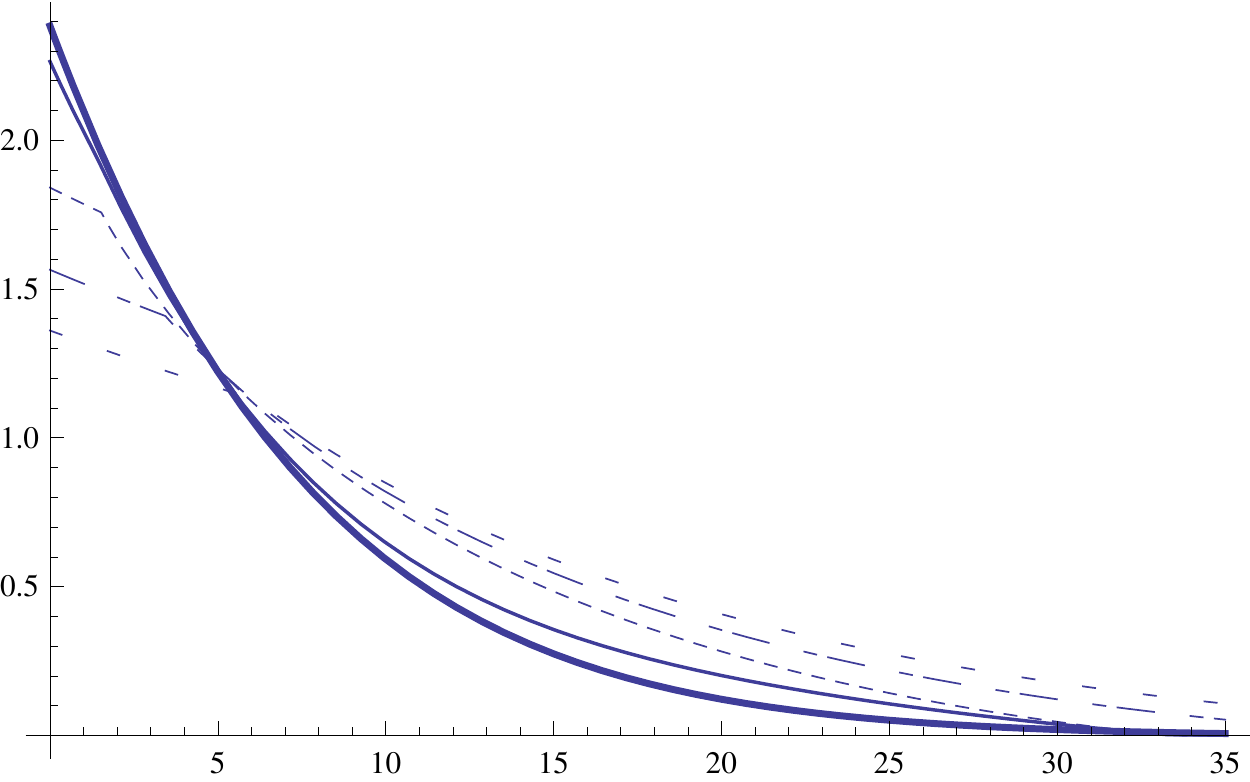}
}
\\
%{\hspace{-2.0cm} $Q\rightarrow$keV}
\subfloat[]
{
\rotatebox{90}{\hspace{0.0cm} $dR/dQ\rightarrow$kg/(y keV)}
\includegraphics[height=.17\textheight]{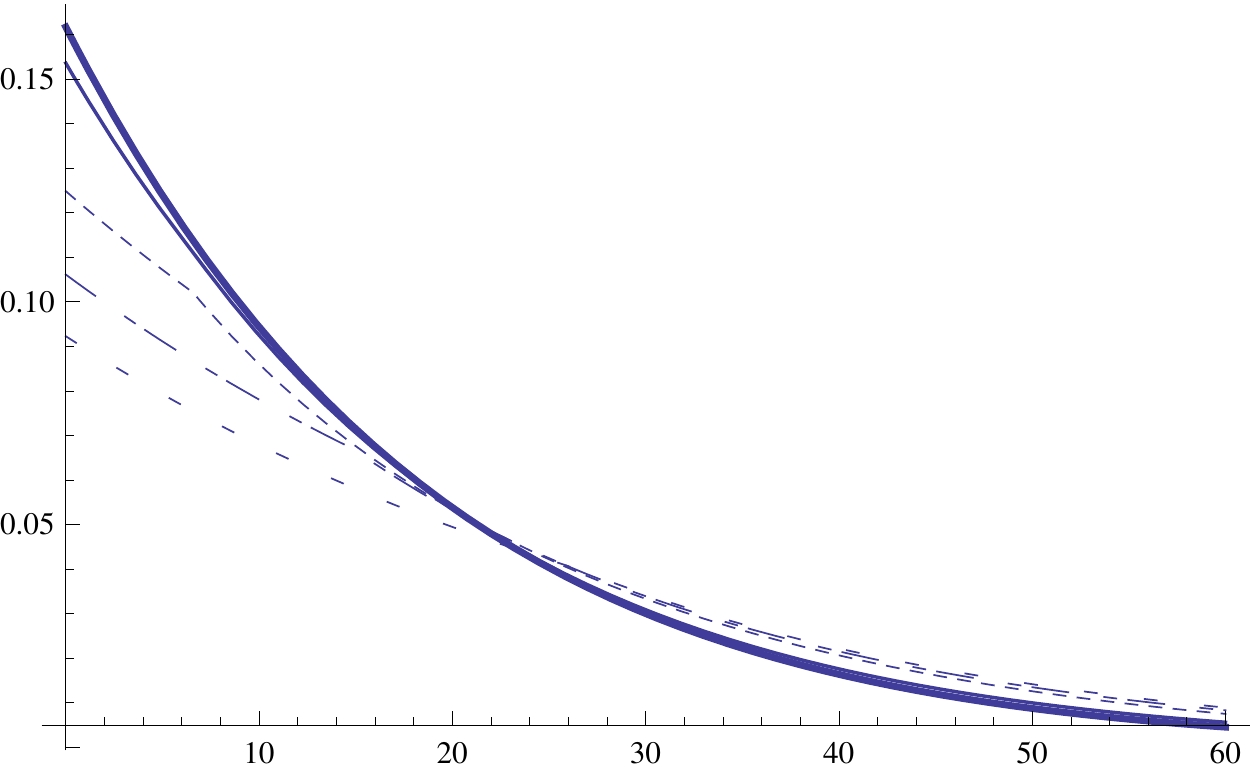}
}
\subfloat[]
{
\rotatebox{90}{\hspace{0.0cm} $dR/dQ\rightarrow$kg/(y keV)}
\includegraphics[height=.17\textheight]{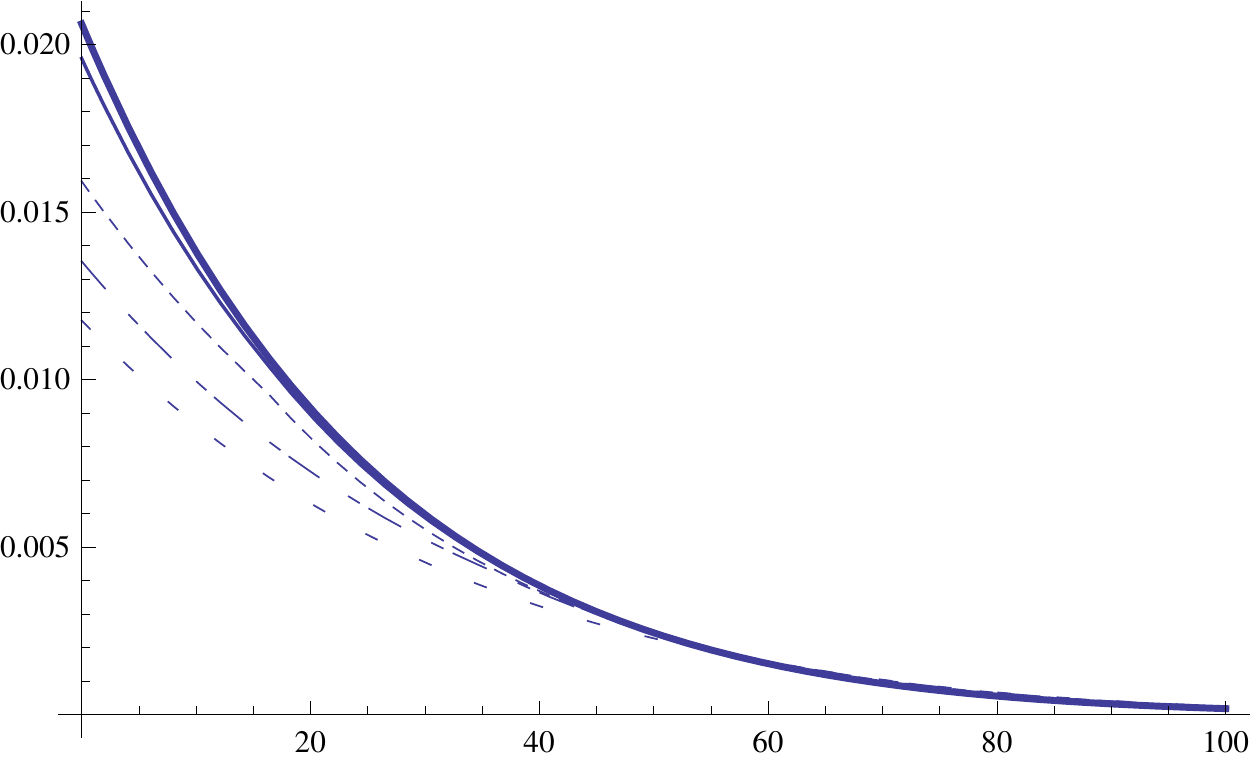}
}
\\
{\hspace{-2.0cm} $Q\rightarrow$keV}
\caption{ The same as in Fig. \ref{fig:dRdQ127} for a scalar WIMP assuming a nucleon cross section $(50/m_{\chi})^2 \times 10^{-8}$pb. }
 \label{fig:dRdQSc127}
\end{center}
\end{figure}

\begin{figure}
\begin{center}
\subfloat[]
{
\rotatebox{90}{\hspace{0.0cm} $d\tilde{H}/dQ\rightarrow$kg/(y keV)}
\includegraphics[height=.17\textheight]{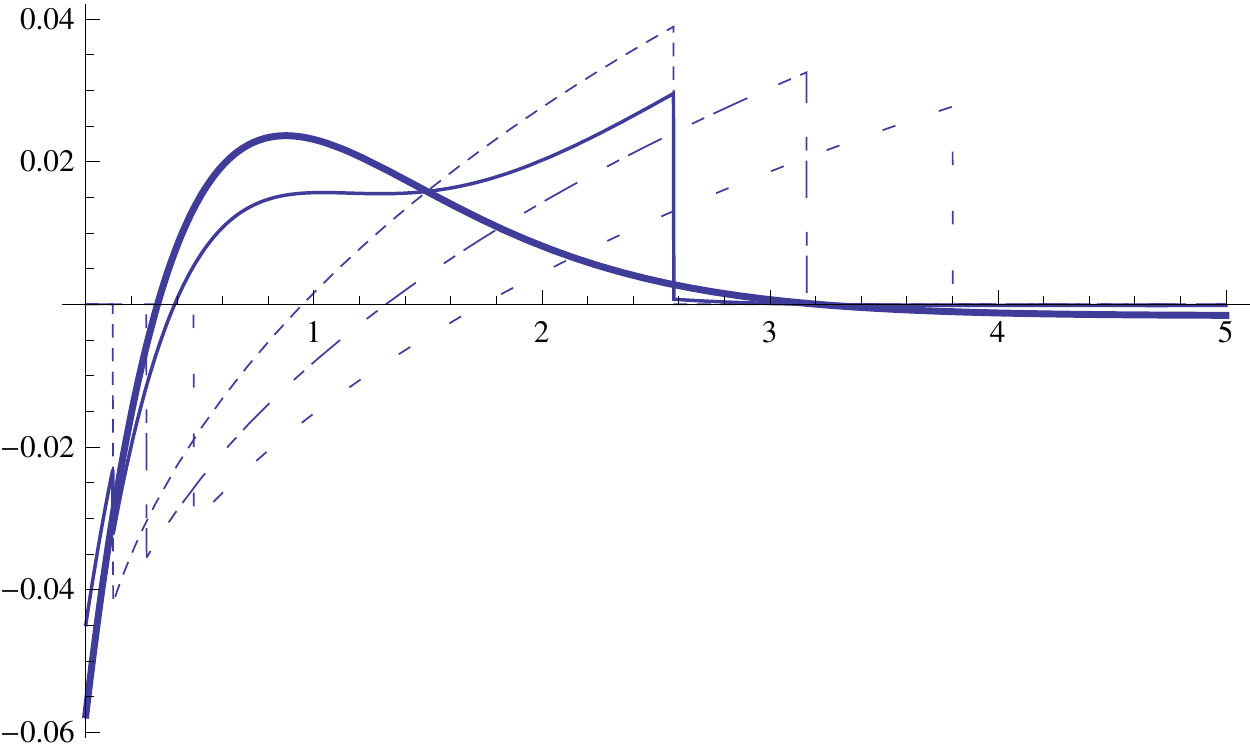}
}
\subfloat[]
{
\rotatebox{90}{\hspace{0.0cm} $d{\tilde H}/dQ\rightarrow$kg/(y keV)}
\includegraphics[height=.17\textheight]{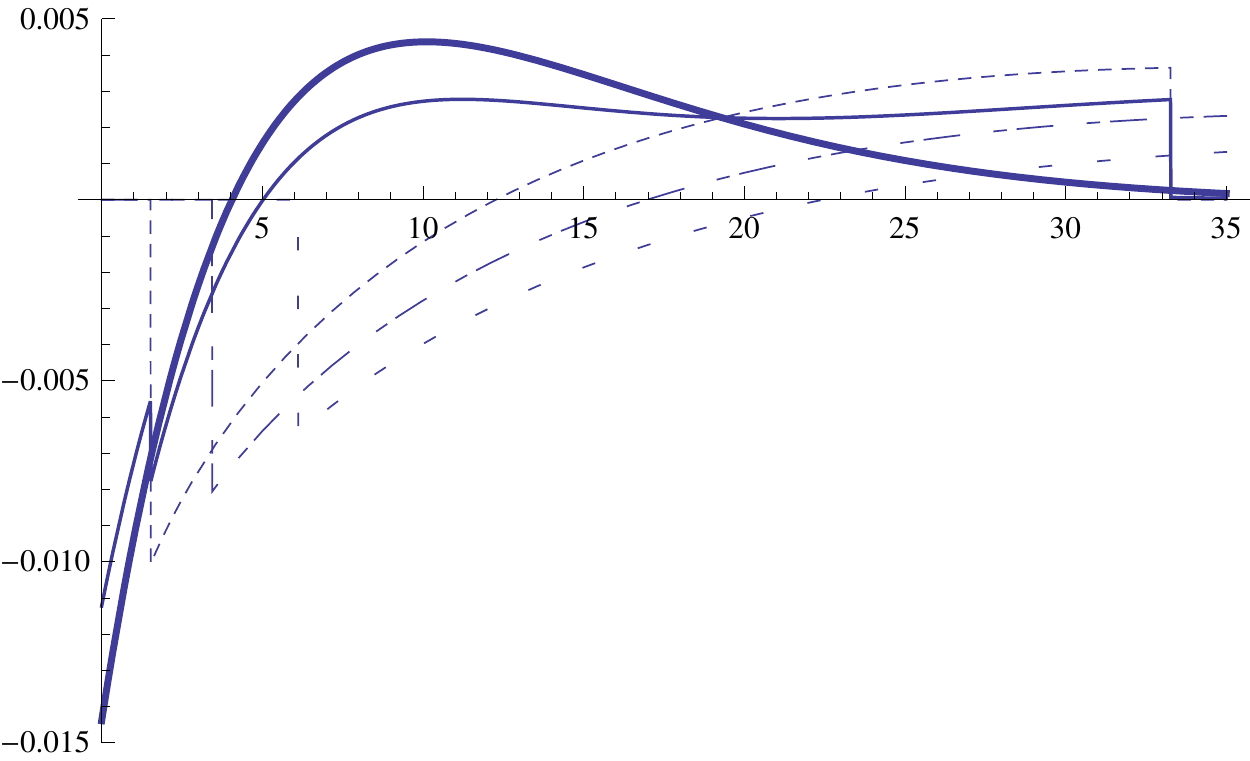}
}
\\
%{\hspace{-2.0cm} $Q\rightarrow$keV}
\subfloat[]
{
\rotatebox{90}{\hspace{0.0cm} $d\tilde{H}/dQ\rightarrow$kg/(y keV)}
\includegraphics[height=.17\textheight]{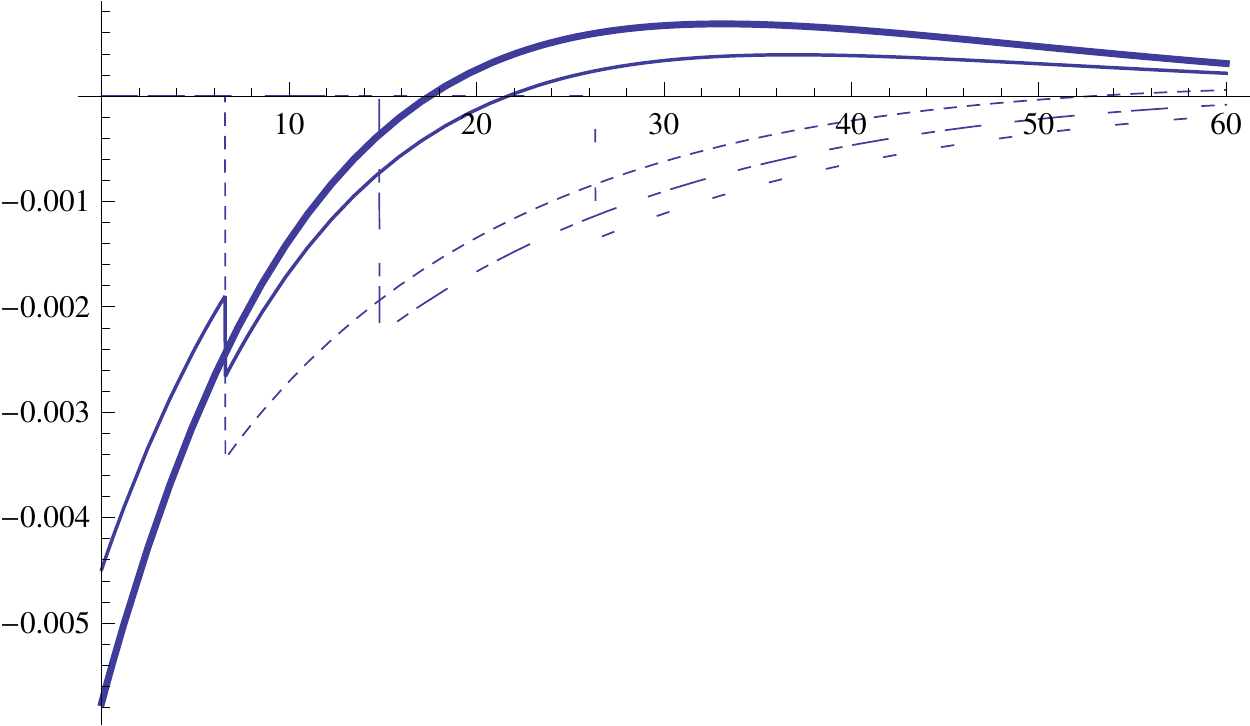}
}
\subfloat[]
{
\rotatebox{90}{\hspace{0.0cm} $d{\tilde H}/dQ\rightarrow$kg/(y keV)}
\includegraphics[height=.17\textheight]{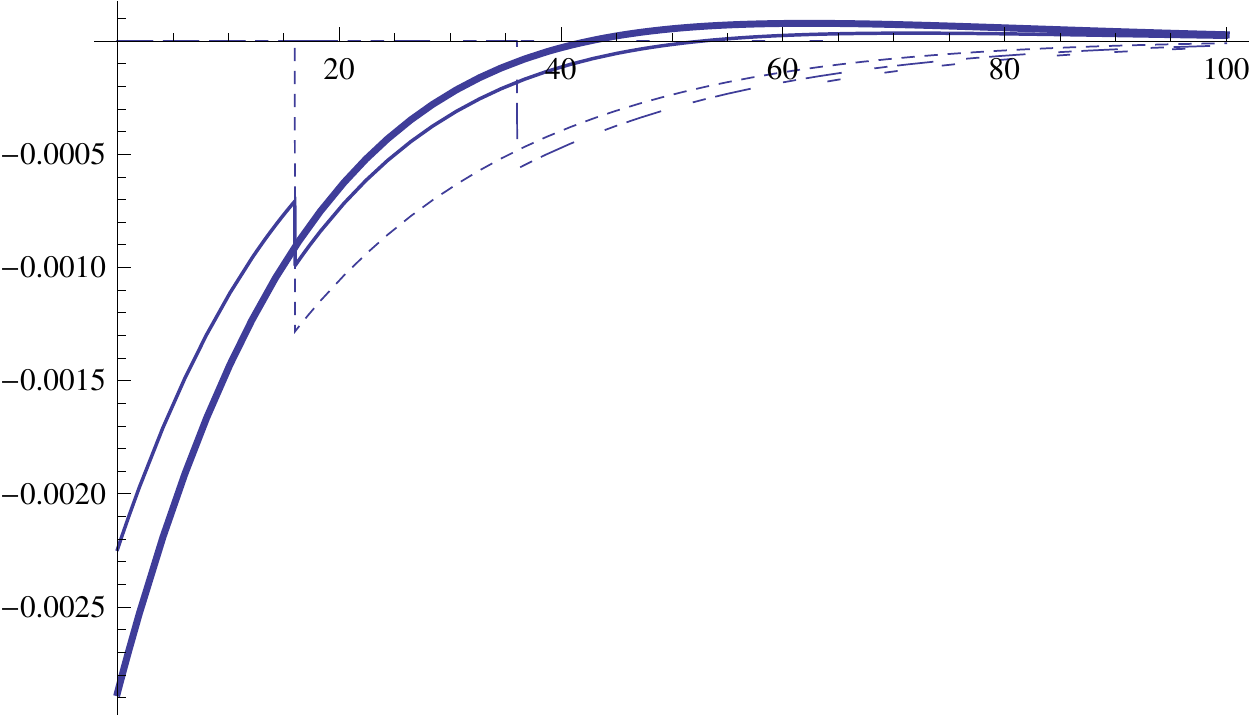}
}
\\
{\hspace{-2.0cm} $Q\rightarrow$keV}
\caption{ The differential rate $\frac{d\tilde{H}}{dQ}$,   as a function of the recoil energy for a heavy target, e.g. $^{127}$I assuming a nucleon cross section of $10^{-8}$pb. Panels (a) (b), (c) and (d) correspond to to 5, 20, 50 and 100 GeV WIMP masses. Otherwise the notation is the same as that of Fig. \ref{fig:flowv}.}
 \label{fig:dHdQ127}
\end{center}
\end{figure}

\begin{figure}
\begin{center}
\subfloat[]
{
\rotatebox{90}{\hspace{0.0cm} $d\tilde{H}/dQ\rightarrow$kg/(y keV)}
\includegraphics[height=.17\textheight]{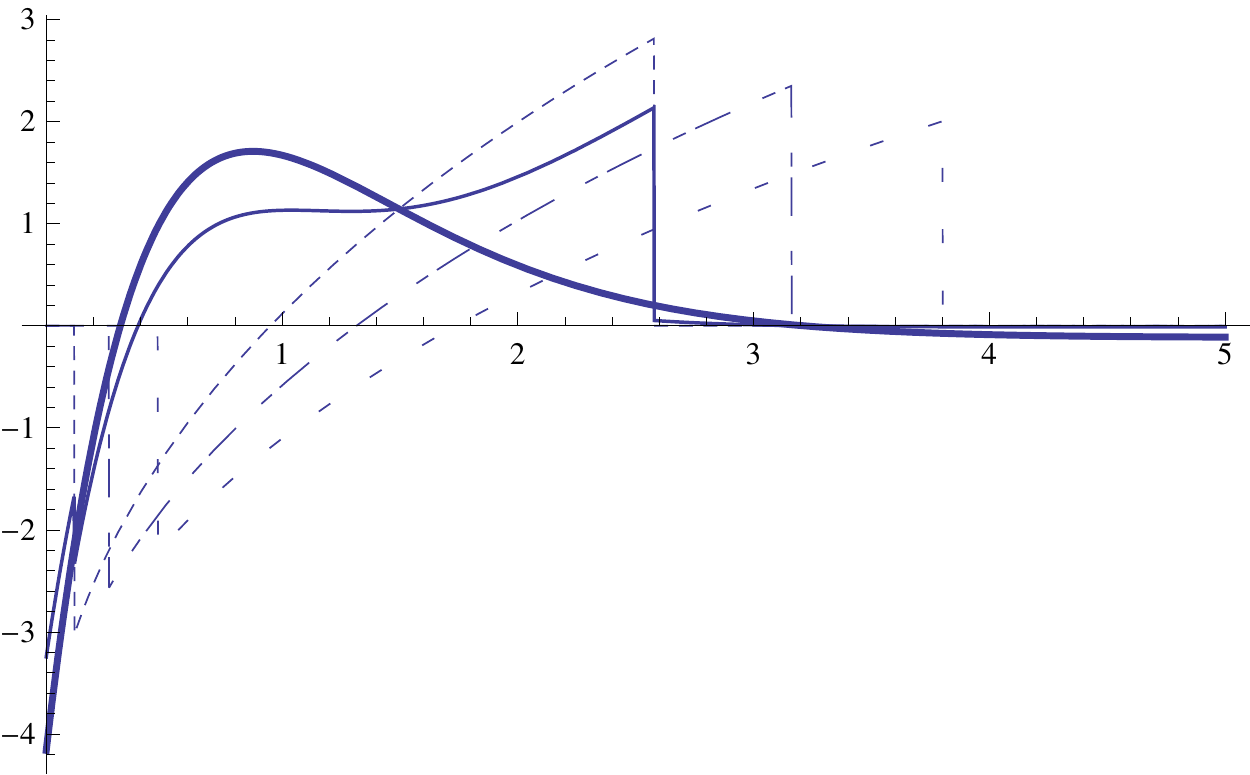}
}
\subfloat[]
{
\rotatebox{90}{\hspace{0.0cm} $d{\tilde H}/dQ\rightarrow$kg/(y keV)}
\includegraphics[height=.17\textheight]{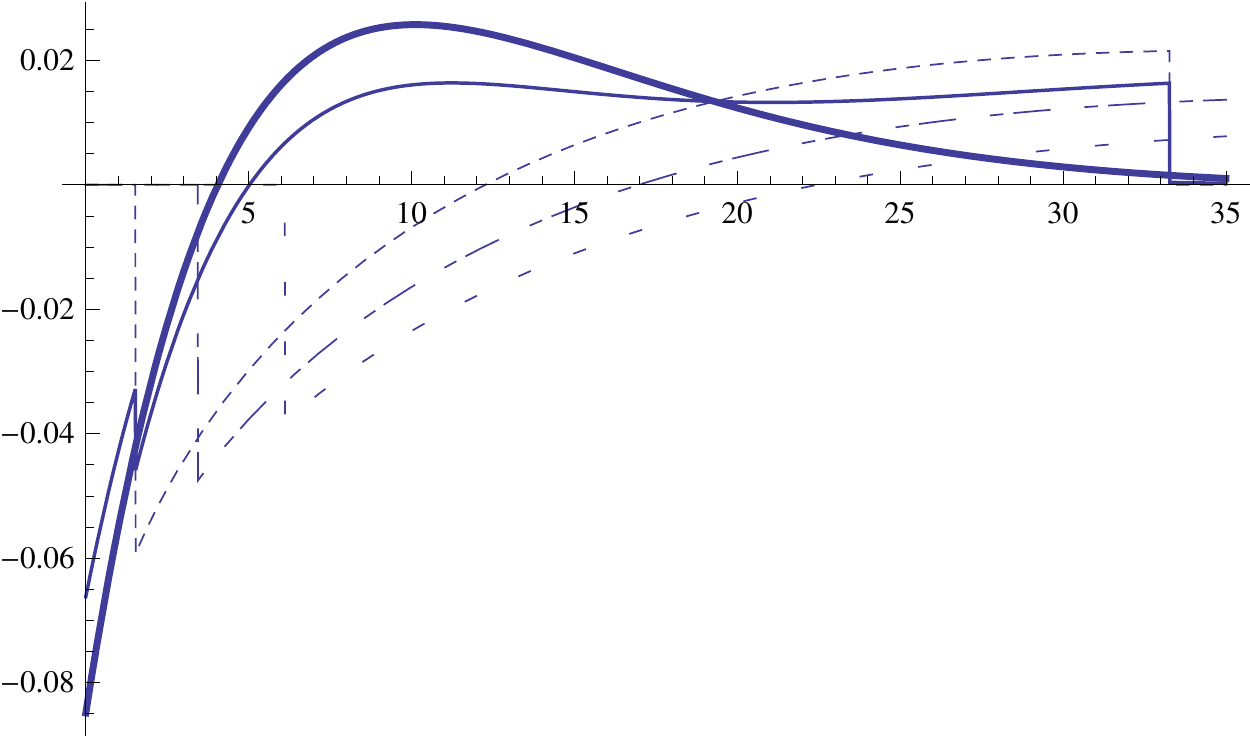}
}
\\
%{\hspace{-2.0cm} $Q\rightarrow$keV}
\subfloat[]
{
\rotatebox{90}{\hspace{0.0cm} $d\tilde{H}/dQ\rightarrow$kg/(y keV)}
\includegraphics[height=.17\textheight]{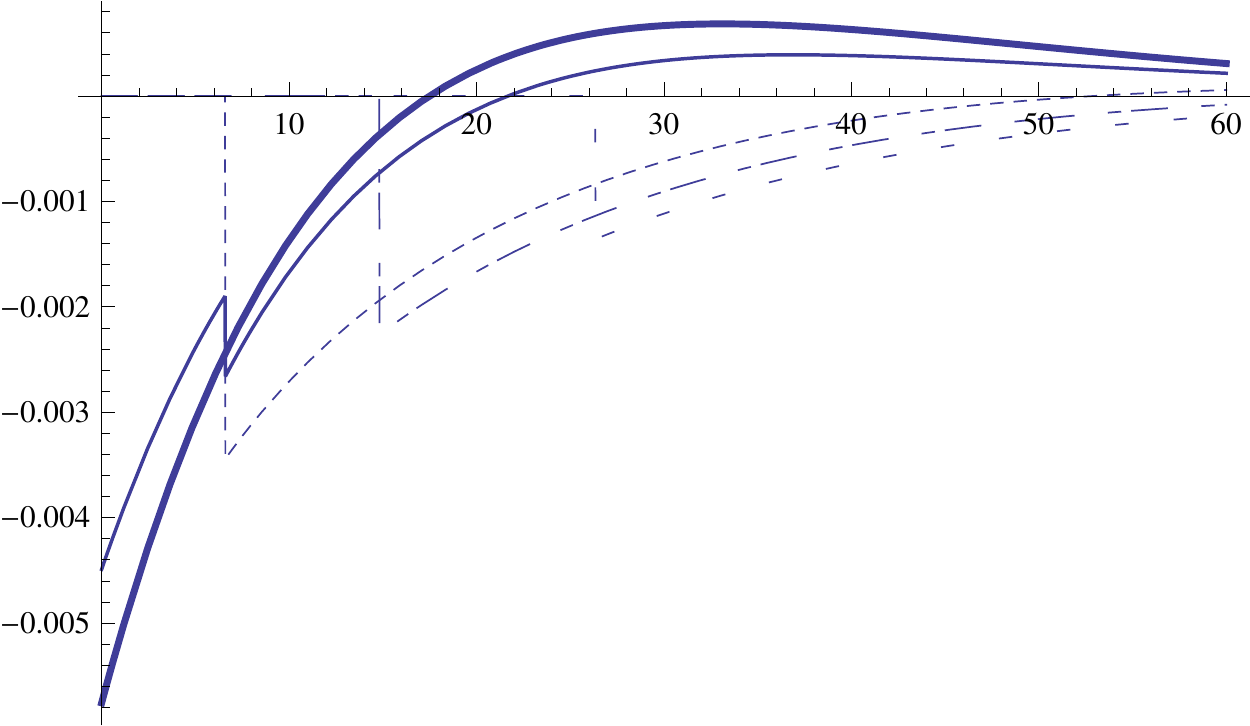}
}
\subfloat[]
{
\rotatebox{90}{\hspace{0.0cm} $d{\tilde H}/dQ\rightarrow$kg/(y keV)}
\includegraphics[height=.17\textheight]{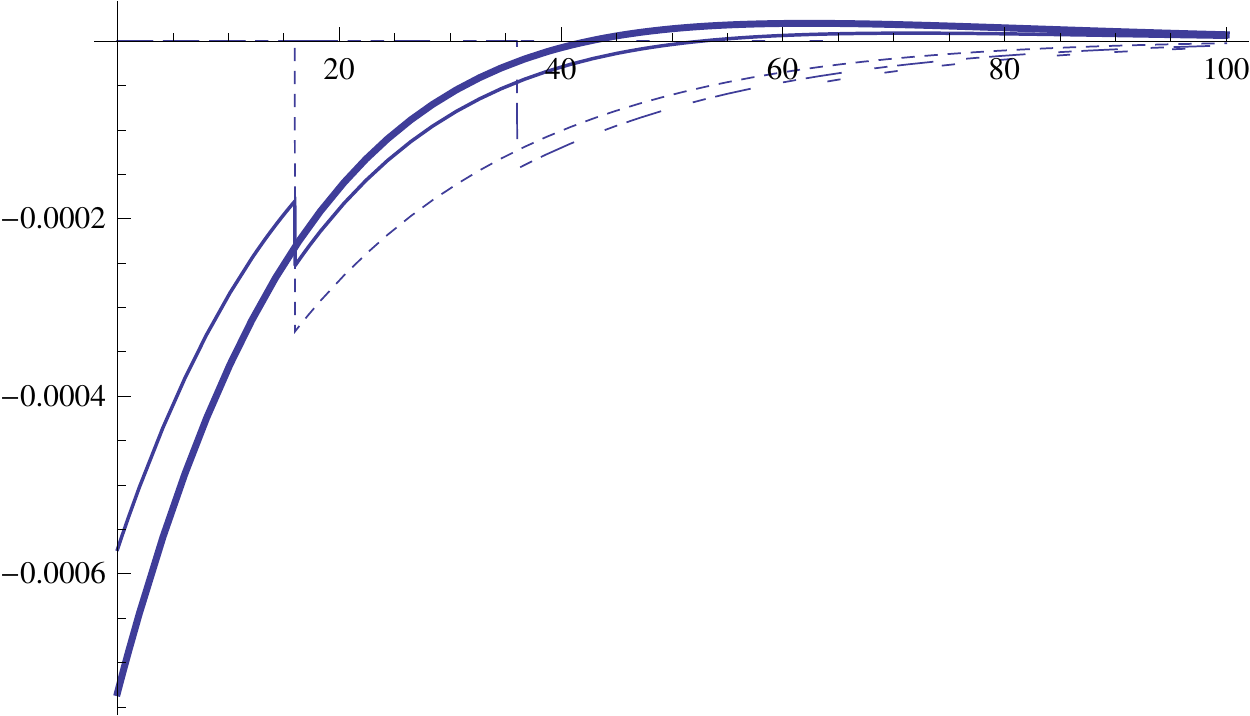}
}
\\
{\hspace{-2.0cm} $Q\rightarrow$keV}
\caption{ The same as in Fig. \ref{fig:dHdQ127} for a scalar WIMP assuming a nucleon cross section $(50/m_{\chi})^2 \times 10^{-8}$pb. }
 \label{fig:dHdQSc127}
\end{center}
\end{figure}

\begin{figure}
\begin{center}
\subfloat[]
{
\rotatebox{90}{\hspace{0.0cm} $dR/dQ\rightarrow$kg/(y keV)}
\includegraphics[height=.17\textheight]{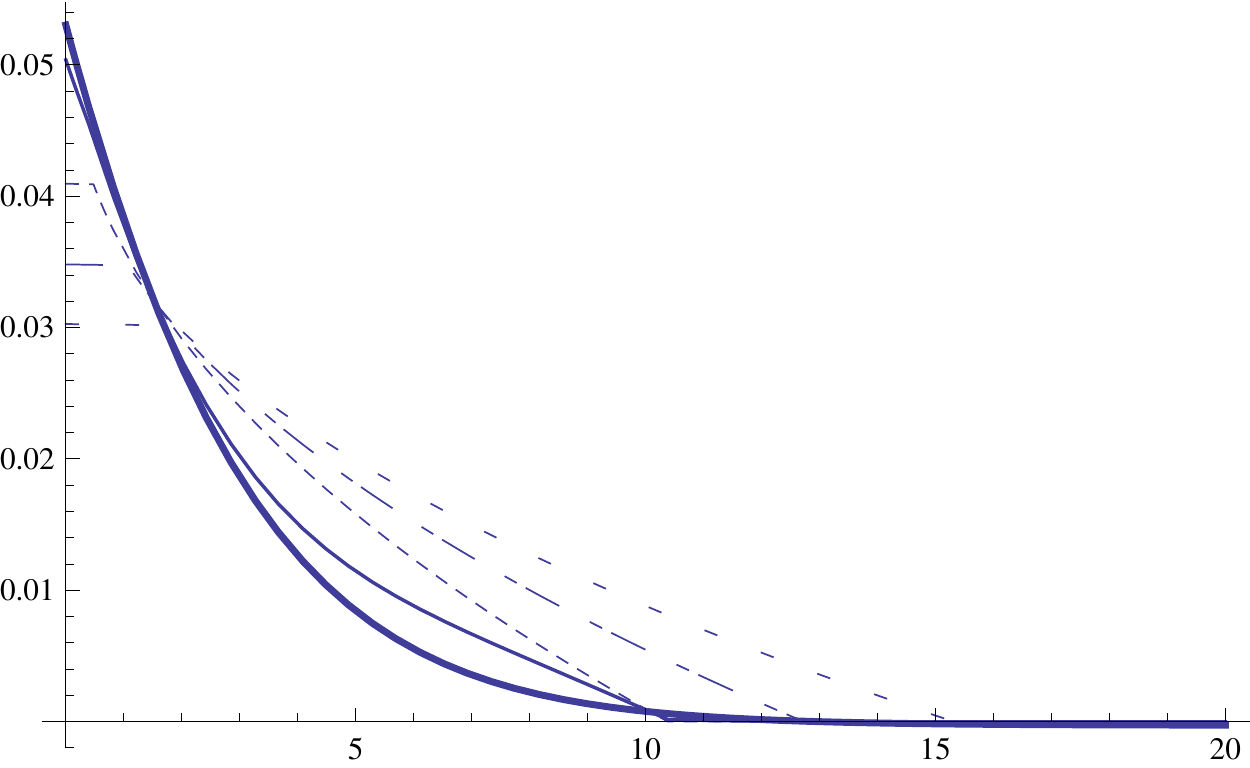}
}
\subfloat[]
{
\rotatebox{90}{\hspace{0.0cm} $dR/dQ\rightarrow$kg/(y keV)}
\includegraphics[height=.17\textheight]{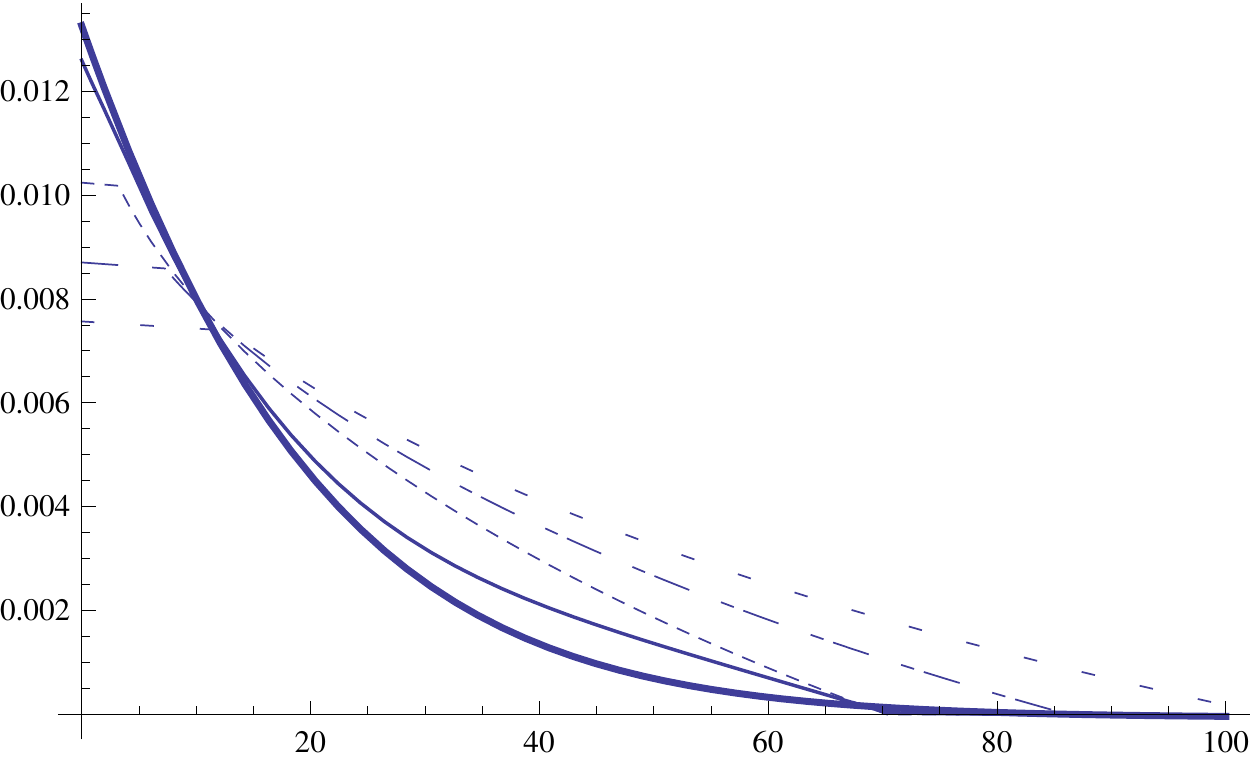}
}
\\
%{\hspace{-2.0cm} $Q\rightarrow$keV}
\subfloat[]
{
\rotatebox{90}{\hspace{0.0cm} $dR/dQ\rightarrow$kg/(y keV)}
\includegraphics[height=.17\textheight]{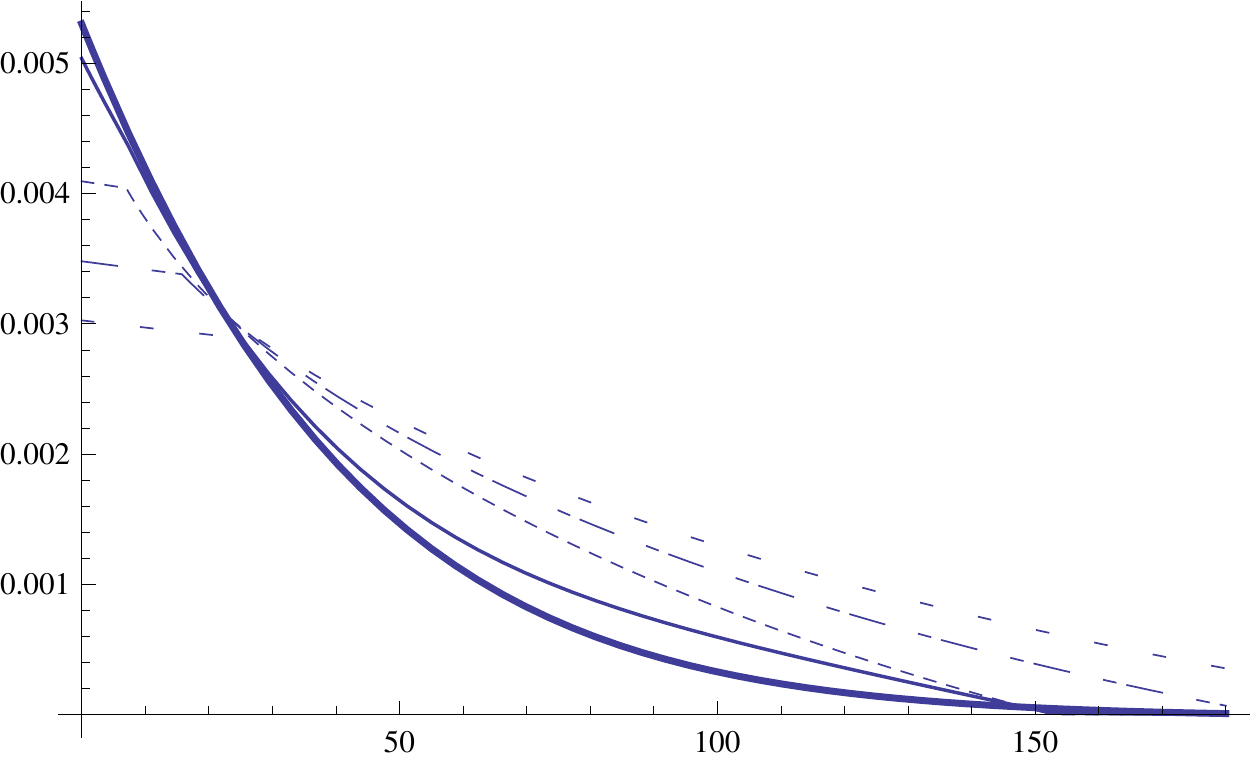}
}
\subfloat[]
{
\rotatebox{90}{\hspace{0.0cm} $dR/dQ\rightarrow$kg/(y keV)}
\includegraphics[height=.17\textheight]{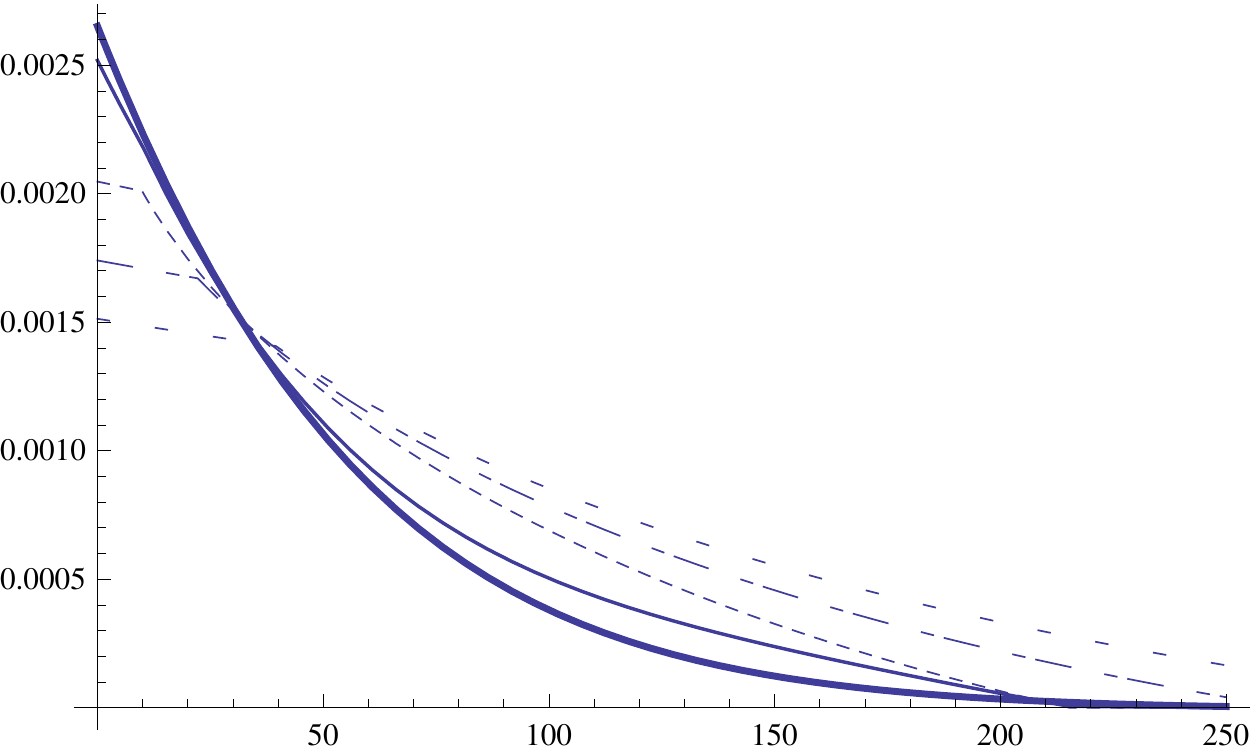}
}
\\
{\hspace{-2.0cm} $Q\rightarrow$keV}
\caption{ The differential rate $\frac{dR}{dQ}$,   as a function of the recoil energy for a light target, e.g. $^{23}$Na assuming a nucleon cross section of $10^{-8}$pb. Panels (a) (b), (c) and (d) correspond to to 5, 20, 50 and 100 GeV WIMP masses. Otherwise the notation is the same as that of Fig. \ref{fig:flowv}.}
 \label{fig:dRdQ23}
\end{center}
\end{figure}

\begin{figure}
\begin{center}
\subfloat[]
{
\rotatebox{90}{\hspace{0.0cm} $dR/dQ\rightarrow$kg/(y keV)}
\includegraphics[height=.17\textheight]{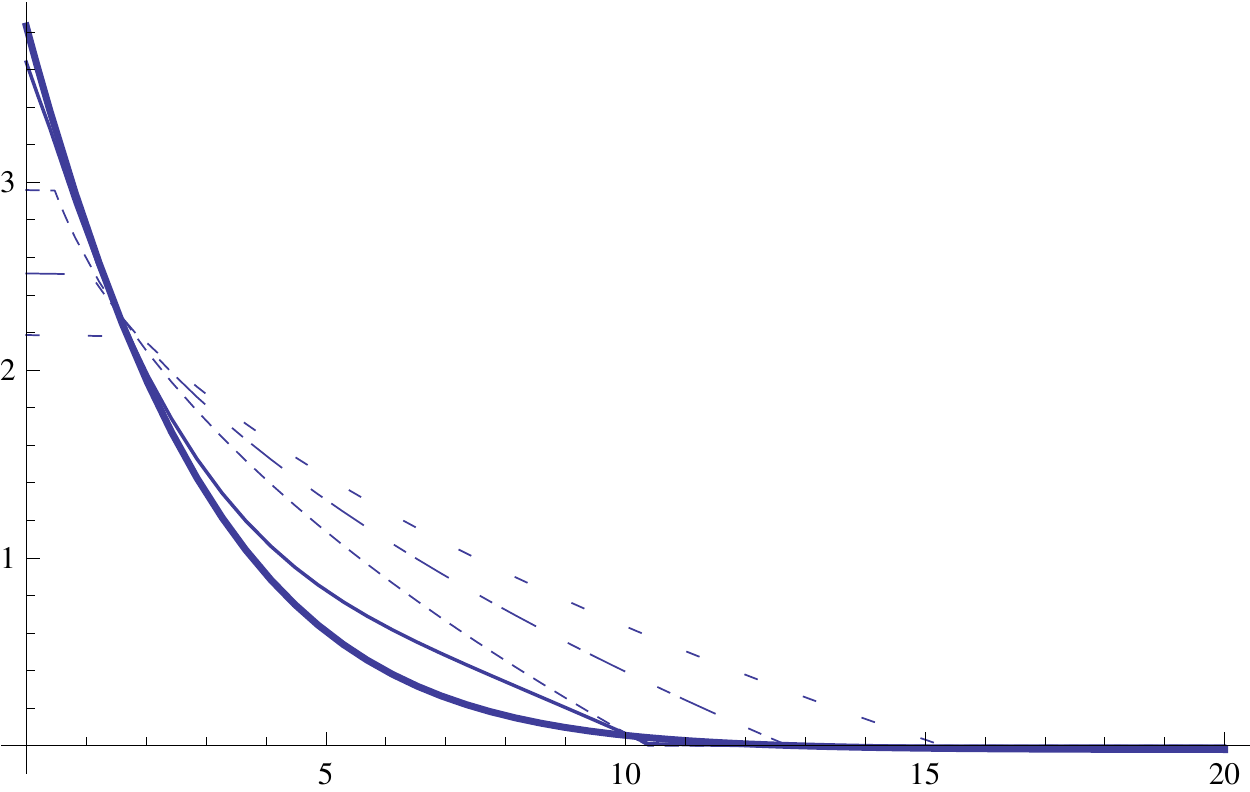}
}
\subfloat[]
{
\rotatebox{90}{\hspace{0.0cm} $dR/dQ\rightarrow$kg/(y keV)}
\includegraphics[height=.17\textheight]{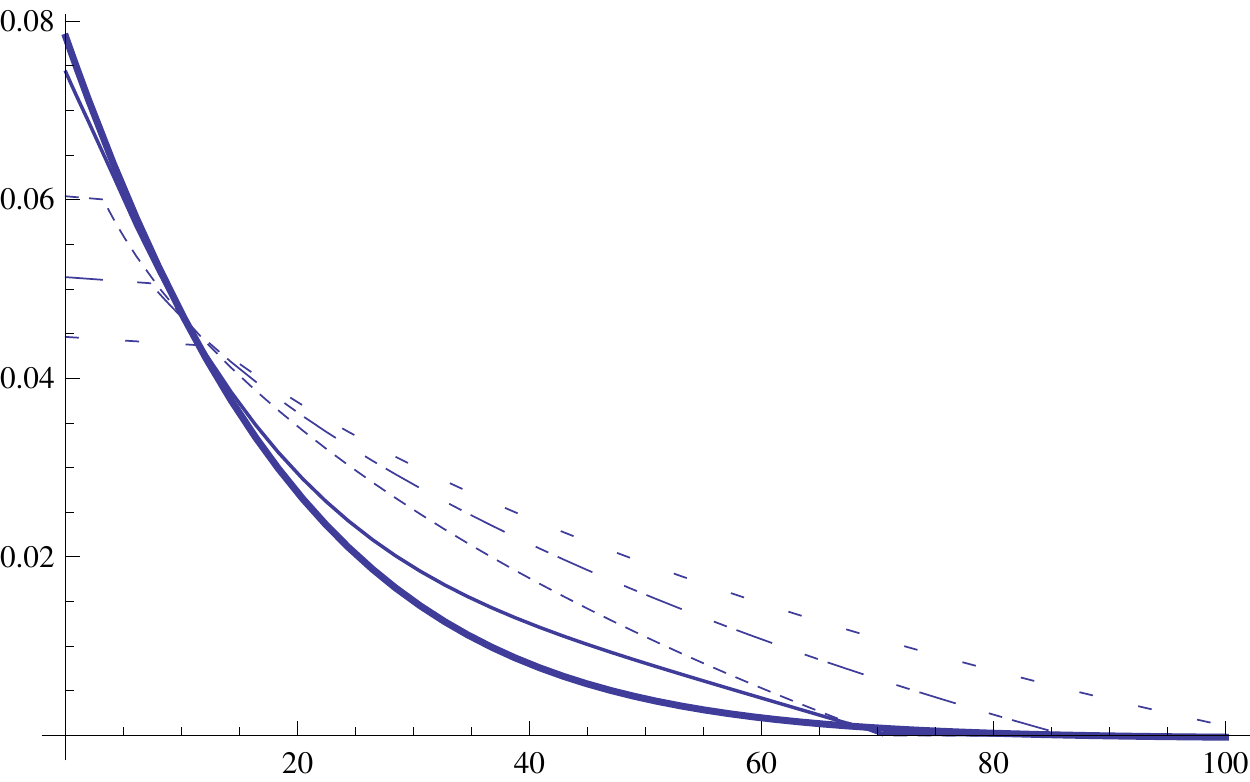}
}
\\
%{\hspace{-2.0cm} $Q\rightarrow$keV}
\subfloat[]
{
\rotatebox{90}{\hspace{0.0cm} $dR/dQ\rightarrow$kg/(y keV)}
\includegraphics[height=.17\textheight]{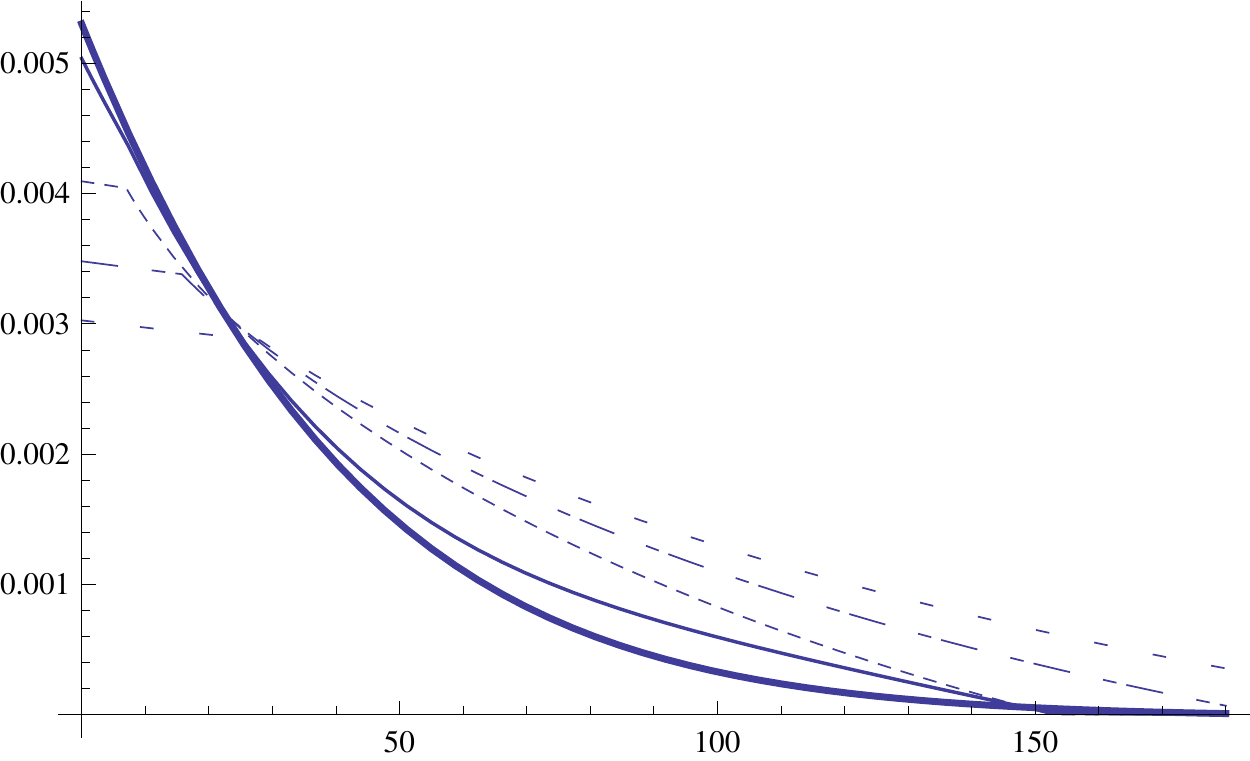}
}
\subfloat[]
{
\rotatebox{90}{\hspace{0.0cm} $dR/dQ\rightarrow$kg/(y keV)}
\includegraphics[height=.17\textheight]{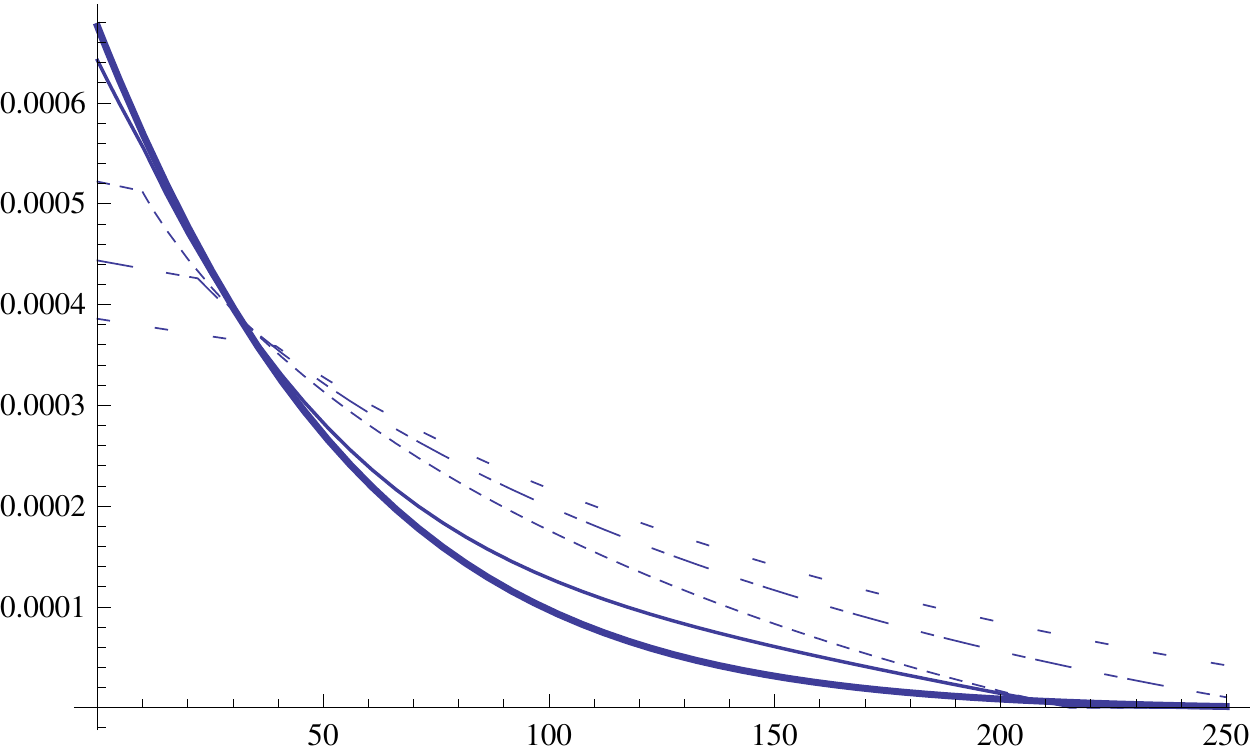}
}
\\
{\hspace{-2.0cm} $Q\rightarrow$keV}
\caption{ The same as in Fig. \ref{fig:dRdQSc127} in the case of a light target,  e.g. $^{23}$Na.}
 \label{fig:dRdQSc23}
\end{center}
\end{figure}

\begin{figure}
\begin{center}
\subfloat[]
{
\rotatebox{90}{\hspace{0.0cm} $d\tilde{H}/dQ\rightarrow$kg/(y keV)}
\includegraphics[height=.17\textheight]{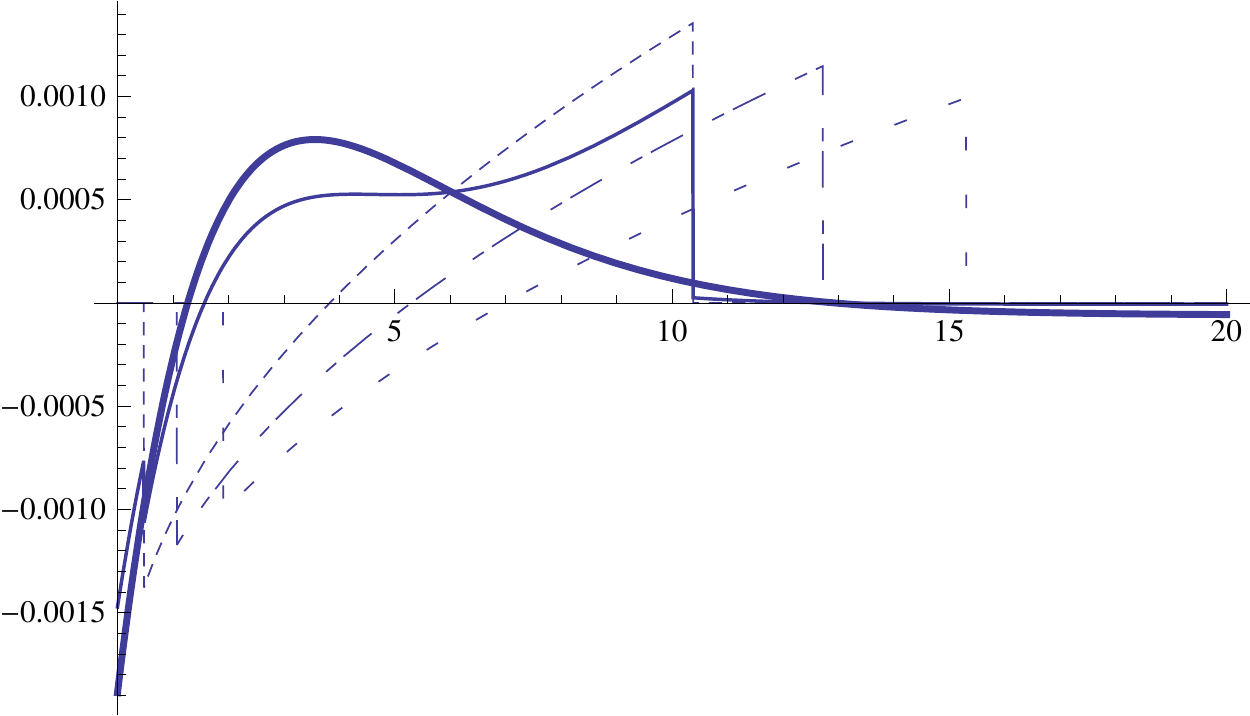}
}
\subfloat[]
{
\rotatebox{90}{\hspace{0.0cm} $d{\tilde H}/dQ\rightarrow$kg/(y keV)}
\includegraphics[height=.17\textheight]{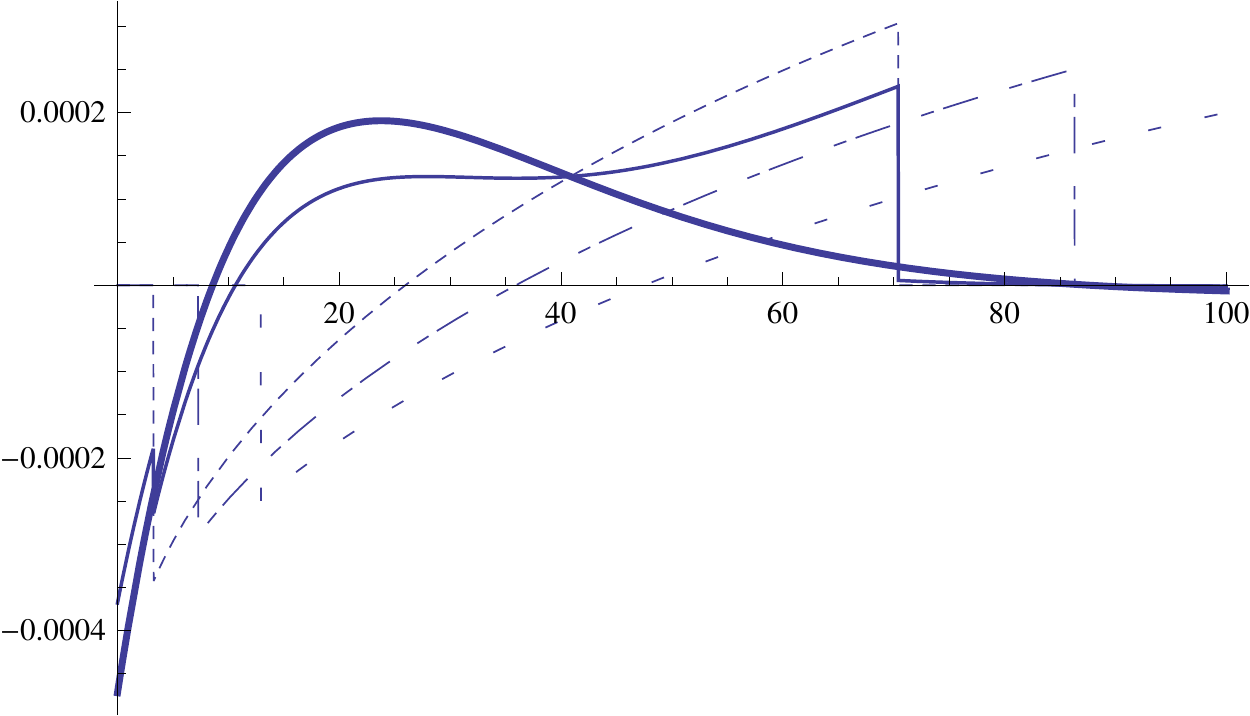}
}
\\
%{\hspace{-2.0cm} $Q\rightarrow$keV}
\subfloat[]
{
\rotatebox{90}{\hspace{0.0cm} $d\tilde{H}/dQ\rightarrow$kg/(y keV)}
\includegraphics[height=.17\textheight]{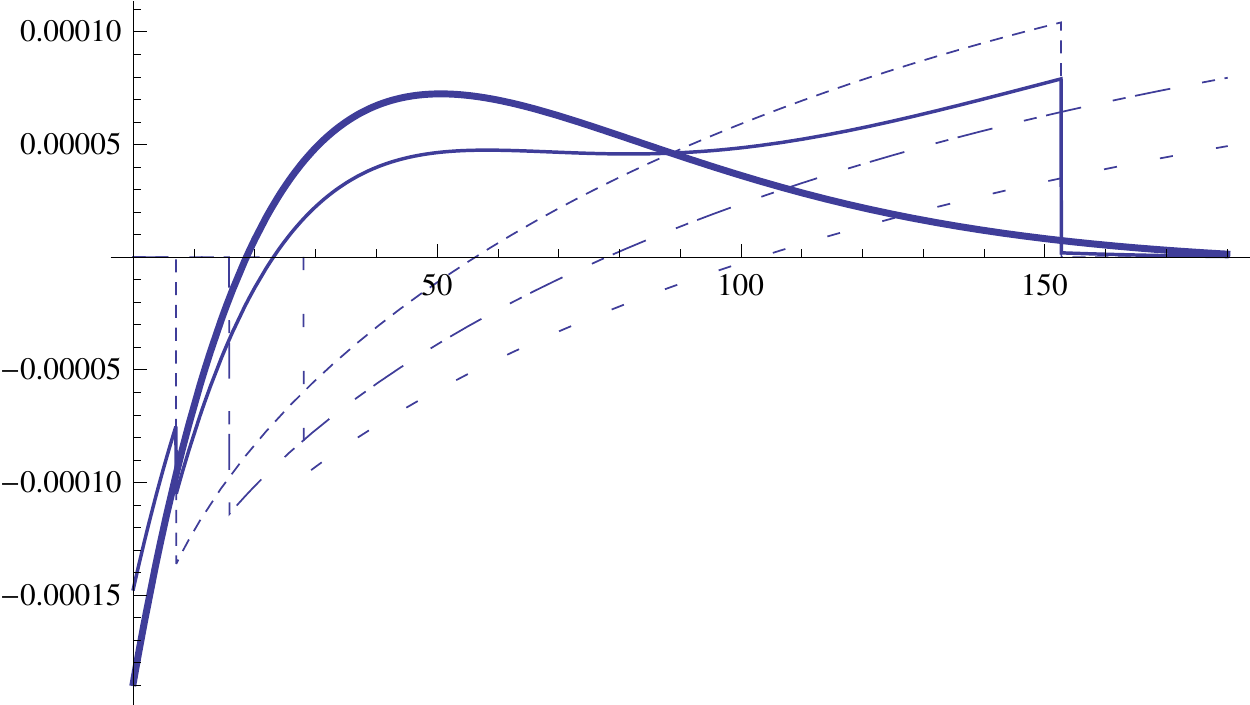}
}
\subfloat[]
{
\rotatebox{90}{\hspace{0.0cm} $d{\tilde H}/dQ\rightarrow$kg/(y keV)}
\includegraphics[height=.17\textheight]{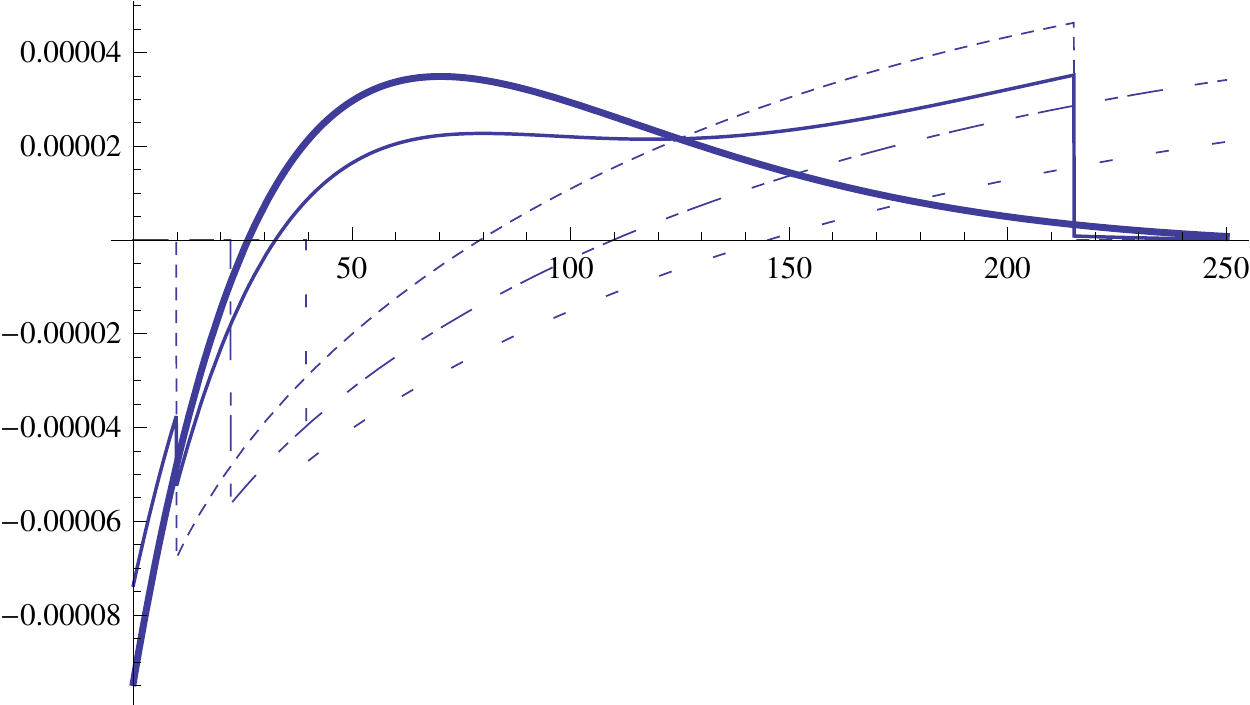}
}
\\
{\hspace{-2.0cm} $Q\rightarrow$keV}
\caption{ The differential rate $\frac{d\tilde{H}}{dQ}$,   as a function of the recoil energy for a light target, e.g. $^{23}$Na assuming a nucleon cross section of $10^{-8}$pb. Panels (a) (b), (c) and (d) correspond to to 5, 20, 50 and 100 GeV WIMP masses. Otherwise the notation is the same as that of Fig. \ref{fig:flowv}.}
 \label{fig:dHdQ23}
\end{center}
\end{figure}

\begin{figure}
\begin{center}
\subfloat[]
{
\rotatebox{90}{\hspace{0.0cm} $d\tilde{H}/dQ\rightarrow$kg/(y keV)}
\includegraphics[height=.17\textheight]{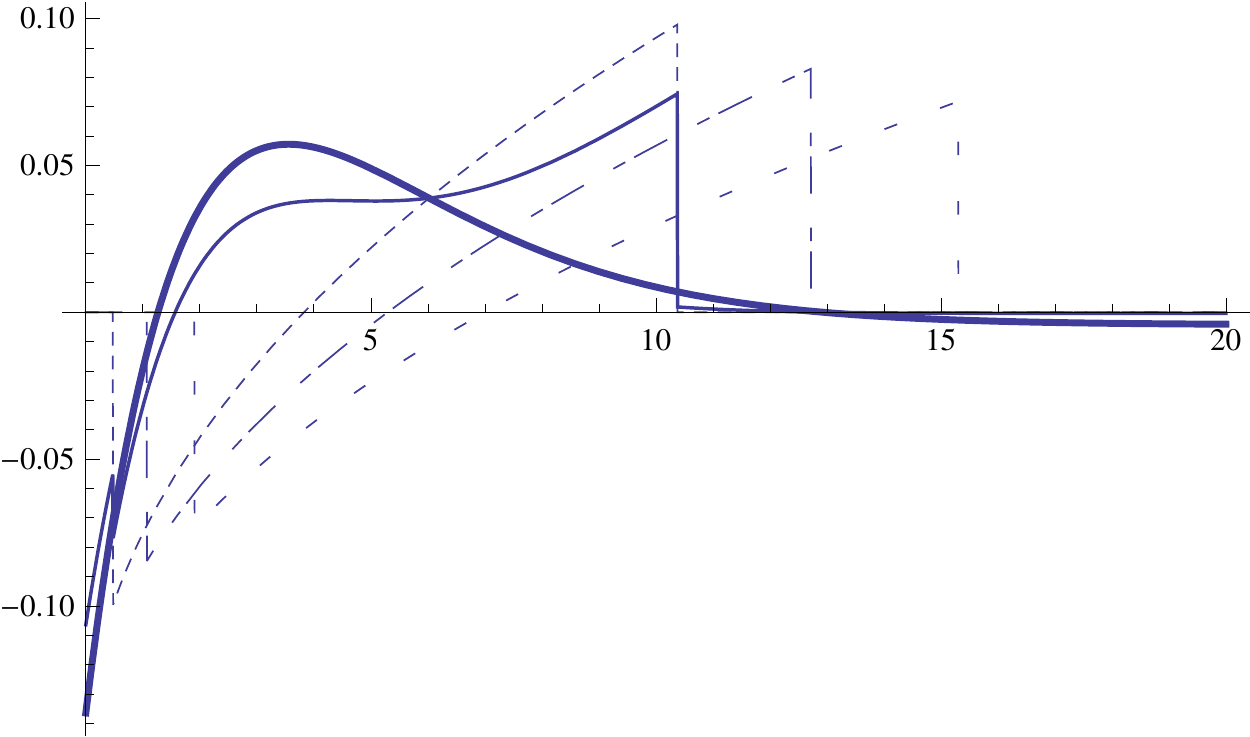}
}
\subfloat[]
{
\rotatebox{90}{\hspace{0.0cm} $d{\tilde H}/dQ\rightarrow$kg/(y keV)}
\includegraphics[height=.17\textheight]{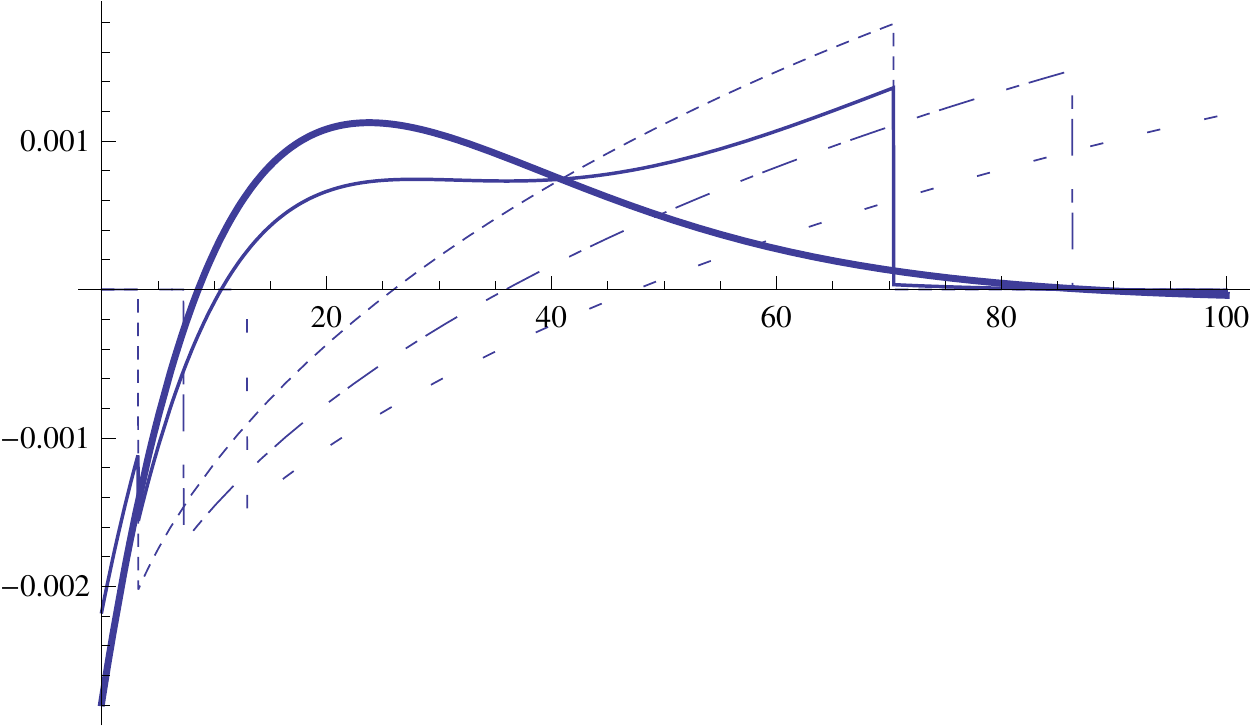}
}
\\
%{\hspace{-2.0cm} $Q\rightarrow$keV}
\subfloat[]
{
\rotatebox{90}{\hspace{0.0cm} $d\tilde{H}/dQ\rightarrow$kg/(y keV)}
\includegraphics[height=.17\textheight]{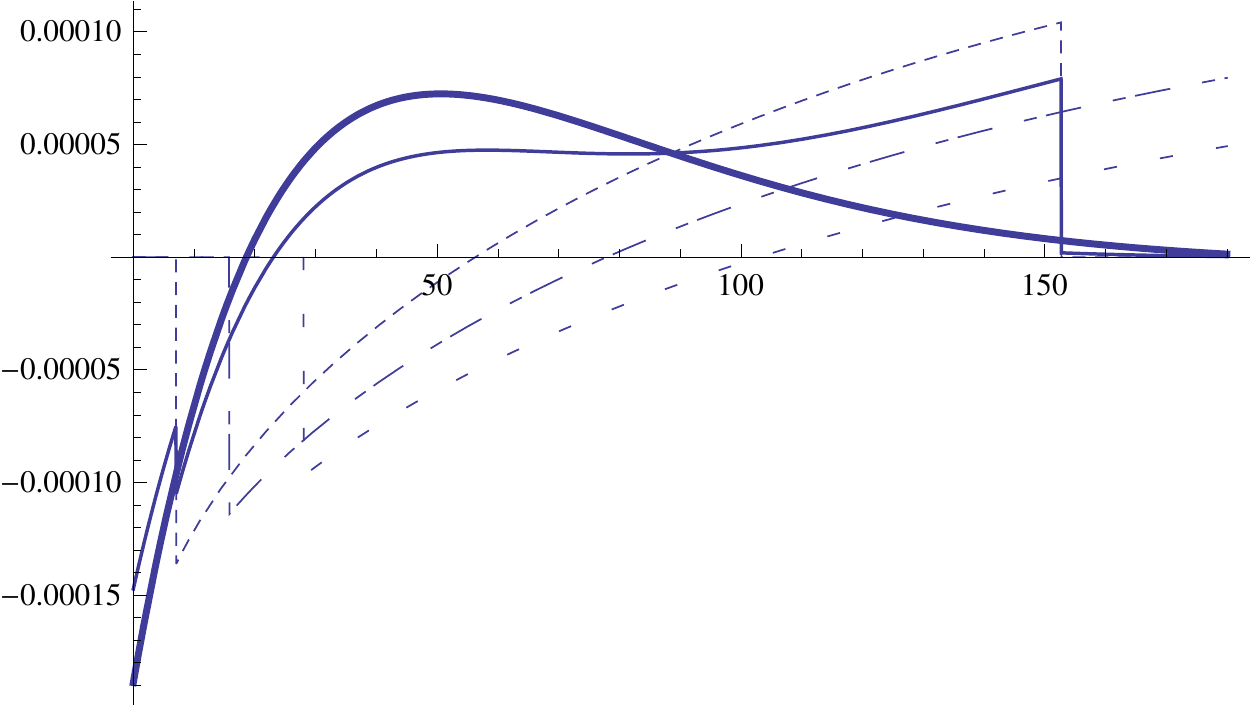}
}
\subfloat[]
{
\rotatebox{90}{\hspace{0.0cm} $d{\tilde H}/dQ\rightarrow$kg/(y keV)}
\includegraphics[height=.17\textheight]{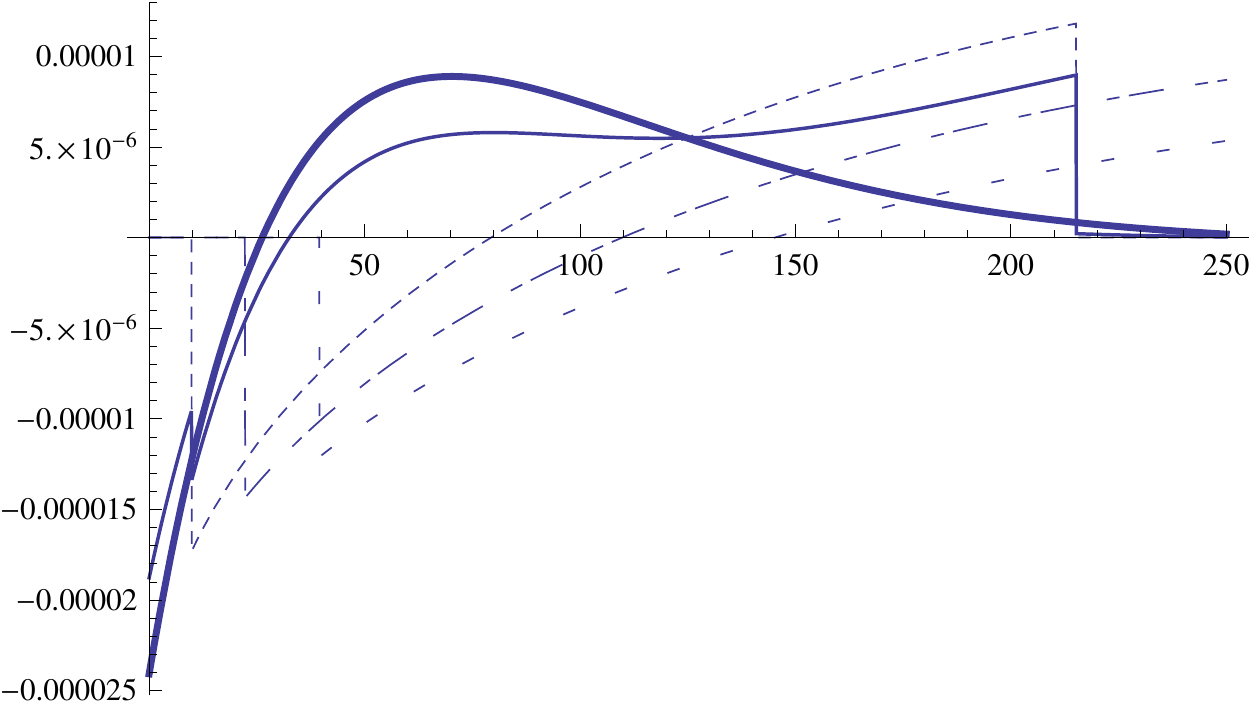}
}
\\
{\hspace{-2.0cm} $Q\rightarrow$keV}
\caption{ The  same as in Fig. \ref{fig:dHdQSc127} in the case of a light target,  e.g. $^{23}$ Na.}
 \label{fig:dHdQSc23}
\end{center}
\end{figure}

\begin{figure}
\begin{center}
\subfloat[]
{
\rotatebox{90}{\hspace{0.0cm} $dR/dQ\rightarrow$kg/(y keV)}
\includegraphics[height=.17\textheight]{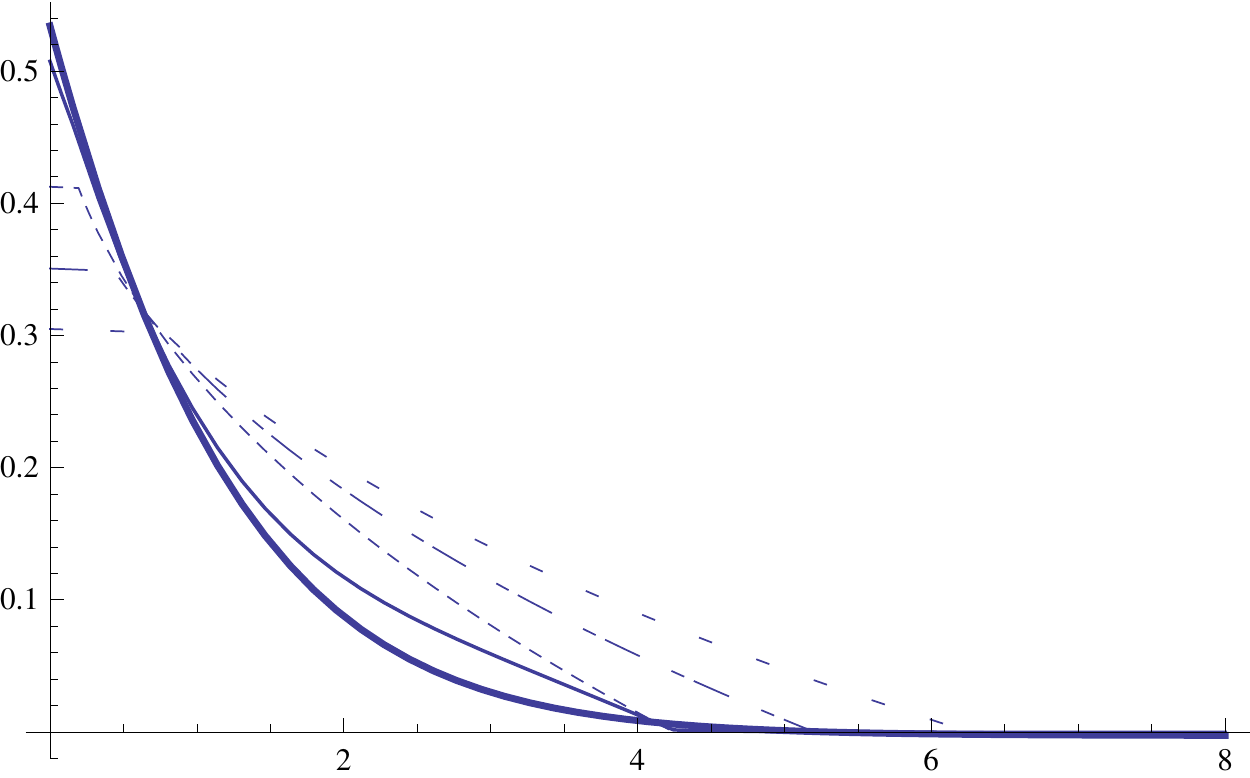}
}
\subfloat[]
{
\rotatebox{90}{\hspace{0.0cm} $dR/dQ\rightarrow$kg/(y keV)}
\includegraphics[height=.17\textheight]{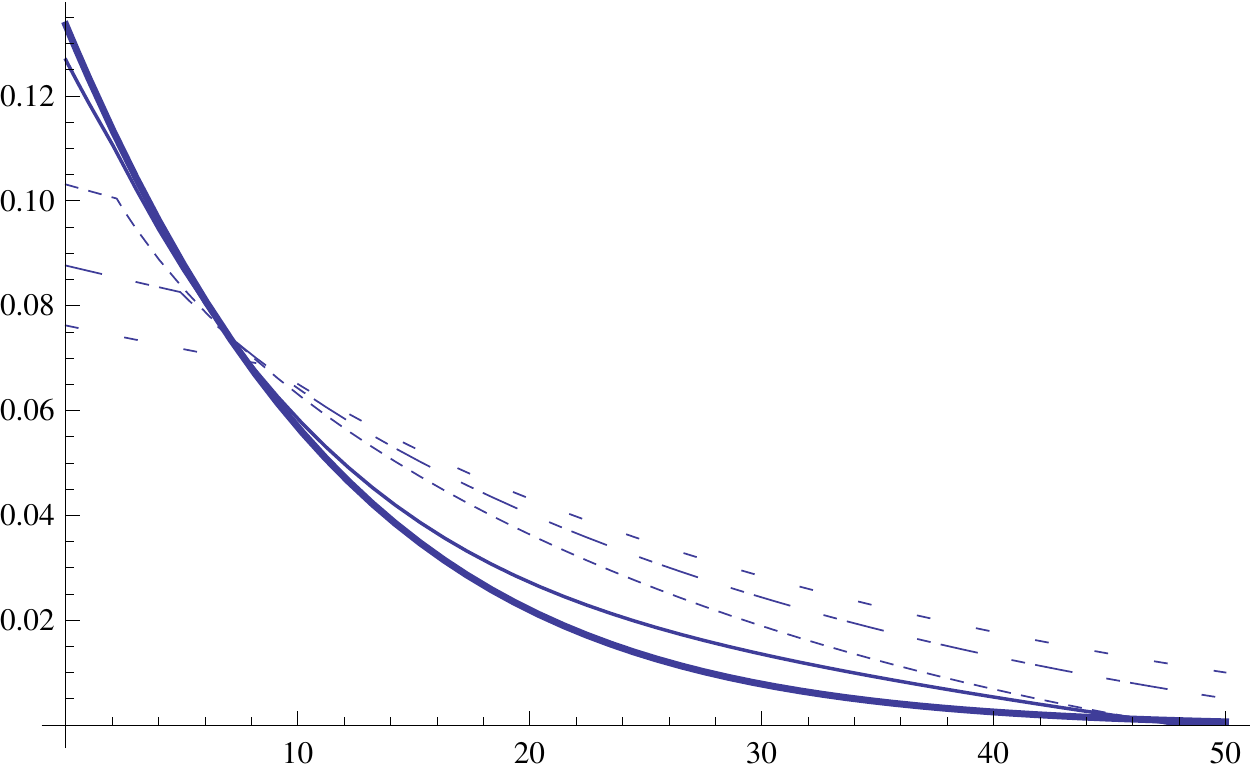}
}
\\
%{\hspace{-2.0cm} $Q\rightarrow$keV}
\subfloat[]
{
\rotatebox{90}{\hspace{0.0cm} $dR/dQ\rightarrow$kg/(y keV)}
\includegraphics[height=.17\textheight]{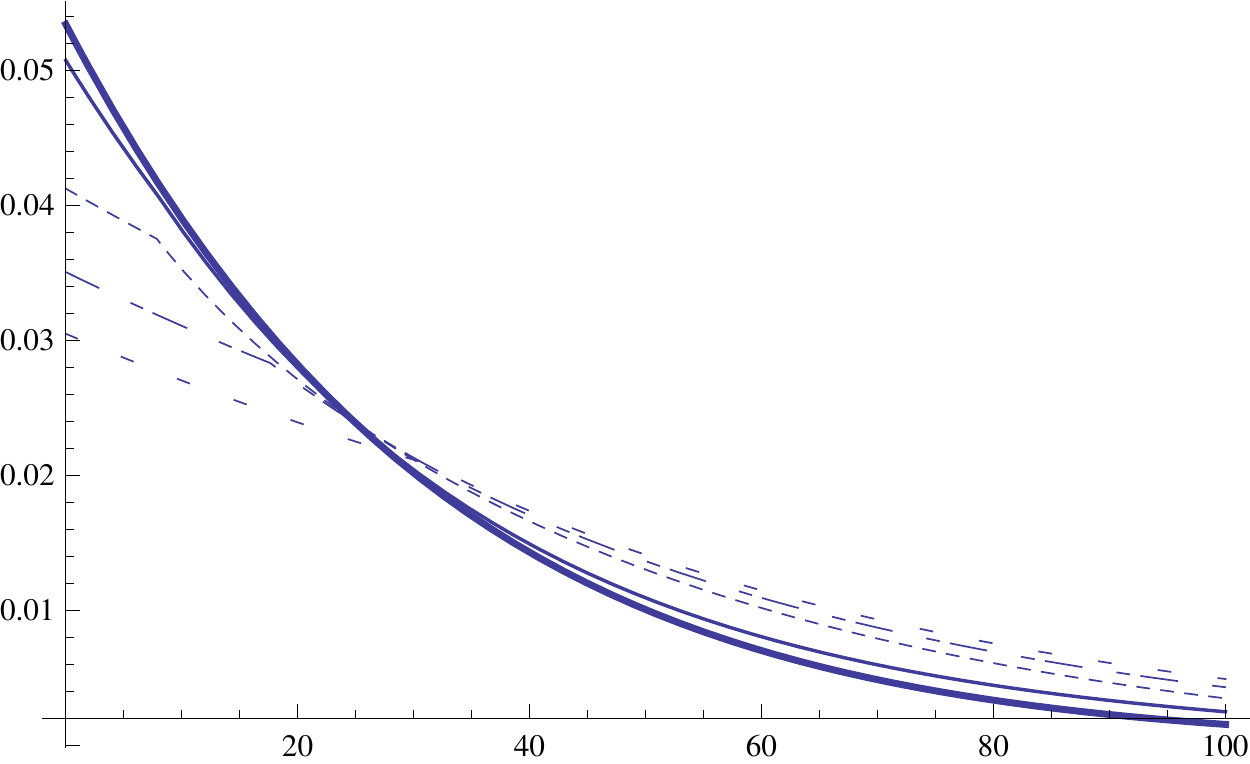}
}
\subfloat[]
{
\rotatebox{90}{\hspace{0.0cm} $dR/dQ\rightarrow$kg/(y keV)}
\includegraphics[height=.17\textheight]{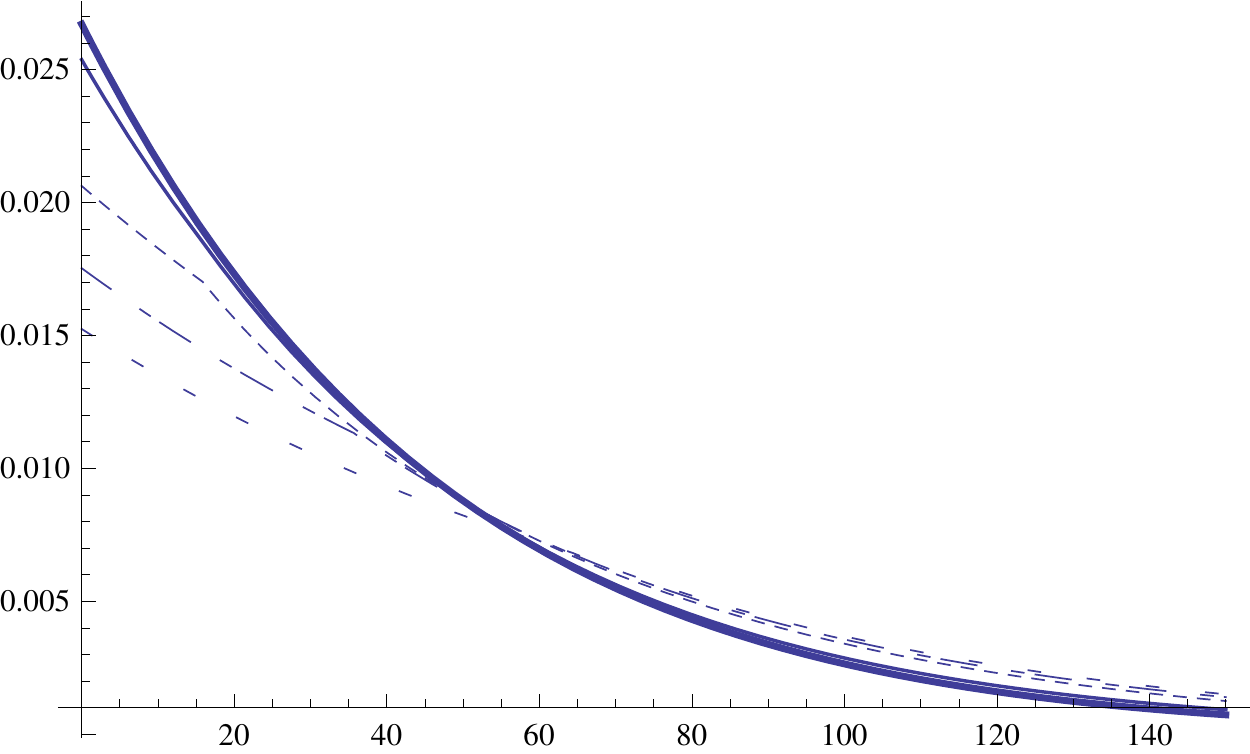}
}
\\
{\hspace{-2.0cm} $Q\rightarrow$keV}
\caption{ The differential rate $\frac{dR}{dQ}$,   as a function of the recoil energy for an intermediate   target, e.g. $^{73}$Ge assuming a nucleon cross section of $10^{-8}$pb. Panels (a) (b), (c) and (d) correspond to to 5, 20, 50 and 100 GeV WIMP masses. Otherwise the notation is the same as that of Fig. \ref{fig:flowv}.}
 \label{fig:dRdQ73}
\end{center}
\end{figure}

\begin{figure}
\begin{center}
\subfloat[]
{
\rotatebox{90}{\hspace{0.0cm} $dR/dQ\rightarrow$kg/(y keV)}
\includegraphics[height=.17\textheight]{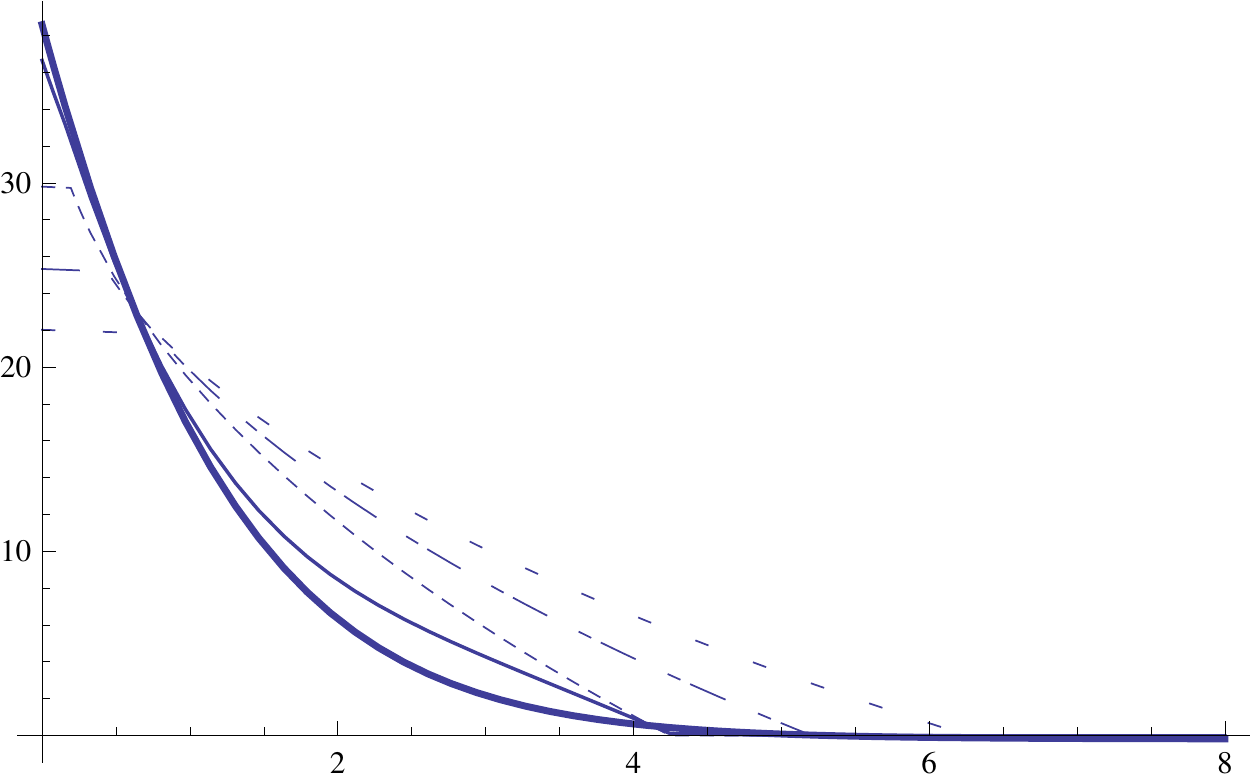}
}
\subfloat[]
{
\rotatebox{90}{\hspace{0.0cm} $dR/dQ\rightarrow$kg/(y keV)}
\includegraphics[height=.17\textheight]{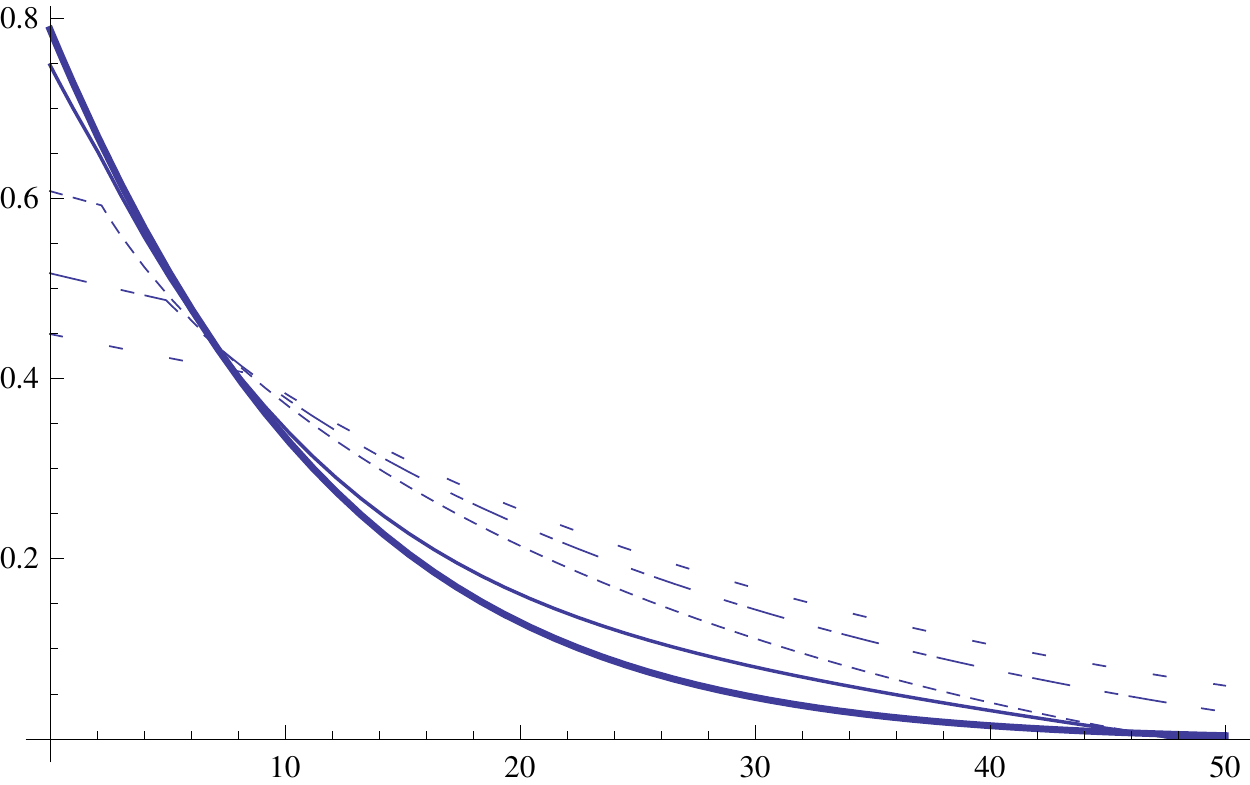}
}
\\
%{\hspace{-2.0cm} $Q\rightarrow$keV}
\subfloat[]
{
\rotatebox{90}{\hspace{0.0cm} $dR/dQ\rightarrow$kg/(y keV)}
\includegraphics[height=.17\textheight]{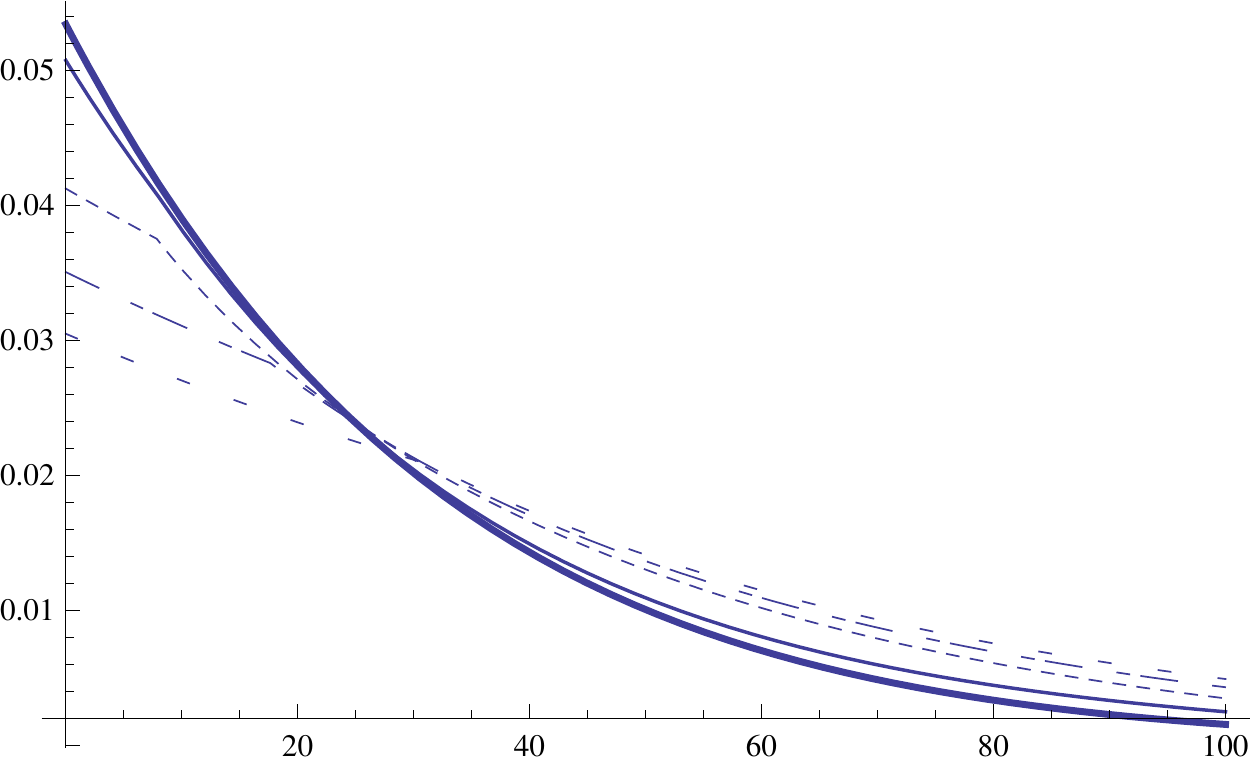}
}
\subfloat[]
{
\rotatebox{90}{\hspace{0.0cm} $dR/dQ\rightarrow$kg/(y keV)}
\includegraphics[height=.17\textheight]{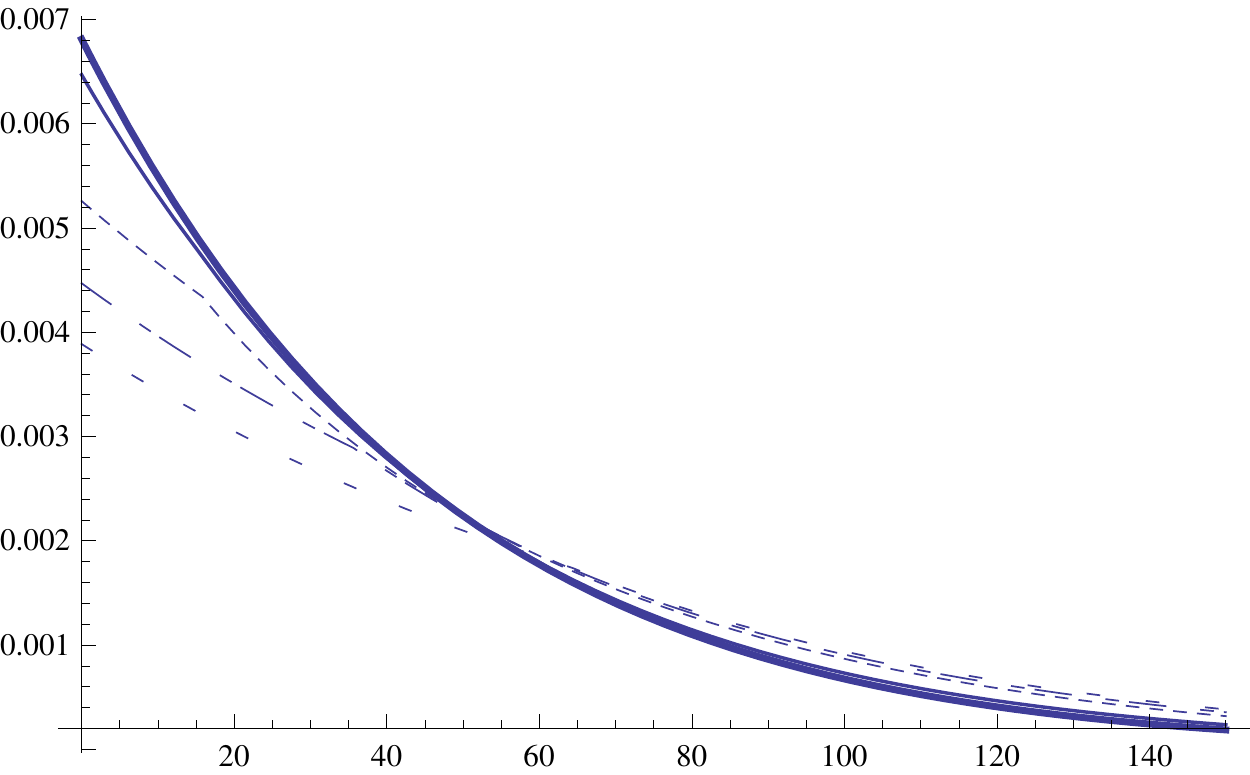}
}
\\
{\hspace{-2.0cm} $Q\rightarrow$keV}
\caption{ The same as in Fig. \ref{fig:dRdQSc127} in the case of an intermediate target, e.g. $^{73}Ge$.}
 \label{fig:dRdQSc73}
\end{center}
\end{figure}

\begin{figure}
\begin{center}
\subfloat[]
{
\rotatebox{90}{\hspace{0.0cm} $d\tilde{H}/dQ\rightarrow$kg/(y keV)}
\includegraphics[height=.17\textheight]{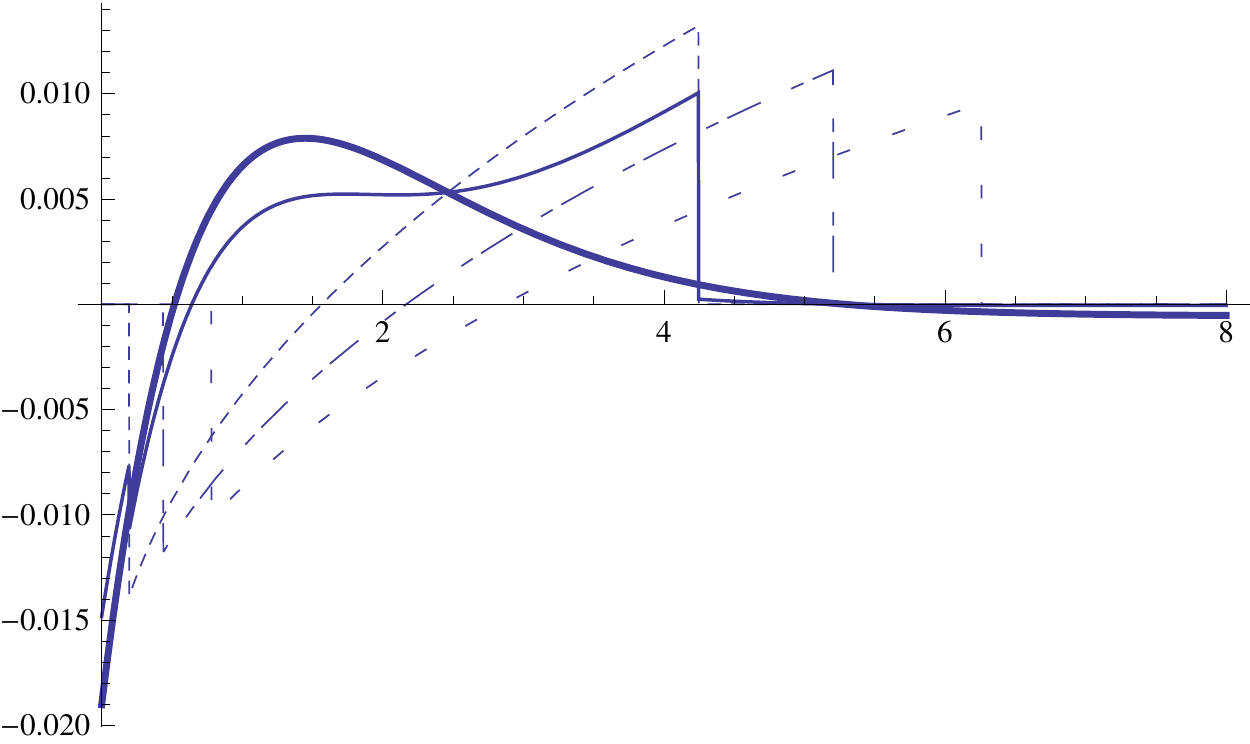}
}
\subfloat[]
{
\rotatebox{90}{\hspace{0.0cm} $d{\tilde H}/dQ\rightarrow$kg/(y keV)}
\includegraphics[height=.17\textheight]{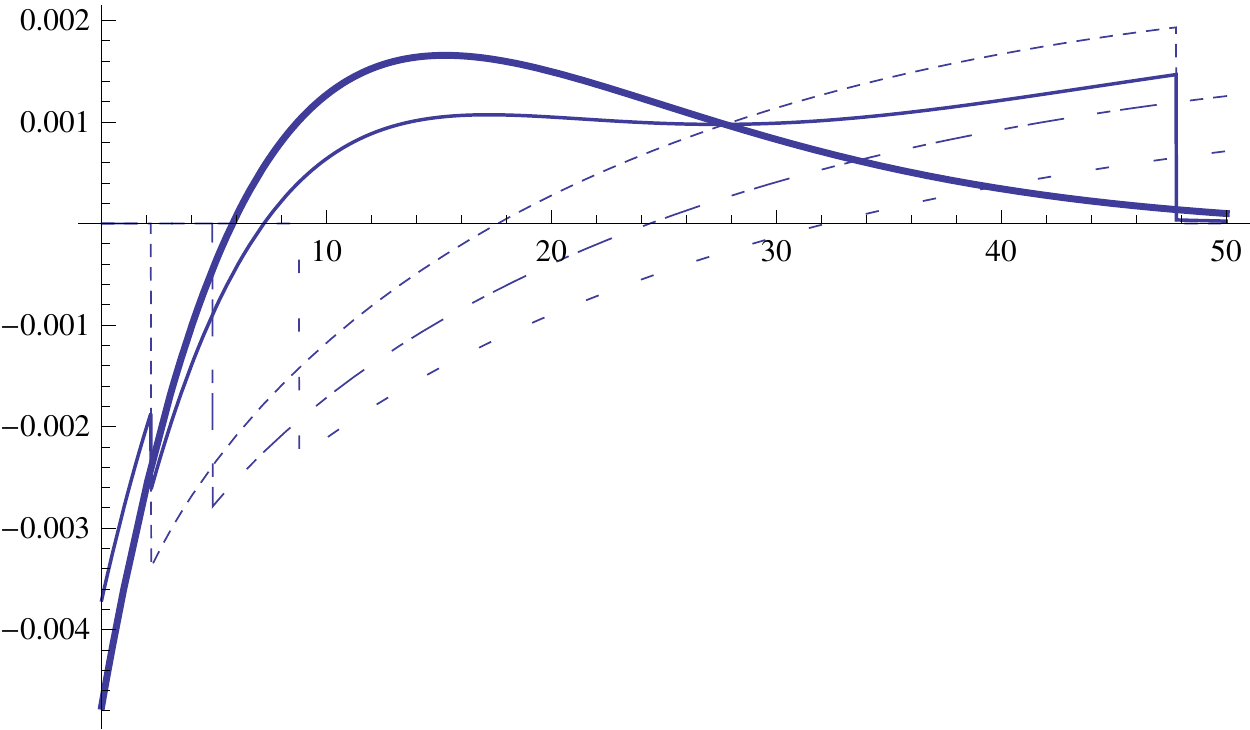}
}
\\
%{\hspace{-2.0cm} $Q\rightarrow$keV}
\subfloat[]
{
\rotatebox{90}{\hspace{0.0cm} $d\tilde{H}/dQ\rightarrow$kg/(y keV)}
\includegraphics[height=.17\textheight]{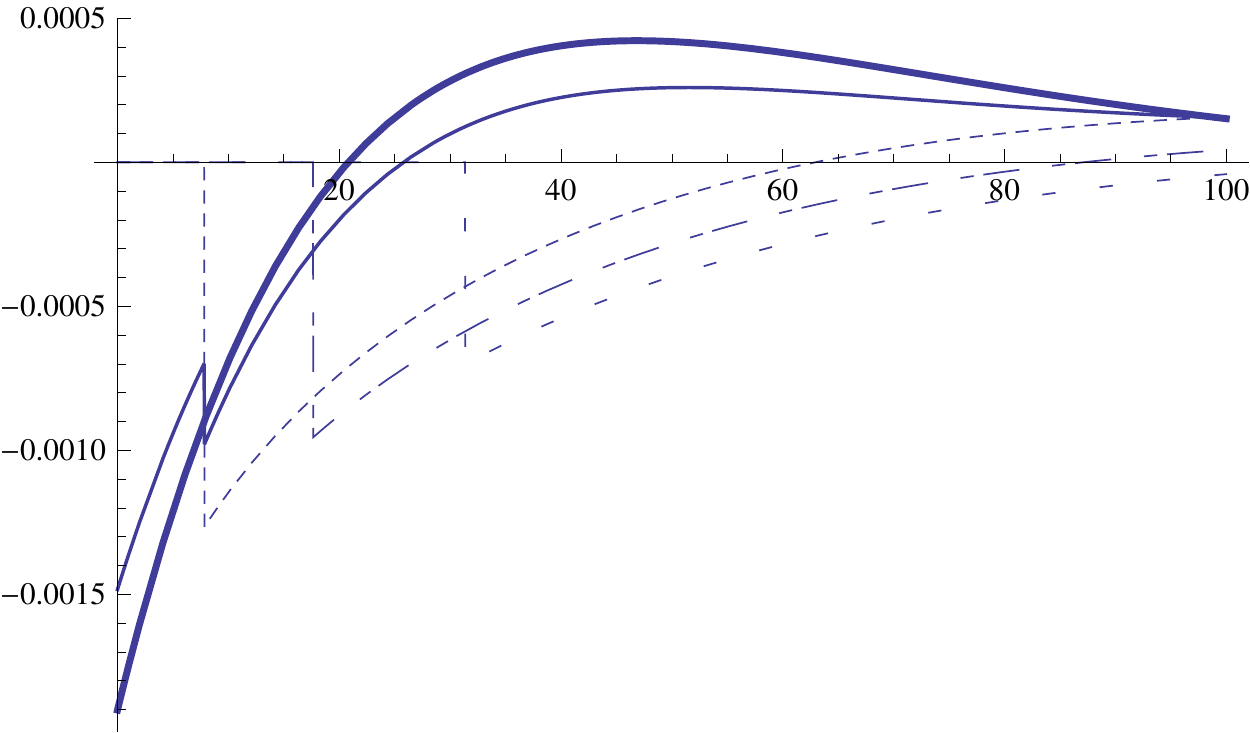}
}
\subfloat[]
{
\rotatebox{90}{\hspace{0.0cm} $d{\tilde H}/dQ\rightarrow$kg/(y keV)}
\includegraphics[height=.17\textheight]{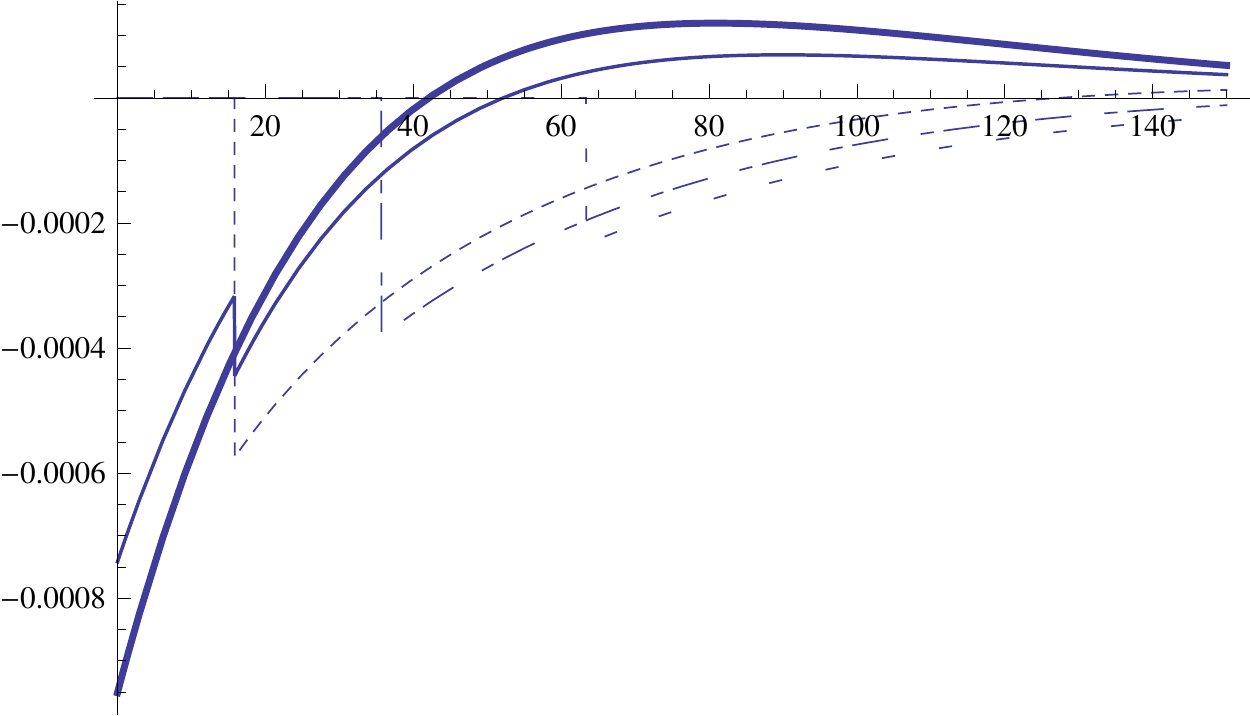}
}
\\
{\hspace{-2.0cm} $Q\rightarrow$keV}
\caption{ The differential rate $\frac{d\tilde{H}}{dQ}$,   as a function of the recoil energy for an intermediate target, e.g. $^{73}$Ge assuming a nucleon cross section of $10^{-8}$pb. Panels (a) (b), (c) and (d) correspond to to 5, 20, 50 and 100 GeV WIMP masses. Otherwise the notation is the same as that of Fig. \ref{fig:flowv}.}
 \label{fig:dHdQ73}
\end{center}
\end{figure}

\begin{figure}
\begin{center}
\subfloat[]
{
\rotatebox{90}{\hspace{0.0cm} $d\tilde{H}/dQ\rightarrow$kg/(y keV)}
\includegraphics[height=.17\textheight]{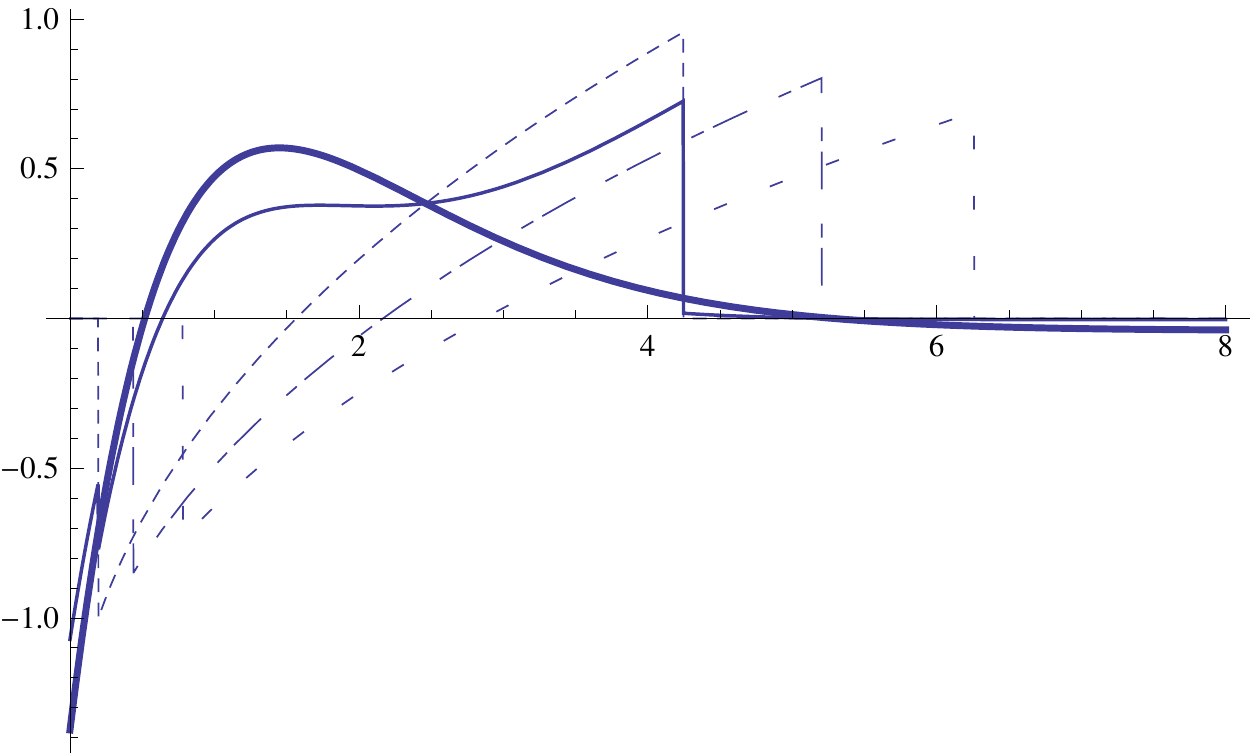}
}
\subfloat[]
{
\rotatebox{90}{\hspace{0.0cm} $d{\tilde H}/dQ\rightarrow$kg/(y keV)}
\includegraphics[height=.17\textheight]{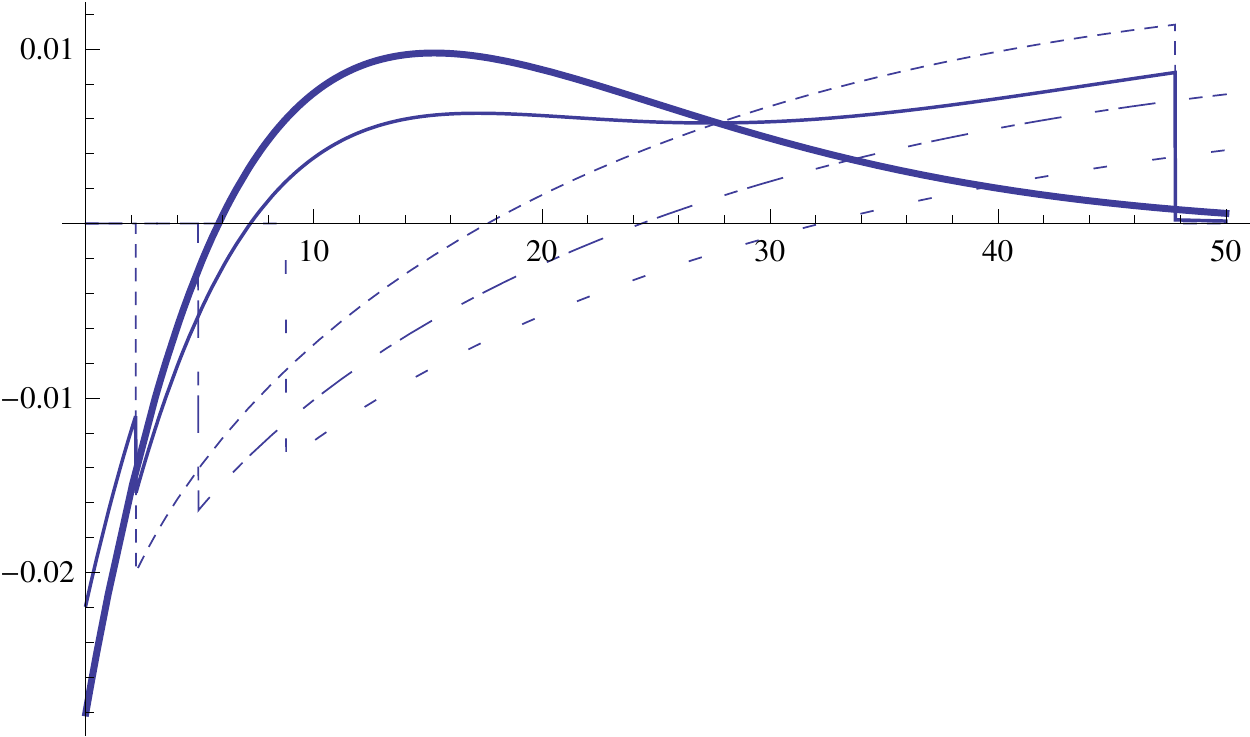}
}
\\
%{\hspace{-2.0cm} $Q\rightarrow$keV}
\subfloat[]
{
\rotatebox{90}{\hspace{0.0cm} $d\tilde{H}/dQ\rightarrow$kg/(y keV)}
\includegraphics[height=.17\textheight]{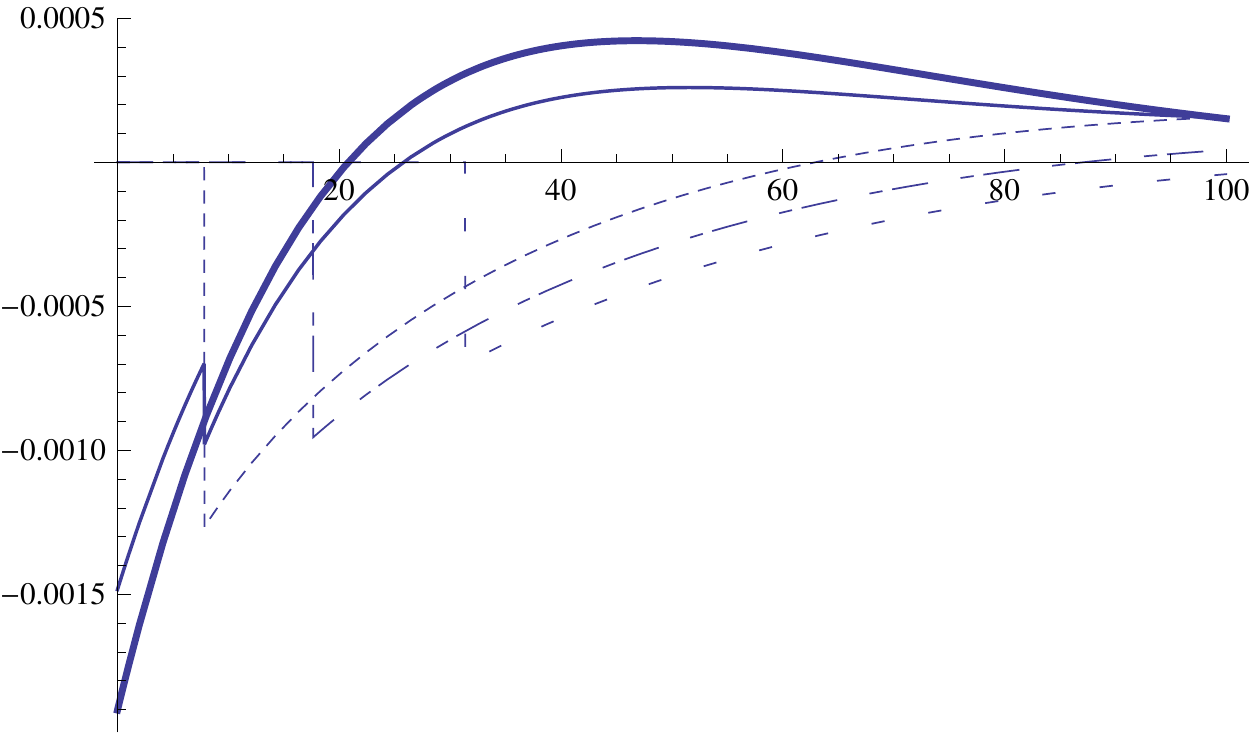}
}
\subfloat[]
{
\rotatebox{90}{\hspace{0.0cm} $d{\tilde H}/dQ\rightarrow$kg/(y keV)}
\includegraphics[height=.17\textheight]{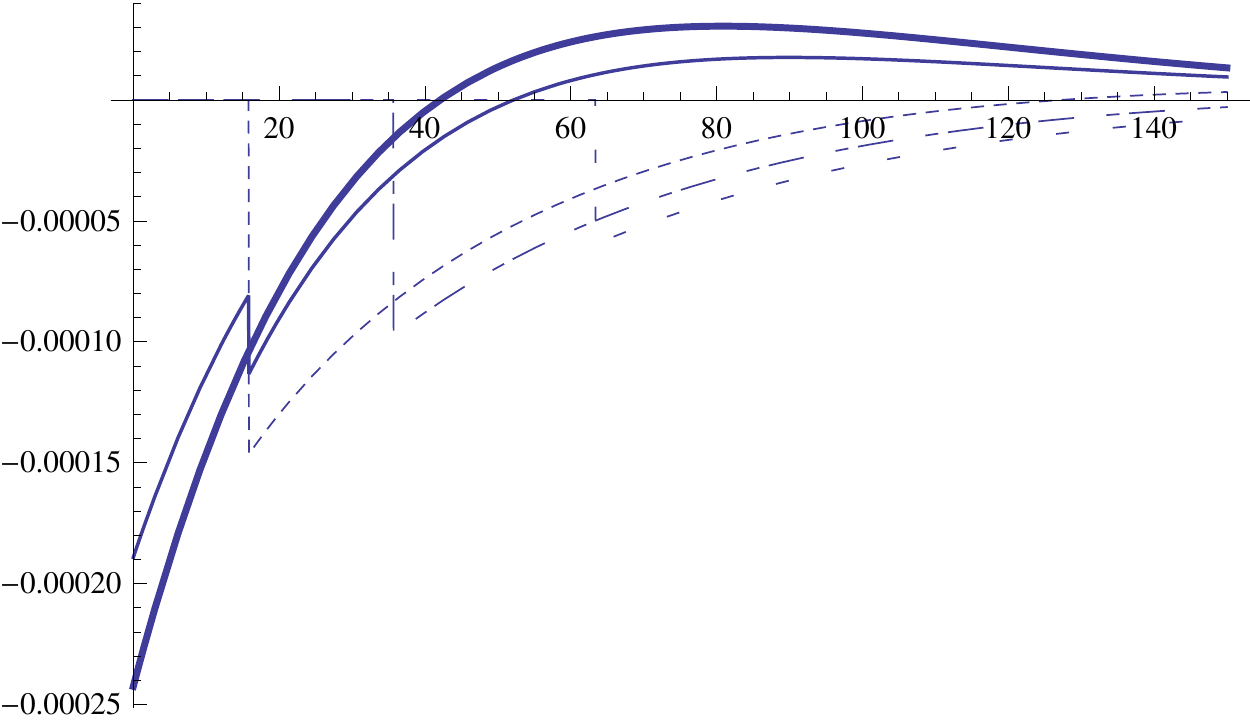}
}
\\
{\hspace{-2.0cm} $Q\rightarrow$keV}
\caption{ The same as in Fig. \ref{fig:dHdQSc127} in the case of an intermediate target, e.g. $^{73}$Ge.}
\label{fig:dHdQSc73}
\end{center}
\end{figure}

We will apply the above formalism in the case of I and Na, which are components of the  target NAI used in the DAMA experiment \cite{DAMA1,DAMA11} and Ge employed, e.g, by the CoGeNT experiment \cite{CoGeNT11}. The results for the Xe target \cite{XENON10} are similar to those for I and for the $^{19}$F target \cite{PICASSO09,PICASSO11} are similar to those for Na . 
The differential rates $\frac{dR}{dQ}|_A$ and  $\frac{d\tilde{H}}{dQ}|_A$, for each component ($A=127$ and $A=23$) and for $A=73$  are exhibited in Fig. \ref{fig:dRdQ127}-\ref{fig:dHdQSc73}. 
%Following the practice of the DAMA experiment we express the energy transfer is in keVee using the phenomenological quenching factor \cite{LIDHART,SIMON03}. 
The nuclear form factor has been included (for a heavy target, like  $^{127}$I or $^{131}$Xe, its effect is sizable even for an energy transfer\cite{JDV12n} of 10 keV, see Fig. \ref{fig:FFsq}).

By comparing the plots of the differential event rates of scalar WIMPs to the standard ones we find that the shapes are the same, but for low mass the scalar WIMPs lead to much larger event rates. So we will restrict the discussion on the shape of these plots to the results obtained for standard WIMPs.

 The introduction of debris flows  makes a small contribution at low  energy transfers. As expected\cite{spergel12} it tends to increase the differential rate at high energy transfers. This is particularly true for light small WIMP-nucleus reduced mass (see Figs \ref{fig:dRdQ127}, \ref{fig:dRdQ23} and \ref{fig:dRdQ73}). One, however, does not see any particular signature in the shape of the resulting curve. Furthermore the event rate in this region is about five times  smaller than the maximum. One, however, observes an interesting pattern concerning the time varying (modulated) part of the rate (see Figs \ref{fig:dHdQ127}, \ref{fig:dHdQ23} and \ref{fig:dHdQ73}). For a heavy target, like $^{127}$I or $^{131}$Xe, it is not surprising that, for WIMPs with relatively large mass, the modulation becomes negative, i.e. the rate becomes minimum in June 3nd, for all models considered here. For low WIMP masses, however, the sign of the modulation due to the flows is opposite to that of the M-B distribution. Thus the use of the light target nucleus $^{19}$F, combined with the low
detection threshold of 1.7 keV for recoil nuclei, makes PICASSO particularly
sensitive to low mass dark matter particles and gives it also some leverage in
the low mass region of the spin independent sector. The present stage of the
experiment\cite{PICASSO12} is approaching the sensitivity to challenge or confirm the claims
of seasonal modulations by the DAMA\cite{DAMA11} and CoGeNT\cite{CoGeNT11} experiments.
A similar situation arises in the case of an intermediate target, like $^{73}$Ge. Here the M-B distribution yields a negative value only for very low energy transfers. The situation becomes most interesting in the case of a light target, see Fig. \ref{fig:dHdQ23}. Here, with the possible exception of quite low energy transfers, which perhaps are below or very near threshold, the M-B distribution yields  a positive modulation amplitude, i.e. a maximum on June 3nd, while the result of debris flows is to cause a change in sign as one moves to  high energy transfers. Also in this case the modulation amplitude tends to increase as the energy transfer increase, while the corresponding contribution due to the M-B distribution tends to decrease. We should remark though that the total rate (average+modulated) tends to decrease at high momentum transfers. We should also stress that we have presented here the absolute modulate rate (events per kg target per year). The relative modulated amplitude (the ratio of the time varying rate divided by the time averaged) maybe larger.
 
 The above results, as we will see in the next section,  have important implications in the total event rates.


 %$Q=3$ keV in Fig. \ref{fig:Hcosa}.
Sometimes, as is the case for the DAMA experiment, the target has many components. In such cases the above formalism can be applied as follows:
\beq
\frac{dR}{dQ}|_A\rightarrow \sum_i X_i\frac{dR}{dQ}|_{A_i},\quad u\rightarrow u_i,\quad X_i=\mbox{the fraction of the component } A_i\mbox{ in the target}
\eeq
We will not, however, pursue such an analysis.
 %I reality for low energy transfer the  $^{23}$Na component becomes relevant. For the light system the %effect of the form factor becomes negligible. In this case Eq. \ref{drdu} must be modified so that:
%$$
%\frac{1}{A} \cdots \mu^2_r(A)A^2\rightarrow \frac{1}{A_1+A_2} \cdots(A_1^2\mu^2_r(A_1)+A_2^2\mu^2_r(A_2)),
%$$
%$$\frac{dt}{du}\rightarrow\frac{1}{\mu^2_r(A_1)A_1^2+\mu^2_r(A_2)A_2^2}\left %[A_1^2\mu^2_r(A_1)\frac{dt_1}{du}+A_2^2\mu^2_r(A_2)\frac{dt_2}{du}\right ]
%$$
%$$
%\frac{dh}{du}\rightarrow \frac{1}{\mu^2_r(A_1)A_1^2+\mu^2_r(A_2)A_2^2}\left %[A_1^2\mu^2_r(A_1)\frac{dh_1}{du}+A_2^2\mu^2_r(A_2)\frac{dh_2}{du}\right ]
%$$
%\barr
%\frac{1}{A} \cdots \mu^2_r(A)A^2\rightarrow \frac{1}{A_1+A_2} \cdots(A_1^2\mu^2_r(A_1)+A_2^2\mu^2_r(A_2)),\quad %&&\frac{dt}{du}\rightarrow\frac{1}{\mu^2_r(A_1)A_1^2+\mu^2_r(A_2)A_2^2}\left [A_1^2\mu^2_r(A_1)\frac{dt_1}{du}+A_2^2\mu^2_r(A_2)\frac{dt_2}{du}\right ] \nonumber\\
%&&\frac{dh}{du}\rightarrow \frac{1}{\mu^2_r(A_1)A_1^2+\mu^2_r(A_2)A_2^2}\left [A_1^2\mu^2_r(A_1)\frac{dh_1}{du}+A_2^2\mu^2_r(A_2)\frac{dh_2}{du}\right ]\nonumber\\
%\earr
%On the other hand Eq. \ref{dhduH} is understood as:
%\beq 
%H(u)=\frac{A_1^2\mu^2_rA_1^2 \frac{dh_1}{du}+A_2^2\mu^2_r(A_2) \frac{dh_2}{du}}
%{A_1^2\mu^2_r(A_1) \frac{dt_1}{du}+A_2^2 \mu^2_r(A_2)\frac{dt_2}{du}}
%\eeq
%where the subscript 1 refers to one component, e.g. $^{127}$I, and 2 refers to the other component, e.g.  $^{23}$Na. 
\section{Some results on total rates}

For completeness and comparison we will briefly present our results on the total rates. Integrating the %differential rates discussed in the previous section we obtain the total time averaged rate $R_0$, the total %modulated rate $\tilde{H}$ and the relative modulation amplitude $h$  given by:
differential rates discussed in the previous section we obtain the total rate $R$, adding the corresponding time averaged rate $R_0$ and the total modulated rate $\tilde{H}$, given by:
\beq
R=R_0+\tilde{H}=\frac{\rho_{\chi}}{m_{\chi}}\frac{m_t}{A m_p}  \left ( \frac{\mu_r}{\mu_p} \right )^2 \sqrt{<\upsilon^2>} A^2 \sigma_n t\left (1+h \cos{\alpha}\right ) ,
\label{Eq:Trates}
\eeq
with
\beq
t=\int_{Q_{th}/Q_0(A)}^{(y_{\mbox{\tiny max}}/a)^2}\frac{dt}{du}du,\quad h=\frac{1}{t}\int_{Q_{th}/Q_0(A)}^{(y_{\mbox{\tiny max}}/a)^2}\frac{dh}{du}du.
\label{Eq:thfac}
\eeq
 $ y_{\mbox{\tiny max}}$ is the maximum velocity allowed by the distribution and $Q_{th}(A)$ is the energy cut off imposed by the detector.
 
\begin{figure}
\begin{center}
\subfloat[]
{
\rotatebox{90}{\hspace{0.0cm} $R_0\rightarrow$kg/y}
\includegraphics[height=.17\textheight]{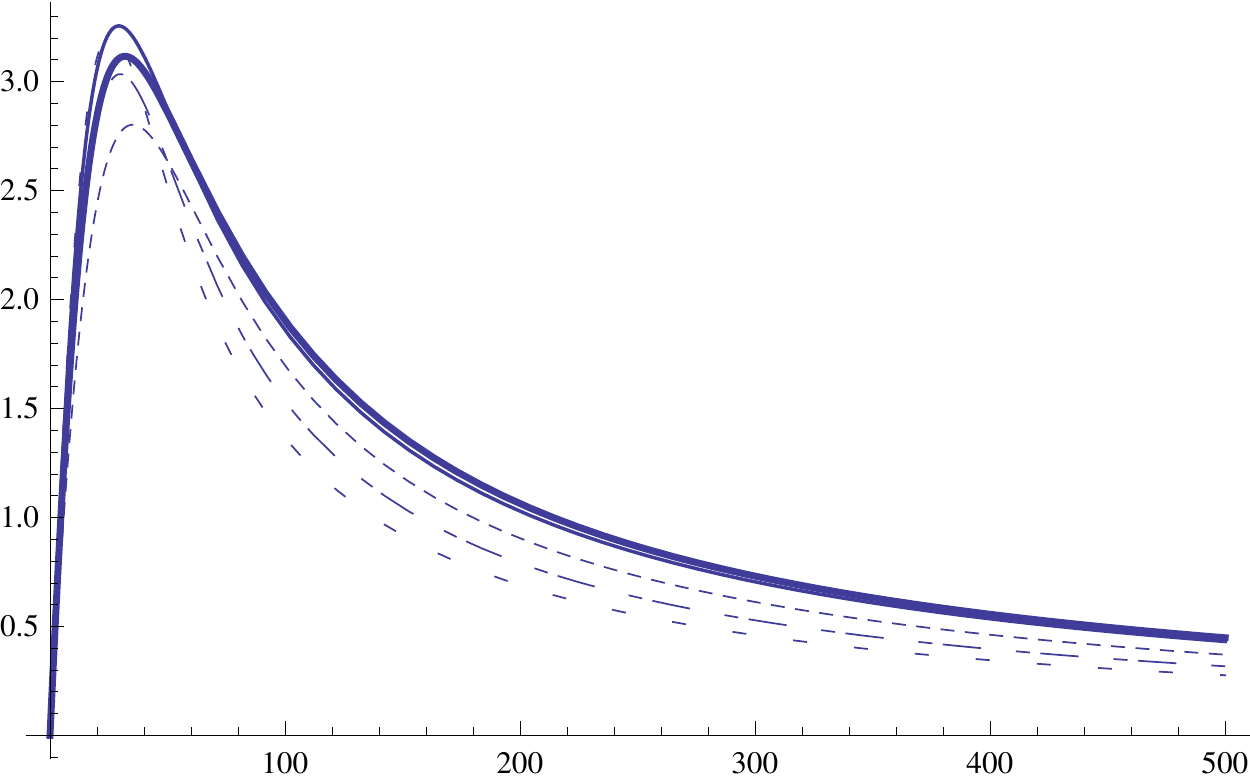}
}
\subfloat[]
{
\rotatebox{90}{\hspace{0.0cm} $R_0\rightarrow$kg/y}
\includegraphics[height=.17\textheight]{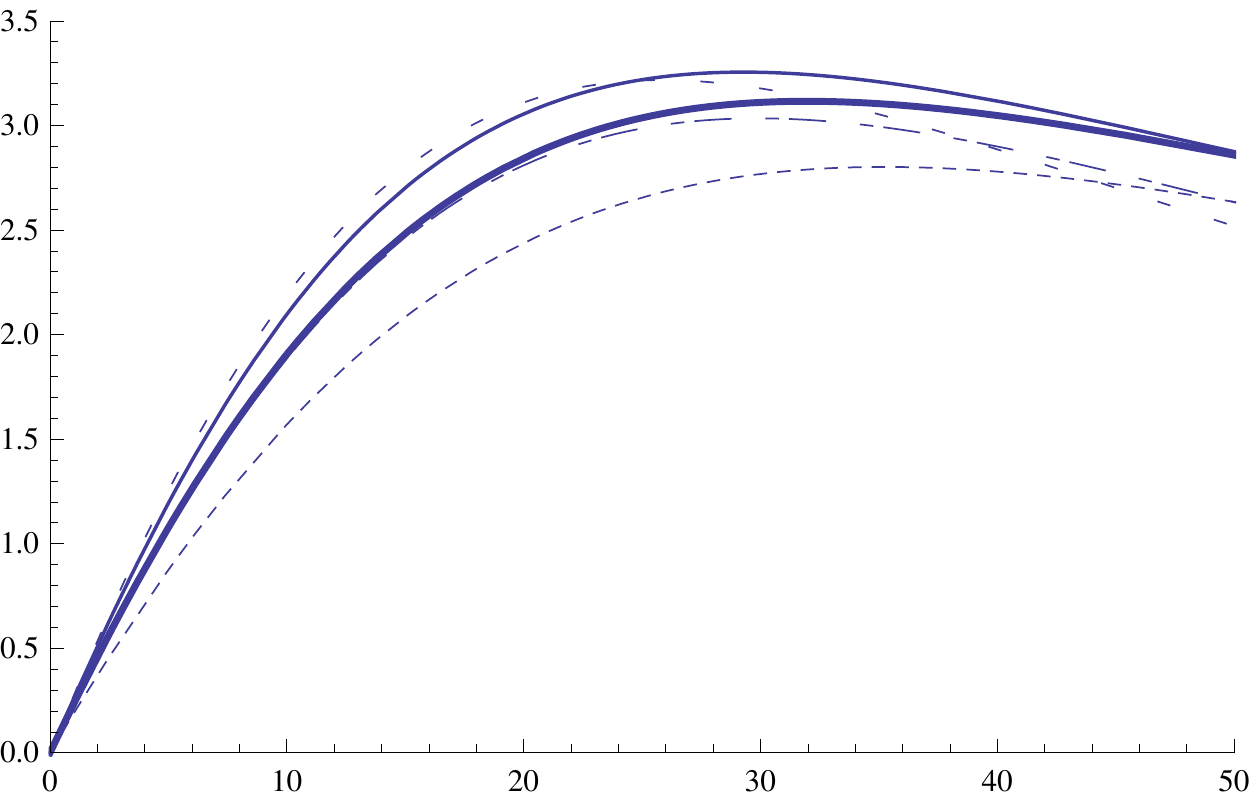}
}
\\
%{\hspace{-2.0cm} $Q\rightarrow$keV}
\subfloat[]
{
%\rotatebox{90}{\hspace{0.0cm} $h\rightarrow$}
\rotatebox{90}{\hspace{0.0cm} $R_0\rightarrow$kg/y}
\includegraphics[height=.17\textheight]{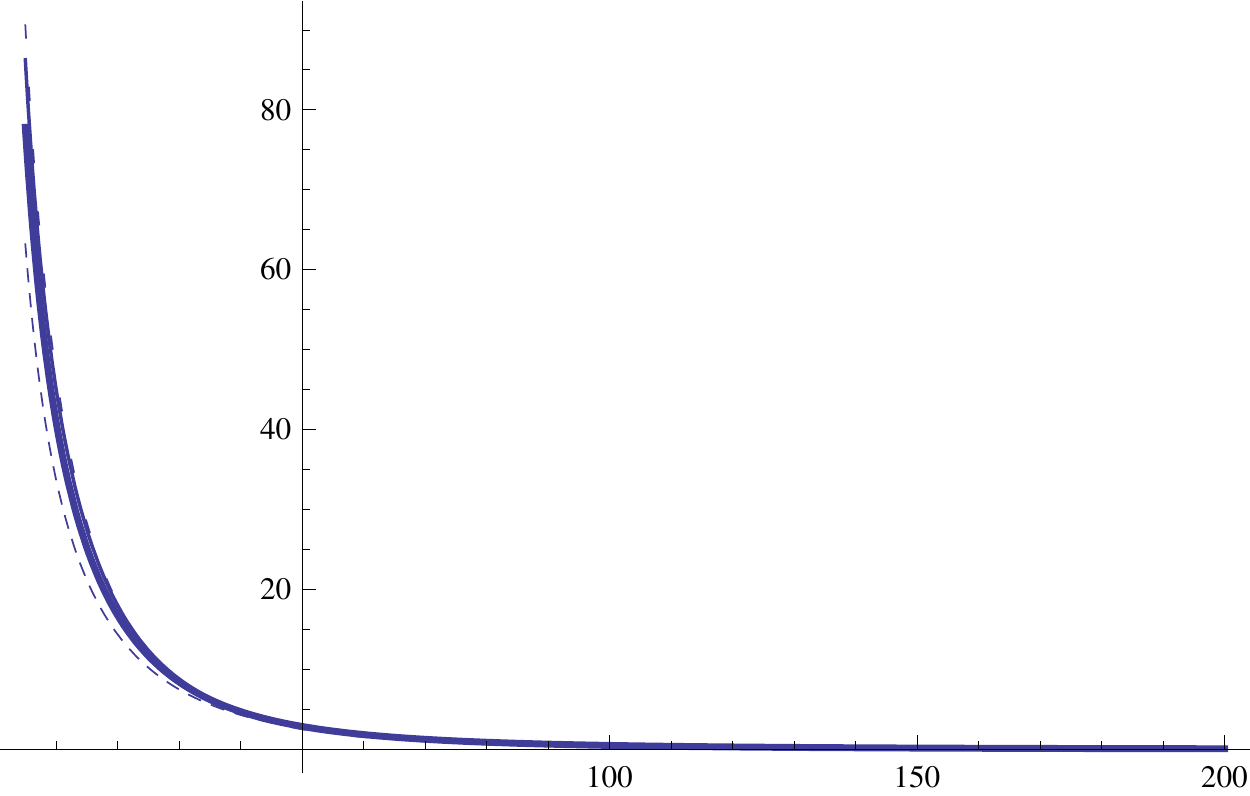}
}
\subfloat[]
{
%\rotatebox{90}{\hspace{0.0cm} $h\rightarrow$}
\rotatebox{90}{\hspace{0.0cm} $R_0\rightarrow$kg/y}
\includegraphics[height=.17\textheight]{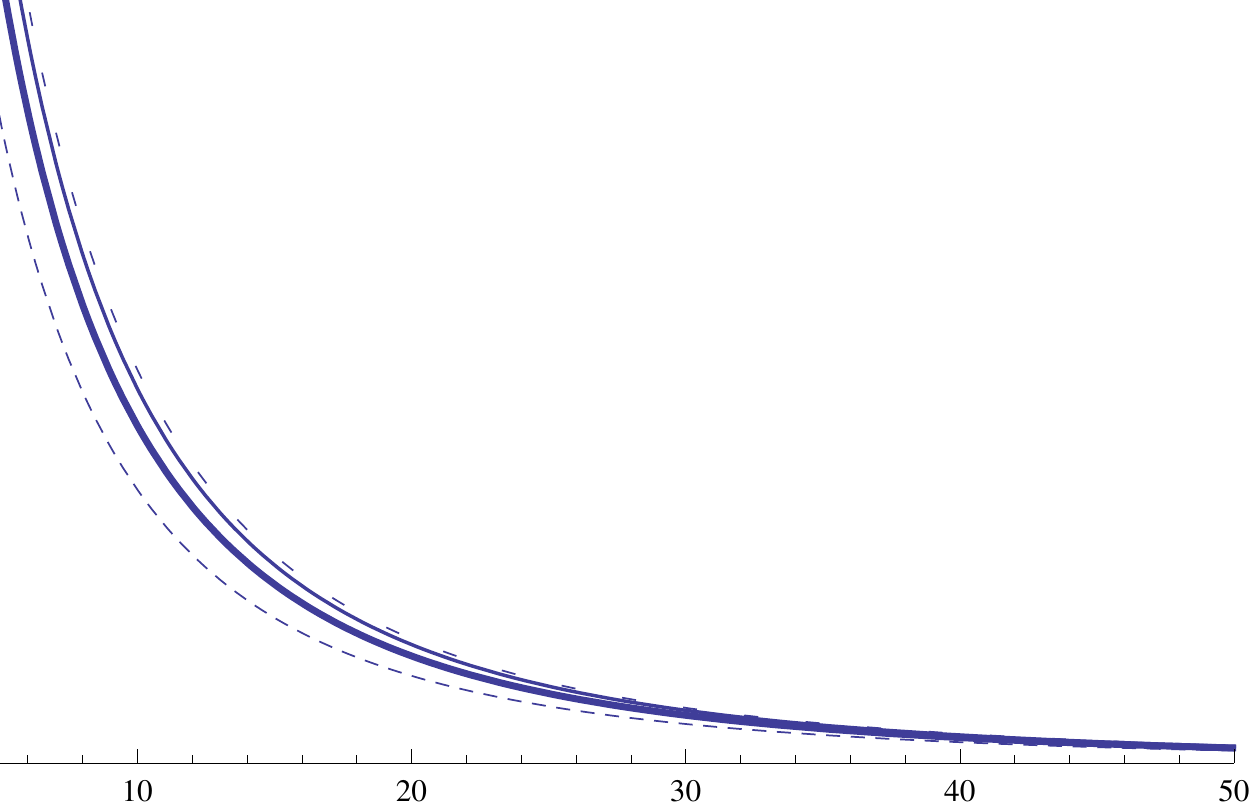}
}\\
\subfloat[]
{
\rotatebox{90}{\hspace{0.0cm} $h\rightarrow$}
\includegraphics[height=.17\textheight]{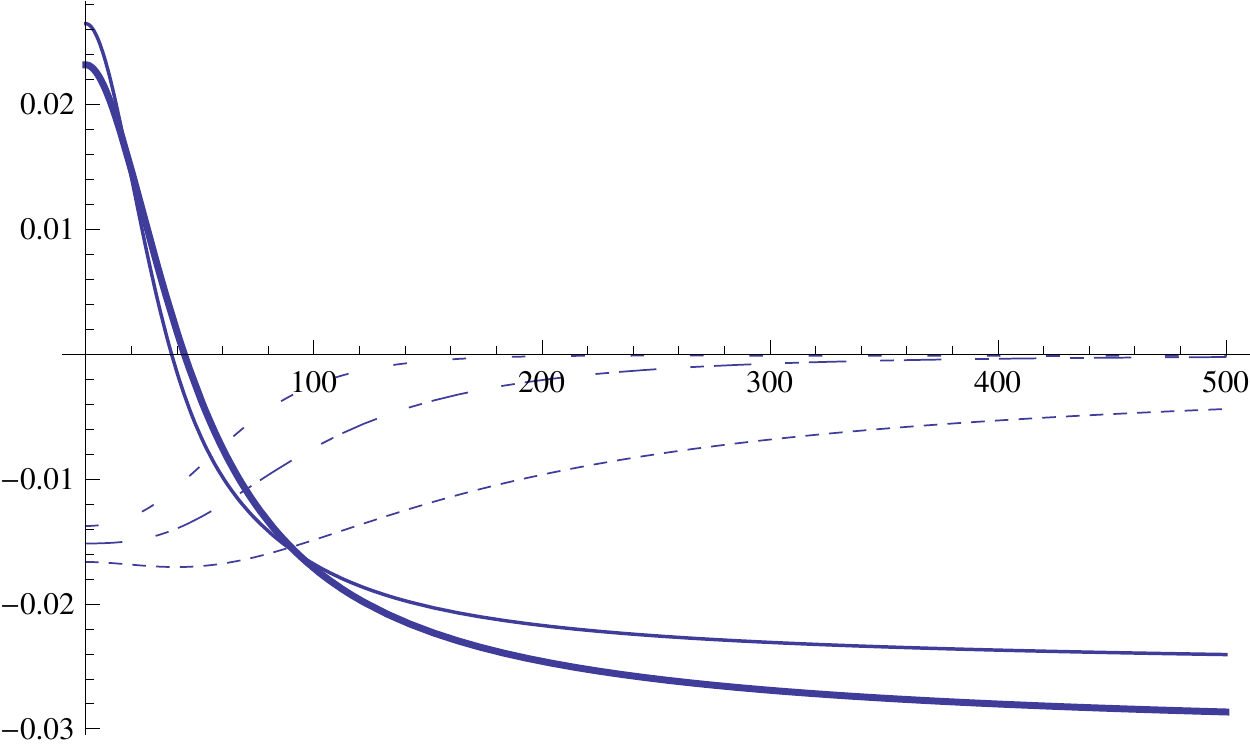}
}
\subfloat[]
{
\rotatebox{90}{\hspace{0.0cm} $h\rightarrow$}
\includegraphics[height=.17\textheight]{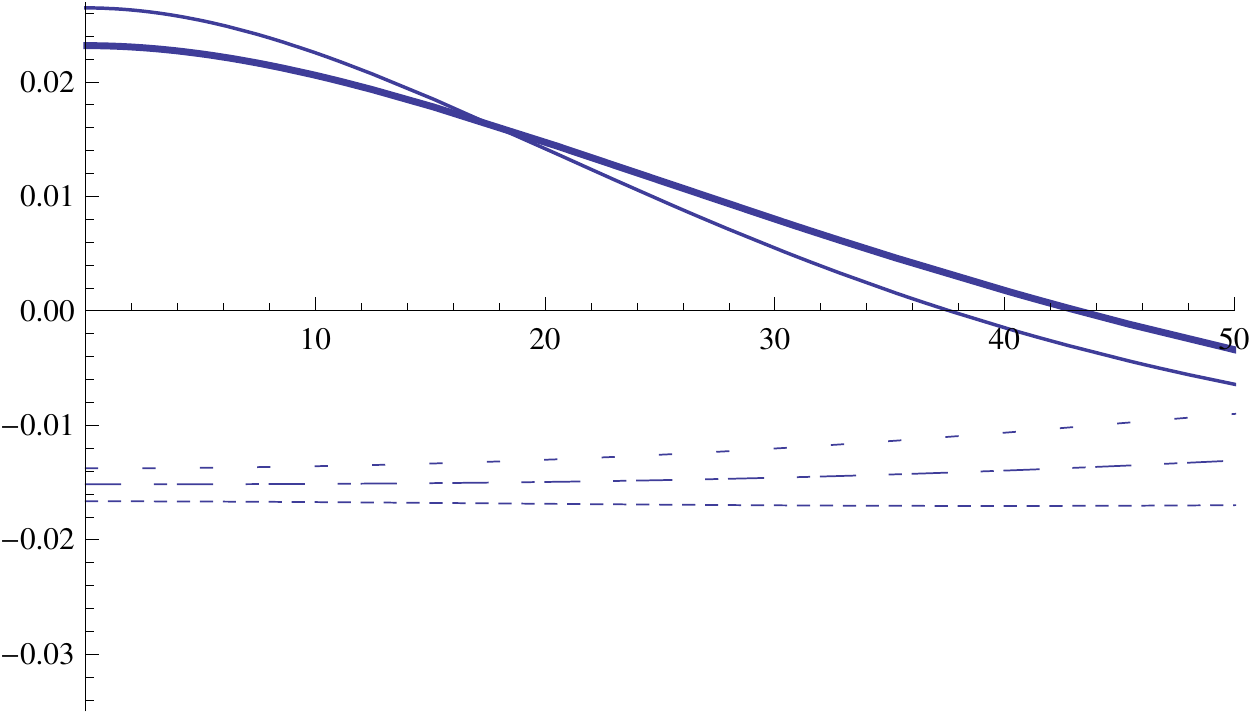}
}
\\
{\hspace{-2.0cm} $m_{\mbox{\tiny{\tiny  WIMP}}}\rightarrow$GeV}
\caption{ The total rate $R_0$ for usual WIMP (top panels) and and the scalar WIMP (middle panels) and the relative modulation h (bottom panels)  as a function of the WIMP mass in GeV in the case of a heavy target $^{127}$I at zero threshold. The panels on the right are a restriction of those on the left to smaller masses. Otherwise the notation is the same as that of Fig. \ref{fig:flowv}.}
\label{fig:Rh131}
\end{center}
\end{figure}

\begin{figure}
\begin{center}
\subfloat[]
{
\rotatebox{90}{\hspace{0.0cm} $R_0\rightarrow$kg/y}
\includegraphics[height=.17\textheight]{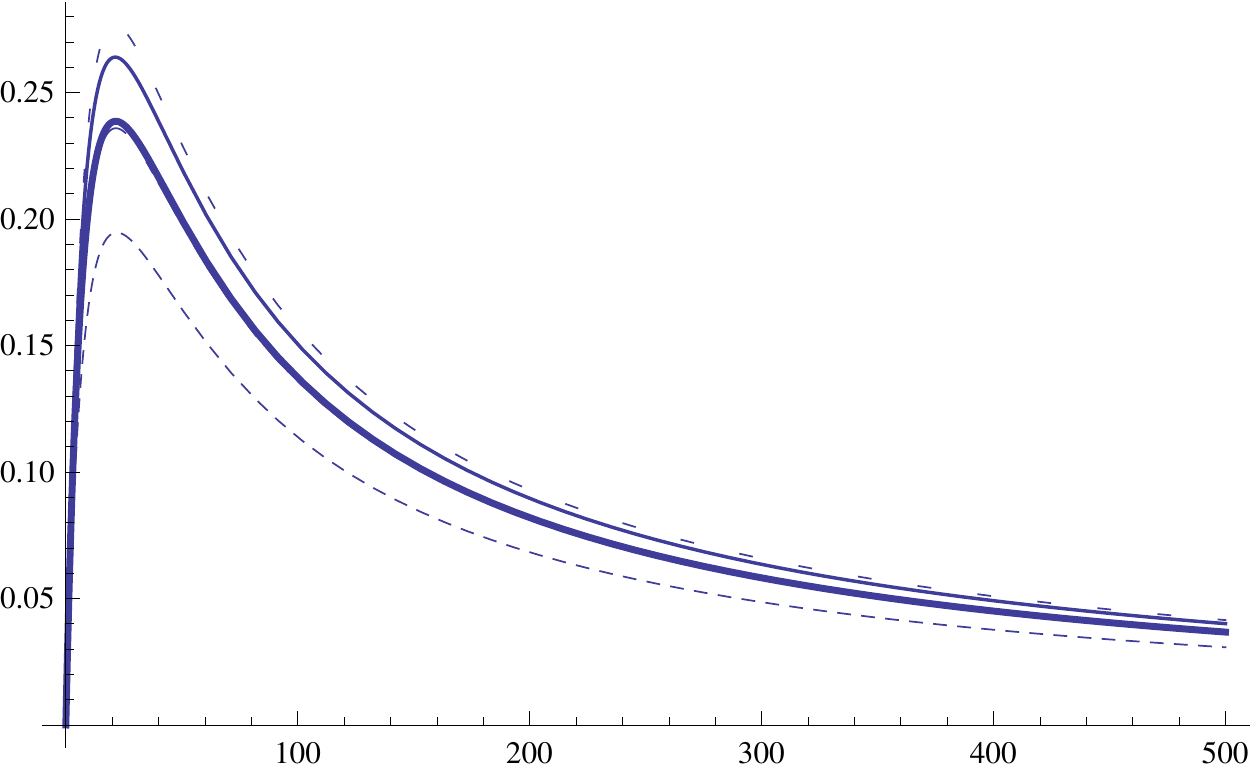}
}
\subfloat[]
{
\rotatebox{90}{\hspace{0.0cm} $R_0\rightarrow$kg/y}
\includegraphics[height=.17\textheight]{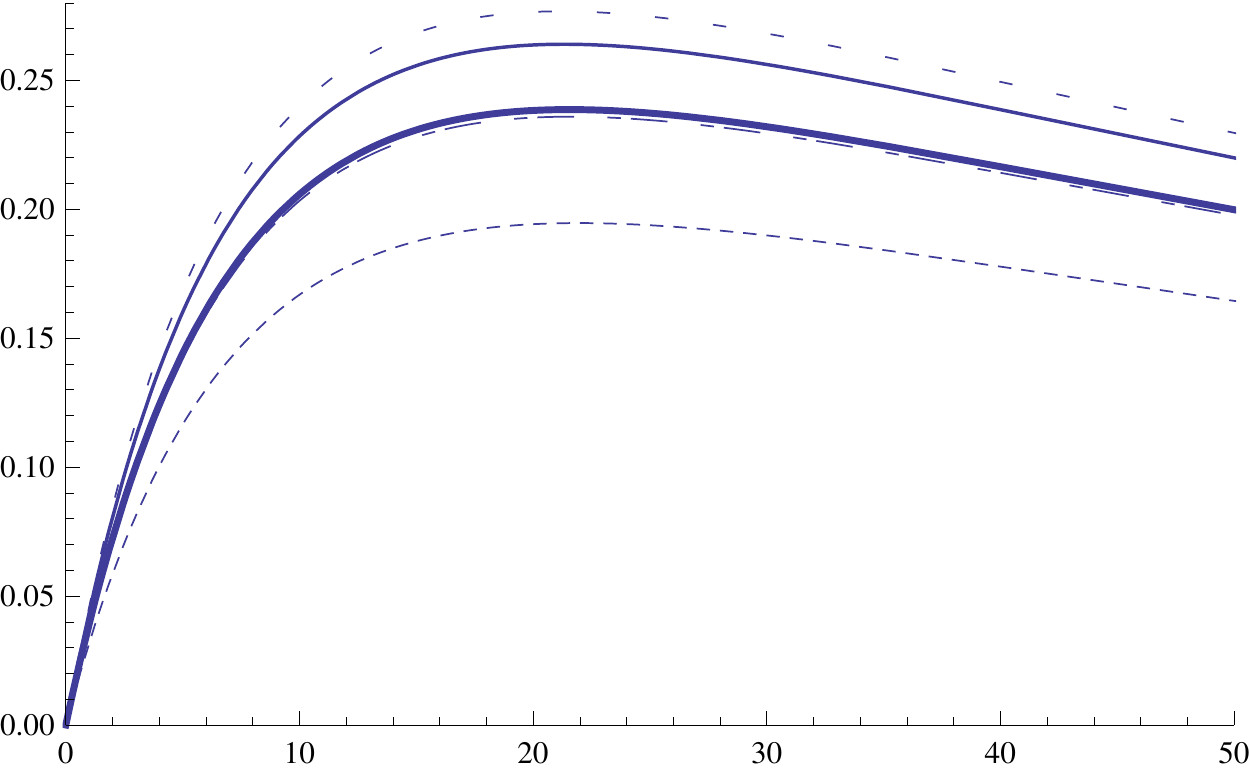}
}
\\
\subfloat[]
{
\rotatebox{90}{\hspace{0.0cm} $R_0\rightarrow$kg/y}
\includegraphics[height=.17\textheight]{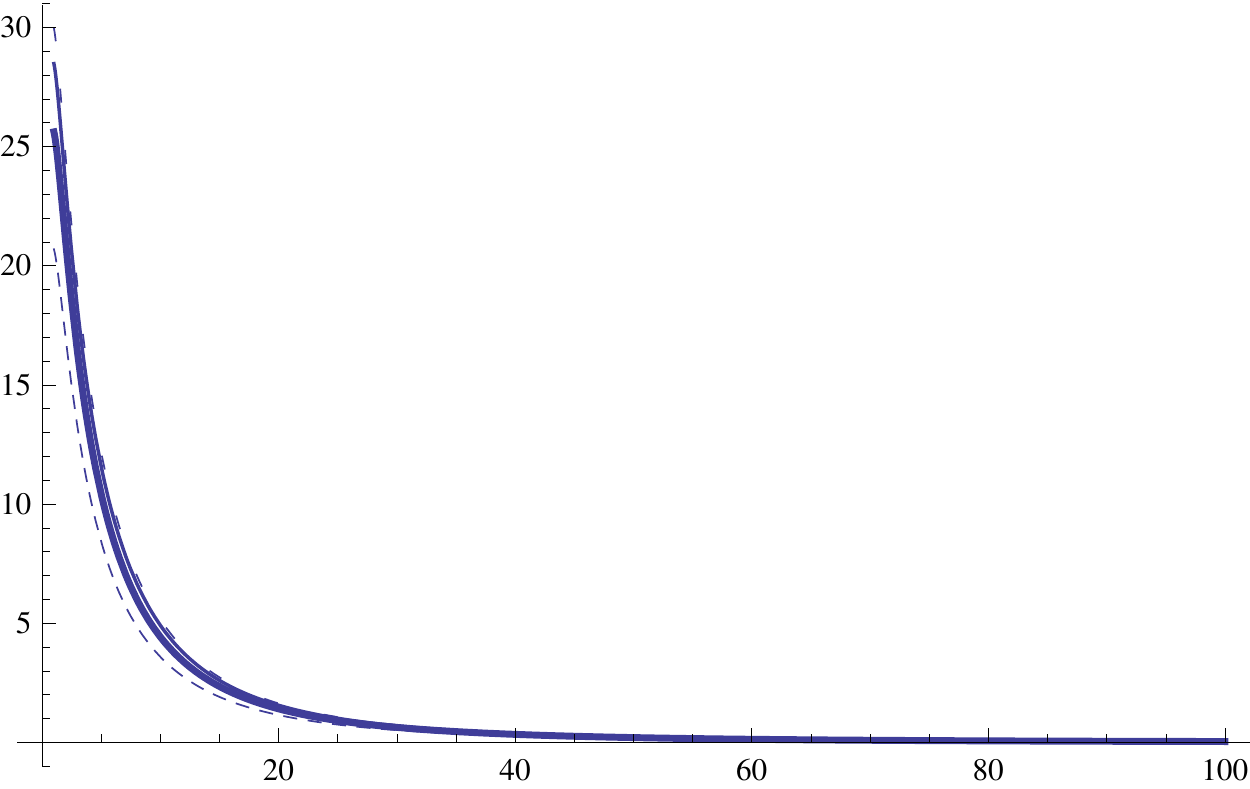}
}
\subfloat[]
{
\rotatebox{90}{\hspace{0.0cm} $R_0\rightarrow$kg/y}
\includegraphics[height=.17\textheight]{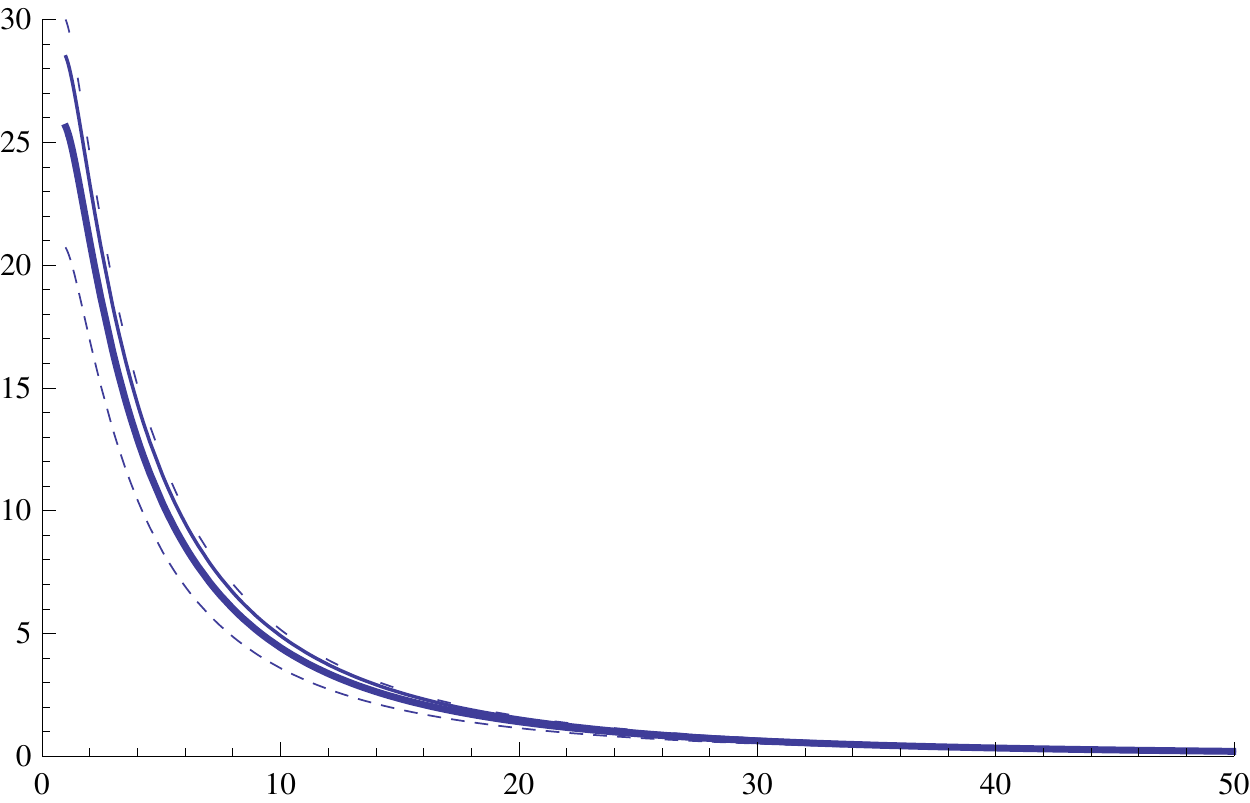}
}
\\
%{\hspace{-2.0cm} $Q\rightarrow$keV}
\subfloat[]
{
\rotatebox{90}{\hspace{0.0cm} $h\rightarrow$}
\includegraphics[height=.17\textheight]{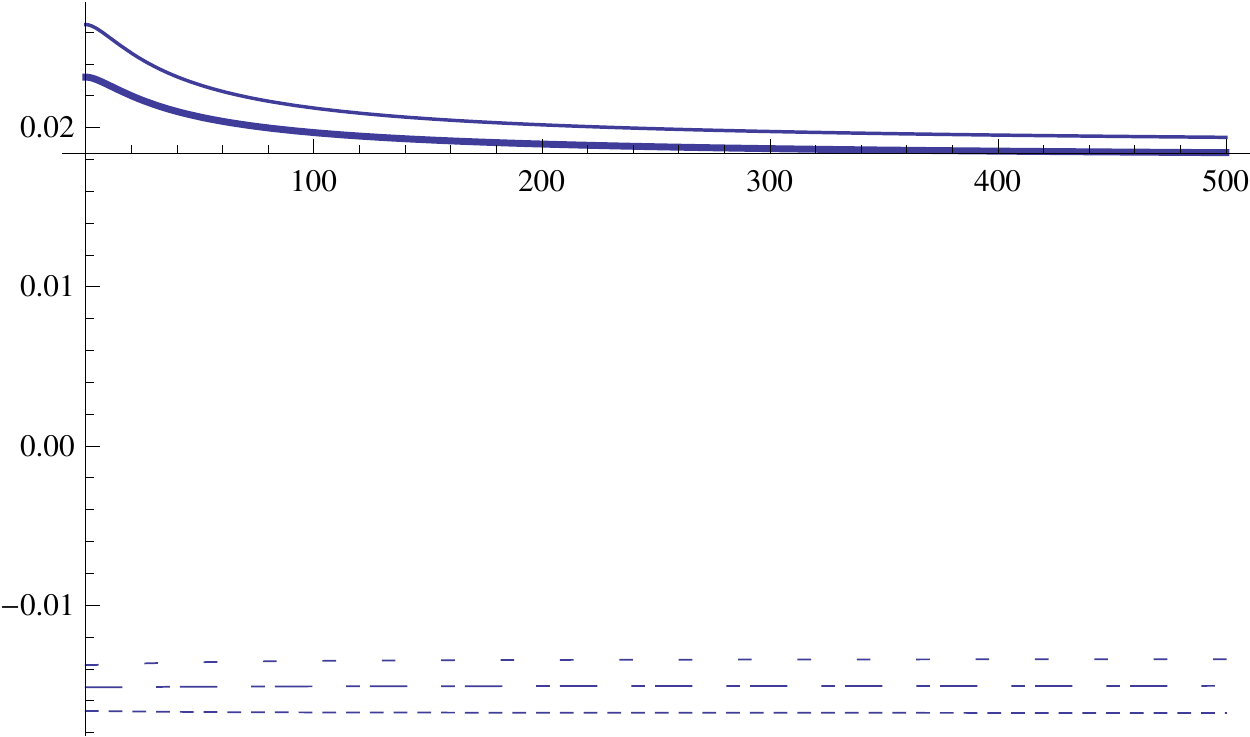}
}
\subfloat[]
{
\rotatebox{90}{\hspace{0.0cm} $h\rightarrow$}
\includegraphics[height=.17\textheight]{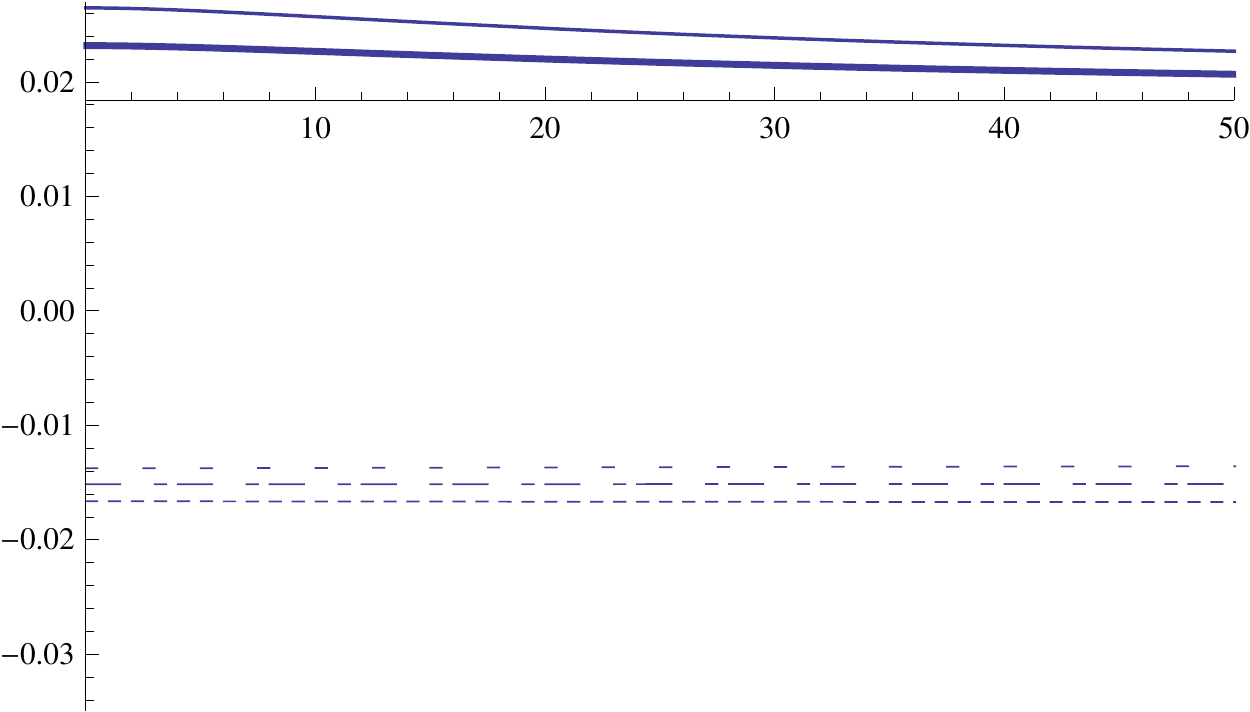}
}
\\
{\hspace{-2.0cm} $m_{\mbox{{\tiny WIMP}}}\rightarrow$GeV}
\caption{ The same as in Fig. \ref{fig:Rh131} for a light target, e,g. $^{23}$Na. }
 \label{fig:Rh23}
\end{center}
\end{figure}

\begin{figure}
\begin{center}
\subfloat[]
{
\rotatebox{90}{\hspace{0.0cm} $R_0\rightarrow$kg/y}
\includegraphics[height=.17\textheight]{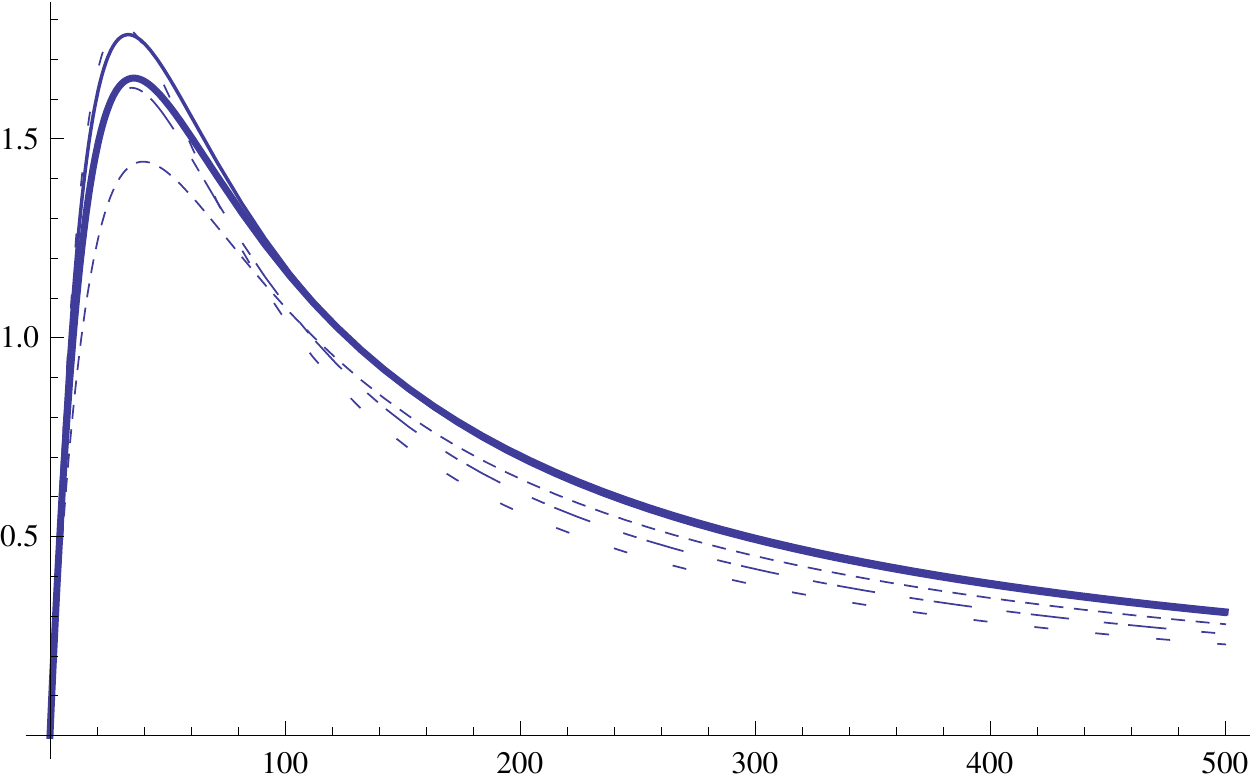}
}
\subfloat[]
{
\rotatebox{90}{\hspace{0.0cm} $R_0\rightarrow$kg/y}
\includegraphics[height=.17\textheight]{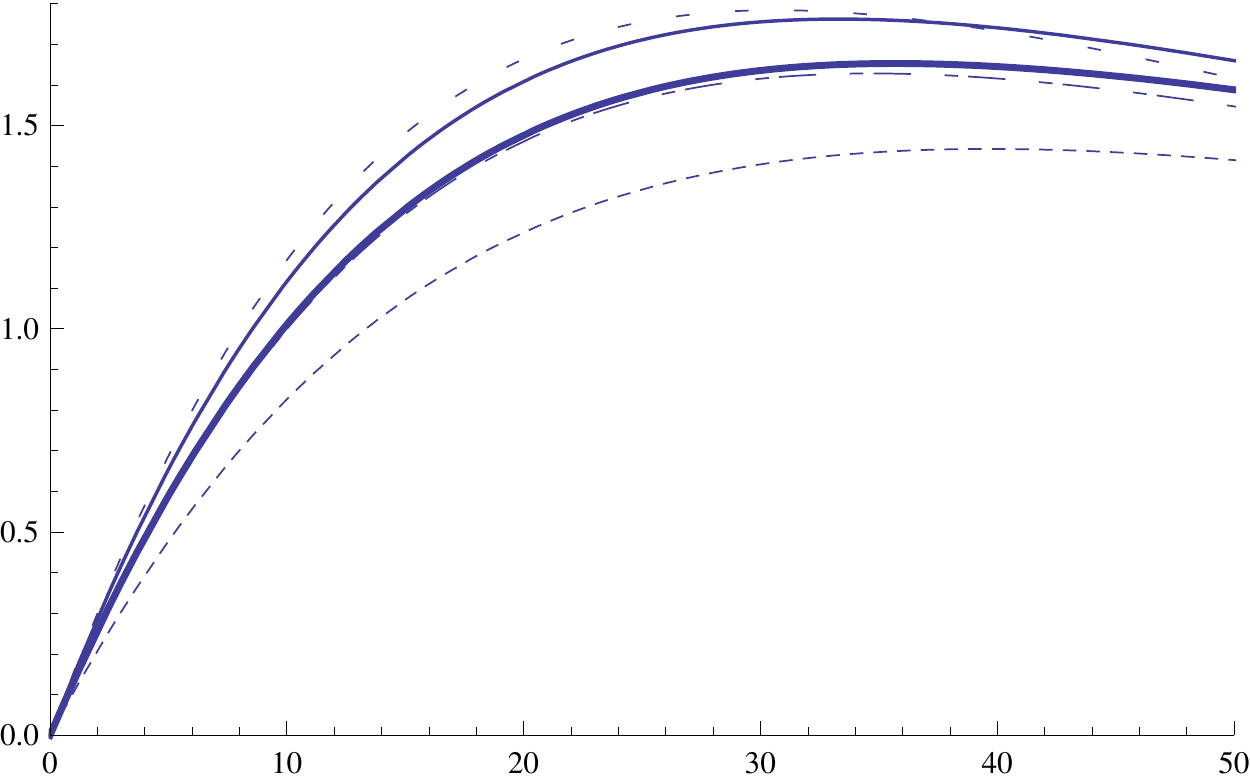}
}
\\
\subfloat[]
{
\rotatebox{90}{\hspace{0.0cm} $R_0\rightarrow$kg/y}
\includegraphics[height=.17\textheight]{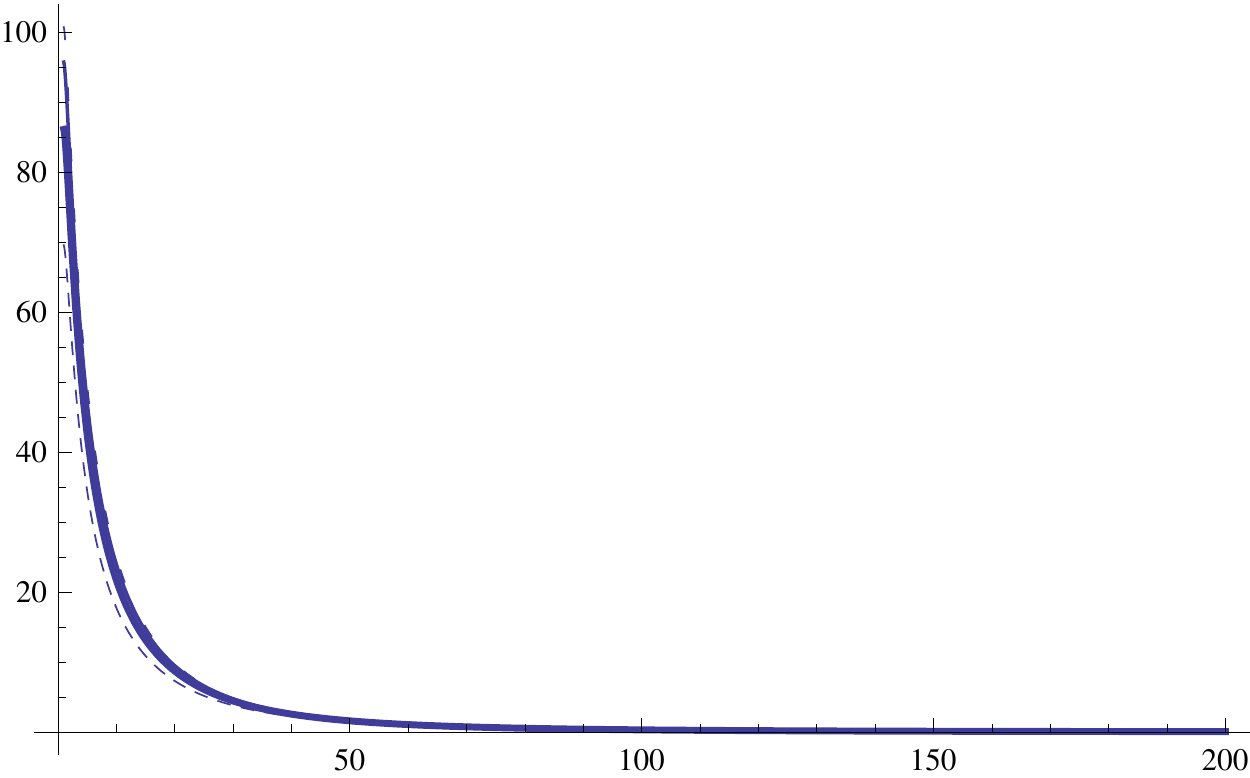}
}
\subfloat[]
{
\rotatebox{90}{\hspace{0.0cm} $R_0\rightarrow$kg/y}
\includegraphics[height=.17\textheight]{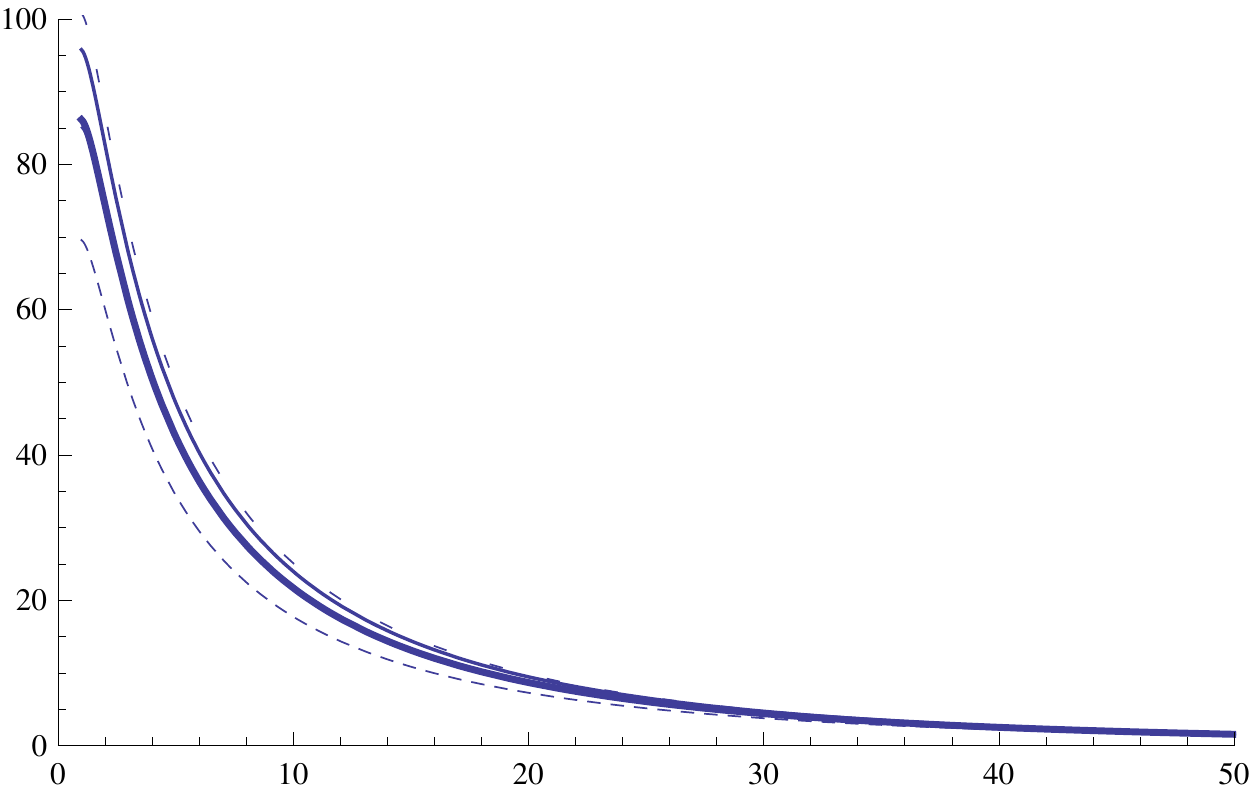}
}
\\

%{\hspace{-2.0cm} $Q\rightarrow$keV}
\subfloat[]
{
\rotatebox{90}{\hspace{0.0cm} $h\rightarrow$}
\includegraphics[height=.17\textheight]{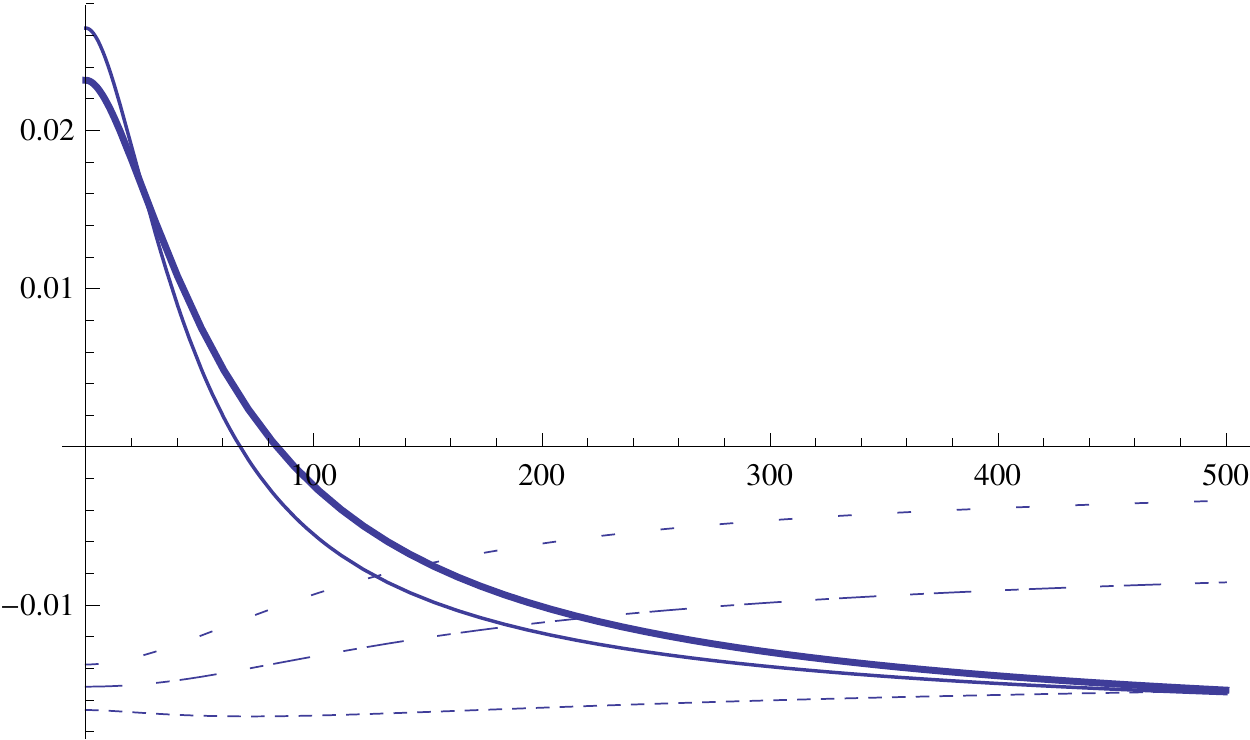}
}
\subfloat[]
{
\rotatebox{90}{\hspace{0.0cm} $h\rightarrow$}
\includegraphics[height=.17\textheight]{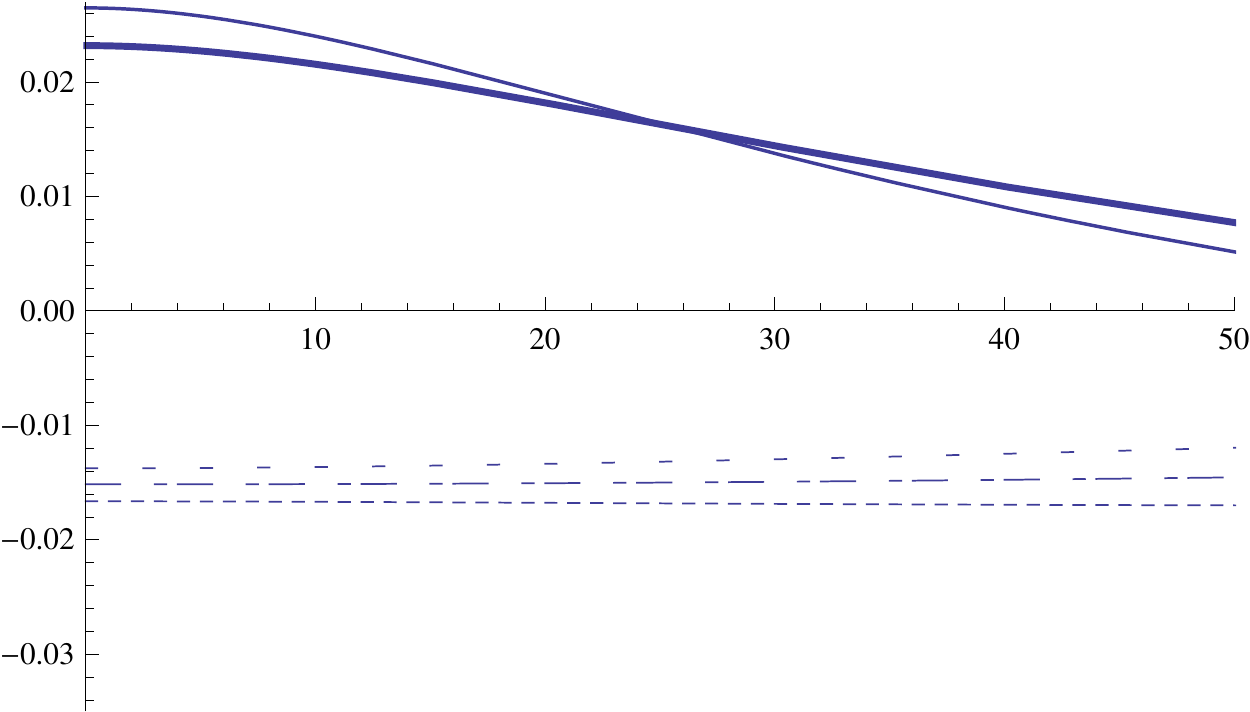}
}
\\
{\hspace{-2.0cm} $m_{\mbox{{\tiny WIMP}}}\rightarrow$GeV}
\caption{  The same as in Fig. \ref{fig:Rh131} for an intermediate target, e,g. $^{73}$Ge.}
 \label{fig:Rh73}
\end{center}
\end{figure}

 The obtained results for quantities $R_0$ and $h$ are exhibited in Figs \ref{fig:Rh131}-\ref{fig:Rh73} assuming a nucleon cross section of $10^{-8}$pb (at $m_{\chi}$=50 GeV for a scalar WIMP). For a standard WIMP in the case of a heavy target the average event rate attains the maximum value of 30 events per kg of target per year at a WIMP mass of 25 GeV, while for heavy WIMPS it eventually falls to about 5 kg/y at 500 GeV(to a good approximation it falls inversely proportional to the WIMP mass above the 200 GeV). For an intermediate target we get 15 kg/y at 25 GeV, with an asymptotic value of 4 kg/y. For a light target the maximum becomes 2.5 kg/y at 20 GeV. Again the asymptotic value at 500 GeV is about 1/5 of the maximum. The situation is very different for a scalar WIMP. At small WIMP masses the event rate becomes huge. The effect will appear less dramatic, if the value of  $10^{-8}$pb is fitted to a much smaller WIMP mass, since it will manifest itself for masses below that choice, but it is there. At high WIMP masses the event rate falls more rapidly with the mass. The relative modulation amplitude, however, being the ratio of the time dependent rate divided by the time averaged rate is the same for both types of WIMPs.\\
 To understand this behavior we should mention that the WIMP mass dependence comes from three sources. 
\begin{itemize}
\item From the momentum transfer, yielding a contribution to the event rate proportional to the square of $\mu_r$ (the WIMP-nucleus reduced mass), which vanishes quadratically for zero WIMP mass.
\item From the WIMP particle density in our vicinity, which is inversely proportional to the mass (from the rotation curves we infer the density, not the number of particles per unit volume). In the limit of large WIMP mass this wins out over the previous one, since the reduced mass  then is essentially the mass of the nucleus. For small WIMP mass the combination of these terms vanishes linearly. 
\item For scalar WIMPS we have the additional  mass dependence coming from the elementary cross section $\sigma_n \propto \left(1+m_{\chi}/m_p \right )^{-2}$ as we have seen. 
\end{itemize}
 We thus conclude that even in the case for a scalar WIMP at  a low mass the total rate is proportional to 
$$ R\propto\mu_r^2 \frac{1}{m_{\chi}}\frac{1}{\left(1+m_{\chi}/m_p \right )^2}\rightarrow 0 \mbox{ as } m_{\chi} \rightarrow 0.$$ 
%where the middle term comes from the WIMP particle density in our vicinity. 
%The net effect of these three factors makes the total event rate diverge as the WIMP mass vanishes.
   
  It is clear that, as far as the time average rates $R_0$ are concerned ,the debris flows do not exhibit any characteristic signature to differentiate them from the standard M-B distribution. The relative modulation  amplitude $h$, however, exhibits a very interesting feature, namely, if caused by the flows, it is negative for all targets, even for the light ones, and in the entire WIMP mass range (minimum in June). On the other hand if it is caused by the M-B distribution it is positive in the case of  light targets regardless of the  WIMP mass. It is also positive  for intermediate/heavy targets, if the WIMPs are relatively light. Then the maximum occurs on June 3nd as expected. It becomes negative only for relatively heavy WIMPs. Thus it is an experimental challenge to measure the small time dependence of the event rate with a relative difference between the maximum and the minimum of $2h\approx4\%$.  From such data on both light and heavy targets, if and when they become available, one may may be able: i)to get a hint about the size of the WIMP mass and ii) infer the existence of flows.

%The results for several WIMP masses are shown in the tables \ref{tab1}-\ref{tab2}.
\section{Discussion and conclusions}
In the present paper we first obtained results on the differential event rates, both modulated and time averaged. We have considered a new type of viable WIMP, namely a scalar WIMP,  motivated by the Big Bounce Scenario of Cosmology (BUS). We then compared the obtained results with the standard WIMP with a nucleon cross section independent of the WIMP mass. We found that:
\begin{itemize}
\item The nucleon cross section is a decreasing function as the WIMP mass increases.\\
This is in line with the predictions of BUS (compare Fig.~\ref{fig:cbplog} and Fig.~\ref{fig:sigmap}).
\item The above mass dependence leads to an increase of the rates at a low WIMP mass.  This may be  good news for the low threshold experiments using light nuclear targets (DM-TPC, NEWAGE, DRIFT,MIMAC etc), which are sensitive to low mass WIMPs.
\item The maximum of the total event rate is shifted to a much lower regime, which may require a lower recoil energy threshold than currently achieved.
\item As far as we know this behavior of the cross section is not excluded by the current data. In fact it may aid the analysis of the experimental data in the low WIMP mass regime even though there is a tendency for model independent analysis of the data, as e.g. in DAMA/LIBRA \cite{DAMAEPJ13}. 
\item It may interesting to draw exclusion plots with this new nucleon cross section and extract the value of $\sigma_0$ entering Eq. (\ref{Eq:sigma0}).
\item It may also  help explaining  the large cross section extracted from the recent CRESST data \cite{CRESST}, if they persist.

\end{itemize}
%focusing our attention on the effects of debris flows. We considered  both the usual types of WIMPs as well as the more exotic scalar WIMPs 
We also examined the sensitivity of the obtained results to the velocity distribution (the nuclear form factor is the same of both types of WIMPs). We considered both a standard M-B velocity distribution and  also models, which  extend it, e.g. debris flows, which  have also been considered~\cite{spergel12}, ~\cite{VergF12} previously. 
 Thus we found out that:
\begin{itemize}
\item The flows indeed enhance the time averaged rates at relatively high energy transfers compared to the M-B distribution, at the expense of the corresponding rates at low energy transfers . All rates, however, fall as the energy transfer increases. This fall is only partially due to the velocity distribution. It is also caused by the nuclear form factor, in particular in the case of heavy targets. Anyway this behavior cannot be exploited to differentiate between them, since the WIMP mass is not known. Thus the time averaged rates do not provide a clear signature to  differentiate the debris flows from the standard M-B distribution.
\item The differential time dependent (modulated) rates provide such a signature, the sign of the modulation amplitude, which determines the position of the maximum. At sufficiently low energy transfer both the M-B and the debris flows favor a negative sign (minimum on June 3nd),  with the flows insisting on such behavior more strongly and exhibiting it all the way to high energy transfers. So if the flows are there this signature may be seen even with detectors, which do not have a very low energy threshold.  
\item For the M-B distribution  this behavior is manifested for an energy which depends on the target and the WIMP mass ( see Figs \ref{fig:dHdQ127},\ref{fig:dHdQSc127},\ref{fig:dHdQ23},\ref{fig:dHdQSc23},\ref{fig:dHdQ73}, 
\ref{fig:dHdQSc73}).  Thus e.g. for a heavy target this recoil energy is 0.5, 5, 20 and 40 keV. This recoil energy  is the same for both types of WIMPs, only  the rate  is different for low WIMP mass.
\item The above behavior is carried over to the total rates. For WIMP flows the maximum is in winter, but for the M-B distribution one finds the usual case (maximum in June 3nd) for low reduced mas but maximum in December  for relatively large reduced mass. 
%while the standard WIMPs favor a positive sign when the target is light or even when the target is heavy but the WIMP is light (maximum on June 3nd).
\end{itemize}

\acknowledgments{
JDV is indebted to the Physics Department of KAIST and IBS for their kind invitation and support and to Professor Yannis K. Semertzidis, Director of CAPP/IBS at KAIST, for useful discussions and his hospitality, while EC would like to thank Changhong Li, Jin U Kang and  Konstantin Savvidy for many useful discussions.
This research project has been supported in parts by the Jiangsu Ministry of Science and Technology under contract BK20131264. We also acknowledge 985 Grants from the Chinese Ministry of Education, and the Priority Academic Program Development for Jiangsu Higher Education Institutions (PAPD).}

\clearpage
\addcontentsline{toc}{section}{References}
\bibliographystyle{JHEP}

\providecommand{\href}[2]{#2}\begingroup\raggedright\begin{thebibliography}{100}

\bibitem{MAXIMA1}
S.~Hanary and {\it et al} {\em Astrophys. J.} {\bf 545} (2000) L5.

\bibitem{MAXIMA2}
J.~Wu and {\it et al} {\em Phys. Rev. Lett.} {\bf 87} (2001) 251303.

\bibitem{MAXIMA3}
M.~Santos and {\it et al} {\em Phys. Rev. Lett.} {\bf 88} (2002) 241302.

\bibitem{BOOMERANG1}
P.~D. Mauskopf and {\it et al} {\em Astrophys. J.} {\bf 536} (2002) L59.

\bibitem{BOOMERANG2}
S.~Mosi and {\it et al} {\em Prog. Nuc.Part. Phys.} {\bf 48} (2002) 243.

\bibitem{DASI02}
N.~W. Halverson et~al. {\em Astrophys. J.} {\bf 568} (2002) 38.

\bibitem{COBE}
G.~F. Smoot and {\it et al}~(COBE~Collaboration) {\em Astrophys. J.} {\bf 396}
  (1992) L1.

\bibitem{flat01}
A.~H. Jaffe and {\it et al} {\em Phys. Rev. Lett.} {\bf 86} (2001) 3475.

\bibitem{SPERGEL}
D.~N. Spergel and {\it et al} {\em Astrophys. J. Suppl.} {\bf 148} (2003) 175.

\bibitem{WMAP06}
D.~Spergel et~al. {\em Astrophys. J. Suppl.} {\bf 170} (2007) 377.
  [arXiv:astro-ph/0603449v2].

\bibitem{PlanckCP13}
The Planck Collaboration, A.P.R. Ade {\it et al}, arXiv:1303.5076
  [astro-ph.CO].

\bibitem{Benne}
D.~P. Bennett and {\it et al} {\em Phys. Rev. Lett.} {\bf 74} (1995) 2867.

\bibitem{UK01}
P.~Ullio and M.~Kamioknowski {\em JHEP} {\bf 0103} (2001) 049.

\bibitem{LS96}
J.~D. Lewin and P.~F. Smith {\em Astropart. Phys.} {\bf 6} (1996) 87.

\bibitem{GOODWIT}
M.~W. Goodman and E.~Witten {\em Phys. Rev. D} {\bf 31} (1985) 3059.

\bibitem{Druck}
A.~Drukier, K.~Freeze, and D.~Spergel {\em Phys. Rev. D} {\bf 33} (1986) 3495.

\bibitem{PSS88}
J.~R. Primack, D.~Seckel, and B.~Sadoulet {\em Ann. Rev. Nucl. Part. Sci.} {\bf
  38} (1988) 751.

\bibitem{GS93}
A.~Gabutti and K.~Schmiemann {\em Phys. Lett. B} {\bf 308} (1993) 411.

\bibitem{RBERNABEI95}
R.~Bernabei {\em Riv. Nouvo Cimento} {\bf 18 (5)} (1995) 1.

\bibitem{ABRIOLA98}
D.~Abriola et~al. {\em Astropart. Phys.} {\bf 10} (1999) 133.
  arXiv:astro-ph/9809018.

\bibitem{HASENBALG98}
F.~Hasenbalg {\em Astropart. Phys.} {\bf 9} (1998) 339. arXiv:astro-ph/9806198.

\bibitem{JDV03}
J.~D. Vergados {\em Phys. Rev. D} {\bf 67} (2003) 103003. hep-ph/0303231.

\bibitem{GREEN04}
A.~Green {\em Phys. Rev. D} {\bf 68} (2003) 023004. ibid: D ${\bf 69}$ (2004)
  109902; arXiv:astro-ph/0304446.

\bibitem{SFG06}
C.~Savage, K.~Freese, and P.~Gondolo {\em Phys. Rev. D} {\bf 74} (2006) 043531.
  arXiv:astro-ph/0607121.

\bibitem{FKLW11}
P. J. Fox, J. Kopp, M. Lisanti and N. Weiner, A CoGeNT Modulation Analysis,
  arXiv:1107.0717 (astro-ph.CO).

\bibitem{SPERGEL88}
D.~Spergel {\em Phys. Rev. D} {\bf 37} (1988) 1353.

\bibitem{DRIFT}
The NAIAD experiment B. Ahmed {\it et al}, Astropart. Phys. {\bf 19} (2003)
  691; hep-ex/0301039\\ B. Morgan, A. M. Green and N. J. C. Spooner, Phys. Rev.
  D {\bf 71} (2005) 103507; astro-ph/0408047.

\bibitem{SHIMIZU03}
Y.~Shimizu, M.~Minoa, and Y.~Inoue {\em Nuc. Instr. Meth. A} {\bf 496} (2003)
  347.

\bibitem{KUDRY04}
V.A. Kudryavtsev, Dark matter experiments at Boulby mine, astro-ph/0406126.

\bibitem{DRIFT2}
B.~Morgan, A.~M. Green, and N.~J.~C. Spooner {\em Phys. Rev. D} {\bf 71} (2005)
  103507. ; astro-ph/0408047.

\bibitem{GREEN05}
B.~Morgan and A.~M. Green {\em Phys. Rev. D} {\bf 72} (2005) 123501.

\bibitem{Green06}
A.~M. Green and B.~Morgan {\em Astropart. Phys.} {\bf 27} (2007) 142. [
  arXiv:0707.1488 (astrp-ph)].

\bibitem{KRAUSS}
C.~Copi, J.~Heo, and L.~Krauss {\em Phys. Lett. B} {\bf 461} (1999) 43.

\bibitem{KRAUSS01}
C.~Copi and L.~Krauss {\em Phys. Rev. D} {\bf 63} (2001) 043507.

\bibitem{Alenazi08}
A.~Alenazi and P.~Gondolo {\em Phys. Rev. D} {\bf 77} (2008) 043532.

\bibitem{Creswick010}
R.J. Creswick and S. Nussinov and F.T. Avignone III, arXiv: 1007.0214
  [astro-ph.IM].

\bibitem{Lisanti09}
Lisanti and J.G. Wacker, arXiv: 0911.1997 [hep-ph].

\bibitem{Giometal11}
F. Mayet {\it et al}, Directional detection of dark matter, arXiv:1001.2983
  (astro-ph.IM).

\bibitem{VEROW06}
J.~Vergados and D.~Owen {\em Phys. Rev.} {\bf D 75} (2007) 043503.

\bibitem{JDV09}
J.~Vergados {\em Astronomical Journal} {\bf 137} (2009) 10. [arXiv:0811.0382
  (astro-ph)].

\bibitem{TETRVER06}
N.~Tetradis, J.~Vergados, and A.~Faessler {\em Phys. Rev.} {\bf D 75} (2007)
  023504.

\bibitem{VSH08}
J.~D. Vergados, S.~H. Hansen, and O.~Host {\em Phys. Rev. D} {\bf D 77} (2008)
  023509.

\bibitem{SIKIVI1}
P.~Sikivie {\em Phys. Rev. D} {\bf 60} (1999) 063501.

\bibitem{SIKIVI2}
P.~Sikivie {\em Phys. Lett. B} {\bf 432} (1998) 139.

\bibitem{Verg01}
J.~D. Vergados {\em Phys. Rev. D} {\bf 63} (2001) 06351.

\bibitem{Green}
A.~M. Green {\em Phys. Rev. D} {\bf 63} (2001) 103003.

\bibitem{Gelmini}
G.~Gelmini and P.~Gondolo {\em Phys. Rev. D} {\bf 64} (2001) 123504.

\bibitem{GREEN02}
A.~M. Green {\em Phys. Rev. D} {\bf 66} (2002) 083003.

\bibitem{KUHLEN10}
M.~Kuhlen et~al. {\em JCAP} {\bf 1002} (2010) 030.

\bibitem{LAWW11}
M.~Lisanti, L.~E. Strigari, J.~G. Wacker, and R.~H. Wechsler {\em Phys. Rev. D}
  {\bf 83} (2011) 023519.

\bibitem{SBWMZ08}
K.~Stewart, J.~Bullock, R.~Wechsler, A.~Maller, and A.~Zenter {\em Astrophys.
  J.} {\bf 683} (2008) 597.

\bibitem{PKB09}
C.~Purcell, S.~Kazantzidis, and J.~Bullock {\em Ap. J. Lett.} {\bf 694} (2009)
  L98.

\bibitem{LisSper11}
M. Lisanti and D.N.Spergel,Dark Matter Debris Flows in the Milky Way,
  arXiv:1105.4166 (astro-ph.CO).

\bibitem{streams11}
A. Natarajan, C. Savage and Katherine Freese, arXiv:1109.0014 (astro-ph.CO) (to
  appear in Phys. Rev. D).

\bibitem{spergel12}
M. Kuhlen, M. Lisanti and D.N. Speregel, Direct Detection of Dark Matter Debris
  Flows, arXiv:1202.0007 (astro-ph.GA).

\bibitem{VergF12}
J.~D. Vergados {\em Phys. Rev. D} {\bf 85} (2013) 123502. arXiv:1202.3105
  [hep-ph].

\bibitem{Li:2014era}
C.~Li, R.~H. Brandenberger, and Y.-K.~E. Cheung, {\it {Big Bounce Genesis}},
  \href{http://xxx.lanl.gov/abs/1403.5625}{{\tt arXiv:1403.5625}}.

\bibitem{Cheung:2014nxi}
Y.-K.~E. Cheung, J.~U. Kang, and C.~Li, {\it {Dark matter in a bouncing
  universe}},  {\em JCAP} {\bf 1411} (2014), no.~11 001,
  [\href{http://xxx.lanl.gov/abs/1408.4387}{{\tt arXiv:1408.4387}}].

\bibitem{Chung:1998ua}
D.~J. Chung, E.~W. Kolb, and A.~Riotto, {\it {Nonthermal supermassive dark
  matter}},  {\em Phys.Rev.Lett.} {\bf 81} (1998) 4048--4051,
  [\href{http://xxx.lanl.gov/abs/hep-ph/9805473}{{\tt hep-ph/9805473}}].

\bibitem{XENON10}
J. Angle {\it et al}, arXiv:1104.3088 [hep-ph].

\bibitem{XENON100.11}
E.~Aprile et~al. {\em Phys. Rev. Lett.} {\bf 107} (2011) 131302.
  arXiv:1104.2549v3 [astro-ph.CO].

\bibitem{CoGeNT11}
C.~Aalseth et~al. {\em Phys. Rev. Lett.} {\bf 106} (2011) 131301. CoGeNT
  collaboration arXiv:10002.4703 [astro-ph.CO].

\bibitem{DAMA1}
R.~Bernabei and Others {\em Eur. Phys. J. C} {\bf 56} (2008) 333. [DAMA
  Collaboration]; [arXiv:0804.2741 [astro-ph]].

\bibitem{DAMA11}
P.~Belli et~al. {\em Phys. Rev. D} {\bf 84} (2011) 05501. the DAMA
  collaboration, arXiv:1106.4667 [astro-ph.GA].

\bibitem{LUX11}
D.C. Malling {it et al}, arXiv:1110.0103((astro-ph.IM)).

\bibitem{CDMSII04}
D.~S.~A. et~al (CDMS~Collaboration) {\em Phys.Rev.Lett.} {\bf 93} (2004)
  211301.

\bibitem{CRESST}
The CRESST Experiment: Recent Results and Prospects, P.Di Stefano, {\it et al},
  arXiv:hep-ex/0011064; The CRESST Collaboration, talk presented at IBS -
  MultiDark Joint Focus Program, Daejeon, s. Korea, 10 – 21 October 2014.

\bibitem{PICASSO09}
S.~Archambault et~al. {\em Phys. Lett. B} {\bf 682} (2009) 185. collaboration
  PICASSO, arXiv:0907.0307 [astro-ex].

\bibitem{PICASSO11}
S.~Archambault et~al. {\em New J. Phys.} {\bf 13} (2011) 043006.
  arXiv:1011.4553 (physics.ins-det).

\bibitem{JDV12n}
J.~D. Vergados {\em Commun. Theor. Phys.} {\bf 57} (2012) 504. arXiv:1108.4768
  (hep-ph).

\bibitem{OikVerMou}
V.~Oikonomou, J.~D. Vergados, and C.~C. Moustakidis {\em Nuc. Phys.} {\bf B
  773} (2007) 19.

\bibitem{Fayet03}
C.~Boehm and P.~Fayet {\em Nucl.Phys. B} {\bf 683} (2004) 29.
  arXiv:hep-ph/0305261.

\bibitem{Ma06}
E.~Ma {\em Phys. Rev. D} {\bf 73} (2006) 077301. arXiv:hep-ph/0601225.

\bibitem{Li:2011nj}
C.~Li, L.~Wang, and Y.-K.~E. Cheung, {\it {Bound to bounce: A coupled
  scalar–tachyon model for a smooth bouncing/cyclic universe}},  {\em
  Phys.Dark Univ.} {\bf 3} (2014) 18--33,
  [\href{http://xxx.lanl.gov/abs/1101.0202}{{\tt arXiv:1101.0202}}].

\bibitem{Li:2013bha}
C.~Li and Y.-K.~E. Cheung, {\it {The scale invariant power spectrum of the
  primordial curvature perturbations from the coupled scalar tachyon bounce
  cosmos}},  {\em JCAP} {\bf 1407} (2014) 008,
  [\href{http://xxx.lanl.gov/abs/1401.0094}{{\tt arXiv:1401.0094}}].

\bibitem{Wands:1998yp}
D.~Wands, {\it {Duality invariance of cosmological perturbation spectra}},
  {\em Phys.Rev.} {\bf D60} (1999) 023507,
  [\href{http://xxx.lanl.gov/abs/gr-qc/9809062}{{\tt gr-qc/9809062}}].

\bibitem{Cai:2011ci}
Y.-F. Cai, R.~Brandenberger, and X.~Zhang, {\it {Preheating a bouncing
  universe}},  {\em Phys.Lett.} {\bf B703} (2011) 25--33,
  [\href{http://xxx.lanl.gov/abs/1105.4286}{{\tt arXiv:1105.4286}}].

\bibitem{Liu:2010fm}
J.~Liu, Y.-F. Cai, and H.~Li, {\it {Evidences for bouncing evolution before
  inflation in cosmological surveys}},  {\em J.Theor.Phys.} {\bf 1} (2012)
  1--10, [\href{http://xxx.lanl.gov/abs/1009.3372}{{\tt arXiv:1009.3372}}].

\bibitem{Cai:2011zx}
Y.-F. Cai, R.~Brandenberger, and X.~Zhang, {\it {The Matter Bounce Curvaton
  Scenario}},  {\em JCAP} {\bf 1103} (2011) 003,
  [\href{http://xxx.lanl.gov/abs/1101.0822}{{\tt arXiv:1101.0822}}].

\bibitem{Li:2014msi}
H.~Li, M.~Li, T.~Qiu, J.~Xia, Y.~Piao, et~al., {\it {What can we learn from the
  tension between PLANCK and BICEP2 data?}},  {\em Sci.China Phys.Mech.Astron.}
  {\bf 57} (2014) 1431--1441.

\bibitem{Quintin:2014oea}
J.~Quintin, Y.-F. Cai, and R.~H. Brandenberger, {\it {Matter Creation in a
  Nonsingular Bouncing Cosmology}},
  \href{http://xxx.lanl.gov/abs/1406.6049}{{\tt arXiv:1406.6049}}.

\bibitem{Wan:2014fra}
Y.~Wan, S.~Li, M.~Li, T.~Qiu, Y.~Cai, et~al., {\it {Single field inflation with
  modulated potential in light of the Planck and BICEP2}},
  \href{http://xxx.lanl.gov/abs/1405.2784}{{\tt arXiv:1405.2784}}.

\bibitem{Cai:2014bea}
Y.-F. Cai, {\it {Exploring Bouncing Cosmologies with Cosmological Surveys}},
  {\em Sci.China Phys.Mech.Astron.} {\bf 57} (2014) 1414--1430,
  [\href{http://xxx.lanl.gov/abs/1405.1369}{{\tt arXiv:1405.1369}}].

\bibitem{Liu:2014tda}
Z.-G. Liu, H.~Li, and Y.-S. Piao, {\it {Pre-inflationary genesis with CMB
  B-mode polarization}},  \href{http://xxx.lanl.gov/abs/1405.1188}{{\tt
  arXiv:1405.1188}}.

\bibitem{Li:2014qwa}
M.~Li, {\it {Generating scale-invariant tensor perturbations in the
  non-inflationary universe}},  \href{http://xxx.lanl.gov/abs/1405.0211}{{\tt
  arXiv:1405.0211}}.

\bibitem{Cai:2014hja}
Y.-F. Cai and Y.~Wang, {\it {Testing quantum gravity effects with latest CMB
  observations}},  {\em Phys.Lett.} {\bf B735} (2014) 108--111,
  [\href{http://xxx.lanl.gov/abs/1404.6672}{{\tt arXiv:1404.6672}}].

\bibitem{Cai:2014xxa}
Y.-F. Cai, J.~Quintin, E.~N. Saridakis, and E.~Wilson-Ewing, {\it {Nonsingular
  bouncing cosmologies in light of BICEP2}},
  \href{http://xxx.lanl.gov/abs/1404.4364}{{\tt arXiv:1404.4364}}.

\bibitem{Hu:2014aua}
B.~Hu, J.-W. Hu, Z.-K. Guo, and R.-G. Cai, {\it {Reconstruction of the
  primordial power spectra with Planck and BICEP2}},
  \href{http://xxx.lanl.gov/abs/1404.3690}{{\tt arXiv:1404.3690}}.

\bibitem{Li:2014cka}
H.~Li, J.-Q. Xia, and X.~Zhang, {\it {Global fitting analysis on cosmological
  models after BICEP2}},  \href{http://xxx.lanl.gov/abs/1404.0238}{{\tt
  arXiv:1404.0238}}.

\bibitem{Xia:2014tda}
J.-Q. Xia, Y.-F. Cai, H.~Li, and X.~Zhang, {\it {Evidence for bouncing
  evolution before inflation after BICEP2}},  {\em Phys.Rev.Lett.} {\bf 112}
  (2014) 251301, [\href{http://xxx.lanl.gov/abs/1403.7623}{{\tt
  arXiv:1403.7623}}].

\bibitem{Cai:2014zga}
Y.-F. Cai and E.~Wilson-Ewing, {\it {Non-singular bounce scenarios in loop
  quantum cosmology and the effective field description}},  {\em JCAP} {\bf
  1403} (2014) 026, [\href{http://xxx.lanl.gov/abs/1402.3009}{{\tt
  arXiv:1402.3009}}].

\bibitem{NoBer08}
M.~Novello and S.~P. Bergliaffa {\em Phys. Rep.} {\bf 463} (2008) 127. arXiv:
  0802.1634 [astro-ph].

\bibitem{Branden12}
R. H. Brandenberger, (2012), arXiv:1206.4196 [astro-ph.CO].

\bibitem{ZeeScal85}
V.~Silveira and A.~Zee {\em Phys. Lett. B} {\bf 161} (1985) 136.

\bibitem{ZeeScal01}
D.~Holz and A.~Zee {\em Phys. Lett. B} {\bf 517} (201) 239.

\bibitem{BentoRos01}
M.~Bento, O.~Berolami, and R.~Rosefeld {\em Phys. lett. B} {\bf 518} (2001)
  276.

\bibitem{BentoBero00}
M.~Bento, O.~Berolami, R.~Rosefeld, and L.~Teodoro {\em Phys. Rev. D} {\bf 62}
  (2000) 041302.

\bibitem{Chen}
T. P. Cheng, {\it Phys. Rev. D} {\bf 38}, 2869 (1988); H-Y. Cheng, {\it Phys.
  Lett. B} {\bf 219}, 347 (1989).

\bibitem{Dree00}
A. Djouadi and M. K. Drees, {\it Phys. Lett. B} {\bf 484}, 183 (2000); S.
  Dawson, {\it Nucl. Phys. B} {\bf 359}, 283 (1991); M. Spira {it et al}, {\it
  Nucl. Phys.} {\bf B453}, 17 (1995).

\bibitem{JDV06}
J. D. Vergados, On The Direct Detection of Dark Matter- Exploring all the
  signatures of the neutralino-nucleus interaction, hep-ph/0601064.

\bibitem{XENON10012}
E.~Aprile et~al. {\em Phys. Rev. Lett.} {\bf 109} (2012) 181301. [XENON100
  Collaboration]; arXiv: 1207.5988 (astro-ph.Co).

\bibitem{MVE05}
C.~C. Moustakidis, J.~D. Vergados, and H.~Ejiri {\em Nucl. Phys. B} {\bf 727}
  (2005) 406. hep-ph/0507123.

\bibitem{XENON14}
E.~Aprile et~al. {\em J. Phys. G: Nucl. Part. Phys.} {\bf 41} (2014) 035201.
  [XENON100 Collaboration]; arXiv:1311.1088 (astro-ph.IM).

\bibitem{JDVDF12}
J.~D. Vergados {\em Phys. Rev. D} {\bf 85} (2012) 123502. arXiv:1202.3105
  (hep-ph).

\bibitem{JDV06a}
J.~D. Vergados {\em Lect. Notes Phys.} {\bf 720} (2007) 69. hep-ph/0601064.

\bibitem{VerMou11}
J.~D. Vergados and C.~C. Moustakidis {\em Eur. J. Phys.} {\bf 9(3)} (2011) 628.
  arXiv:0912.3121 [astro-ph.CO].

\bibitem{DIVA00}
P.~C. Divari, T.~S. Kosmas, J.~D. Vergados, and L.~D. Skouras {\em Phys. Rev.
  C} {\bf 61} (2000) 054612--1.

\bibitem{Ress}
M. T. Ressell {\it et al.}, {\it Phys. Rev. D} {\bf 48}, 5519 (1993); M.T.
  Ressell and D. J. Dean, Phys. Rev. C {\bf 56}, 535 (1997).

\bibitem{PICASSO12}
The PICASSO collaboration: S. Archambault {\it et al}, Private Communication.

\bibitem{DAMAEPJ13}
R.~Bernabei et~al. {\em Eur. Phys. J.} {\bf C 73} (2013) 2648. [DAMA/LIBRA
  phase 1]; arXiv:1308.5109 (astro-GA.).

\end{thebibliography}\endgroup
\providecommand{\href}[2]{#2}\begingroup\raggedright
\endgroup

\end{document}